\documentclass[12pt]{article}
\usepackage[font={footnotesize}]{caption}
\usepackage{amssymb}
\usepackage{amsmath}
\usepackage{xcolor}

\numberwithin{equation}{section}

\usepackage{graphicx}

\newtheorem{theorem}{Theorem}[section]

\newtheorem{remark}[theorem]{Remark}

\newcommand{\sgn}{{\rm sgn}}

\def\d{{\mathrm d}}

 \def\ou{{{\overline u}}}

\def\rhor{{\rho_r}}

\def\barr{\hbox{I\hskip -.5ex R}}

\def\sbarr{\hbox{{\scriptsize I}\hskip -.5ex {\scriptsize R}}}
\usepackage{cancel}

\newcommand{\bs}[1]{{\boldsymbol{#1}}}

\newcommand{\dsl}[1]{{\displaystyle{#1}}}

\begin{document}
\title{Hydrodynamic models and confinement effects \\ by horizontal boundaries}

\author{R.\ Camassa$^1$, G.\ Falqui$^2$, G.\ Ortenzi$^2$,  M.\ Pedroni$^3$ and C. Thomson$^1$\\
\smallskip\\
{\small $^1$University of North Carolina at Chapel Hill, Carolina Center for Interdisciplinary}\\ {\small Applied Mathematics,
Department of Mathematics, Chapel Hill, NC 27599, USA }
\\
{\small $^2$Dipartimento di Matematica e Applicazioni, Universit\`a di Milano-Bicocca, 
Milano, Italy}
\\
{\small $^3$Dipartimento di Ingegneria Gestionale, dell'Informazione e della Produzione,}\\ {\small Universit\`a di Bergamo,
Dalmine (BG),  Italy} }

\maketitle
\abstract{Confinement effects by rigid boundaries in the dynamics of ideal fluids are considered from the perspective of long-wave models and their parent Euler systems, with the focus on the consequences of establishing contacts of material surfaces with the confining boundaries. When contact happens, we show that the model evolution  can lead to the dependent variables developing singularities in finite time. The conditions and the nature of these singularities are illustrated in several cases, progressing from a single layer homogeneous fluid with a constant pressure free surface and flat bottom, to the case of a two-fluid system contained  between two horizontal rigid plates, and finally, through numerical simulations, to the full Euler stratified system. These demonstrate the qualitative and quantitative predictions of the models within a set of examples chosen to illustrate the theoretical results. 

}
\section{Introduction
}

In investigations of the behavior of stratified fluids under gravity with ideal fluid models, it is well known that setups for which constant density level 
sets (isopycnals) intersect non-vertical rigid boundaries can cause several difficulties that need to be addressed, either by careful considerations of the 
mathematical properties of the governing equations or by modifying them to include additional physical effects (see e.g.,~\cite{CCFOP5,CCFOP6}, or Carrier \& Greenspan~\cite{Carrier} and Phillips~\cite{Phillips} for the case of sloped boundaries). With this in mind, under these circumstances, it seems important to assess whether the simplicity 
afforded by the governing equations for ideal fluids can continue to be physically relevant, or whether the assumptions underlying the foundation of these equations need to be modified, and more complex systems become essential.

\begin{figure}[h]
\centering
{\includegraphics[width=10cm]{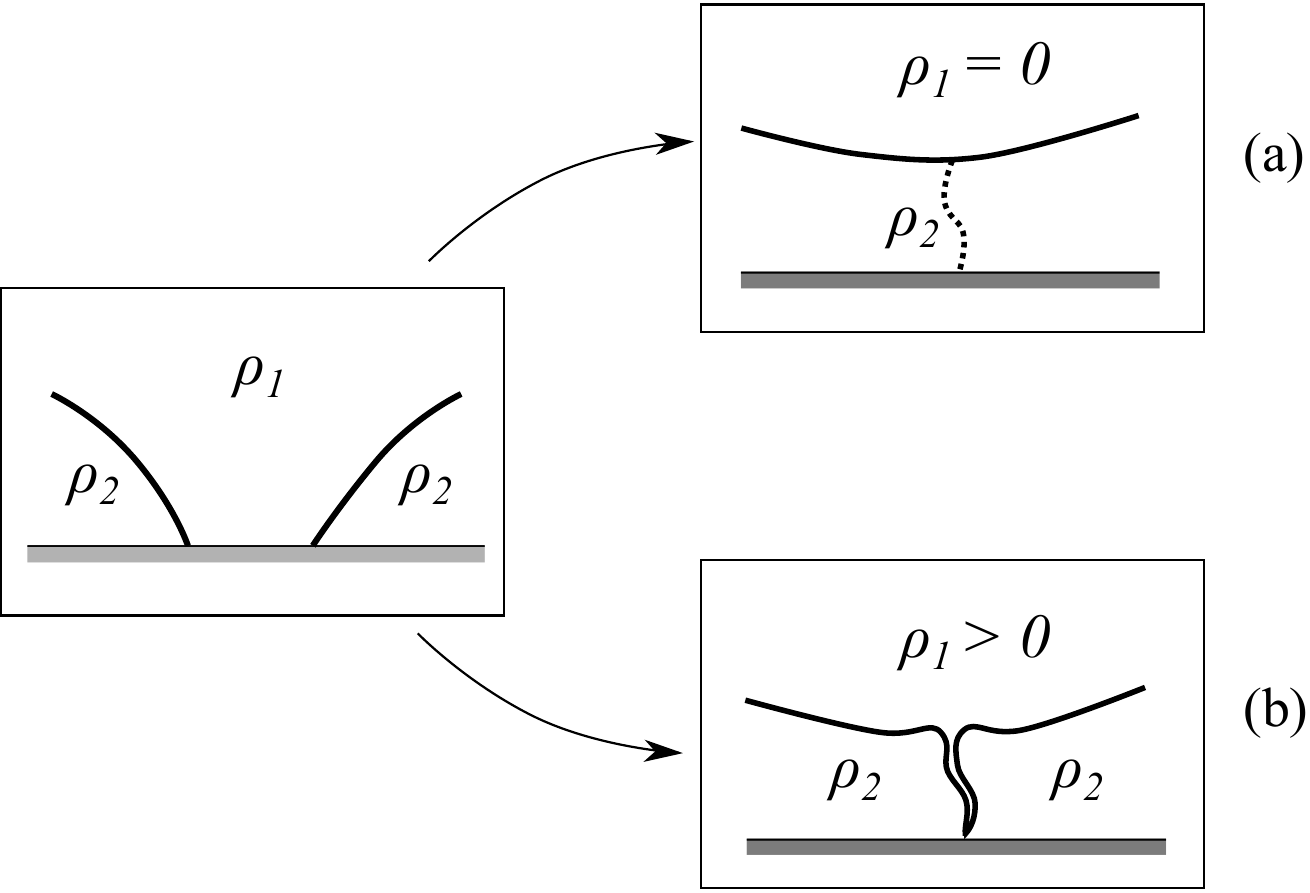}}
\caption{Schematics of the air-water vs. two-layer mapping of the fluid domain's boundary for the ``hole filling" setup. In the air-water case, a portion of 
the boundary of the fluid domain can be viewed as folded back onto itself (a) in finite time and be maintained within the fluid. For the case of two-layer fluids, 
if fold-back contact between smooth portions of the interface  does not occur in finite time, a decreasing but non-zero measure stem (b) connecting the upper layer 
fluid to the bottom floor can persist as time increases.}
\label{aw-vs-2layer-fig}
\end{figure}

To this end, in previous work~\cite{CFOP-proc,CFO}, we have focussed on the consequences that isopycnals 
intersecting horizontal confining surfaces can have on the structure of the incompressible, variable density Euler system, such as its conservation laws and  
Hamiltonian formalism. In this work, we want to shift our attention to issues that concern more specifically the time evolution of data in this class. 
Much of the difficulties here  stem from the mathematical assumptions underlying the foundation of the governing equations, specifically the property that 
``{\it fluid particles on a wall stay on the wall}\," (\cite{Childress}, p.22). When this assumption is accepted, the shrinking or 
opening of ``islands" of density variations along the confining boundaries becomes a rather subtle issue, even in limiting cases such as layered fluids separated 
by sharp interfaces (see figure~\ref{aw-vs-2layer-fig} for a schematic). The latter setup in turn admits the well-known water-wave, or air-water, limit, in which the density of the lighter fluid, air, is assumed to be negligible 
and enters the dynamics only as means of maintaining constant pressure on the water's free surface.  

There are some similarities between the ``splash" and ``splat" singularities studied 
in~\cite{FeffermanCastro,Fefferman1,Fefferman2} (or their backward-time versions) and the focus of this work on the interaction of a fluid's density isoline with rigid boundaries. For instance, the splash example provided in~\cite{Fefferman1}, due to the reflectional symmetry across a vertical line, can be viewed as interaction of the free surface with a vertical wall. In this respect, it is worth mentioning the exact 
free-surface solutions for an Euler fluid evolving under gravity pioneered by John~\cite{John}, by assuming a simple quadratic dependence on the spatial variables of 
the velocity potential. This leads to parabolic interface shapes that collapse 
self-similarly onto their axis for certain time limits. It is remarkable that in our investigation we find a similar idea and behavior playing a prominent role  for models of interfaces. 
In fact, rather than with the full Euler system we will be mostly concerned in this work  with models that can extract the essential elements of confining-boundary/interface interactions. While a direct investigation of Euler equations could in principle be extended to include boundary 
issues for similar setups, much insight can be gained by the study of such reduced models that incorporate boundary conditions and physically relevant properties 
either exactly or approximately, e.g., through layer averaging of the horizontal velocity field. 
Such is the case for the Airy (shallow water) system for the water wave problem or its strongly nonlinear counterpart for the layered 
stratifications~\cite{tabak-2004,EP,CCFOP1,CCFOP2}. 

Of course, as already remarked earlier, a description of fluids neglecting phenomena such as viscosity, diffusivity, and surface tension is realistically limited from a physical perspective. 
However, under suitable assumptions on the magnitude of the corresponding parameters, and on the scales of initial data and evolution times, omitting 
these effects can be expected to capture correctly the qualitative behavior of the fluid, and in special cases even to reproduce it quantitatively.  
Similar considerations can be made within classes of simplified models, for instance with regard to whether neglecting dispersion in wave propagation can be justified. For example, in many setups the evolution supported by the Airy (nondispersive) shallow-water system can be viewed as providing 
a ``skeleton" for the relevant physics, shedding light on where (and how) singularity generation in the form of  shocks may arise, which in turn would signal the need for additional physical mechanisms 
to continue the evolution beyond limiting events; such  may be the situation encountered in the collapsing of dry spots or of density islands. In our opinion, Stoker's words (\cite{Sto}, page XIX) about the relevance of approximate models such as the shallow water equations ``for clarifying some of
the mysteries concerning the dynamical causes" still ring true when applied to the present context.

This paper is essentially organized in two parts. The first concentrates on the dynamics of a single layer fluid (under gravity) and on the interaction of its free surface with a flat bottom boundary. The second part focuses mainly on the case of stratified fluids filling a two-dimensional domain between two flat plates. Both parts present a combination of analytical and numerical approaches to the dynamical questions arising in the study of the time evolution for certain classes of initial conditions.

Specifically, after a brief review in~\S\ref{Euler-2D} of the mathematical and physical setting of a stratified, incompressible Euler fluid between two horizontal boundaries, we state the fundamental kinematic implications of the appropriate no-flux boundary conditions. The peculiar features of shock formation for the Airy's model when the free surface connects to a ``dry spot"  at the bottom of fluid layer is investigated 
in~\S\ref{sezione:Airy}  by means of several families of initial data. We first focus on the simple wave case in~\S\ref{sezione:simpler-catastrophe}.  With the aim of extending this simple wave study to more general setups, 
we then examine, in~\S\ref{sezione:parab-solut}, a class of initial data leading to exact solutions  that are the counterpart of the well-known self-similar solution of the Hopf equation. These are also a convenient starting point for a perturbative approach to the study of contacts with the upper lid of the lower layer fluid in the (nondispersive) two-fluid system when the upper fluid density is small.

Next, in 
\S\ref{sezione:two-fluid}, we look at  how the kinematic properties for the full Euler system transfer to generic quasi-linear systems whose structure is representative of long-wave  models, such as the general two-layer systems.  
The study of these properties is carried out  
in particular by focussing on their behavior near the hyperbolic-elliptic transition curves in the hodograph plane. 

In the two-fluid context of~\S\ref{sezione:two-fluid}, general results 
about the behavior of contact points with hyperbolic-elliptic transition curves are illustrated  by the persistence of an interface contact to a rigid boundary, for as long as singularities do not develop in the system's evolution. 

The general features of our study are then illustrated with a suite of numerical simulations in our last section,~\S\ref{sezione:numerics}. The simulations of both the models, when closed form solutions are not available, and the parent, stratified Euler equations focus on a class of initial conditions and time scales that isolate the consequences of contact phenomena for density isolines. 

\section{Physical setup and equations of motion}
\label{Euler-2D}
 We consider the Euler equations for an ideal incompressible and inhomogeneous 
 fluid, subject to gravity, in two dimensions: 
 \begin{equation}
 \bs{v}_t+\bs{v}\cdot\nabla\bs{v}=-\dsl{\frac{\nabla p}{\rho}}-g\, \bs{k}\, ,  \qquad
 \nabla \cdot \bs{v}=0 \, , \qquad \rho_t+\bs{v}\cdot\nabla\rho=0\, .  
 \label{Eceq}
 \end{equation}
 Here $\bs{v}=(u,w)$ is the velocity field, $\rho$ and $p$ are the density and pressure fields, 
 respectively, $g$ is the constant gravity acceleration, $\{\bs{i},\bs{k}\}$ are the unit directional vectors for, respectively, horizontal  $x$ and  vertical $z$
 coordinates, and all physical variables depend on spatial coordinates $(x,z)$ and time $t$. 
The fluid domains we shall consider here are slabs in the $(x,z)$ plane rigidly confined by horizontal plates of infinite extent
located at $z= 
0$ and $z= 
h$. 
Our study will focus on two-dimensional 
dynamics, though it can be generalized to  fully three-dimensional cases. 
The Euler equations (\ref{Eceq}) are then supplemented by initial data $\bs{v}(x,z,0)=\bs{v}_0(x,z)$,
$\rho(x,z,0)=\rho_0(x,z)$, for some functions $(\bs{v}_0,\rho_0)$, and by
boundary conditions
\begin{equation}\label{boucon}
{\partial p \over \partial z} \to -g \rho, \quad \text{and} \quad \rho\to \rho_b(z), \quad \text{ for } |x|\to\infty, \qquad 
w(x, 0)=w(x,h)=0 \quad \text{for all $x$}\,  .
\end{equation}
The first two conditions in~(\ref{boucon}) impose that  the fluid at the far ends of the channel be in hydrostatic  equilibrium.

\begin{figure}[h]
\centering
{\includegraphics[width=10cm]{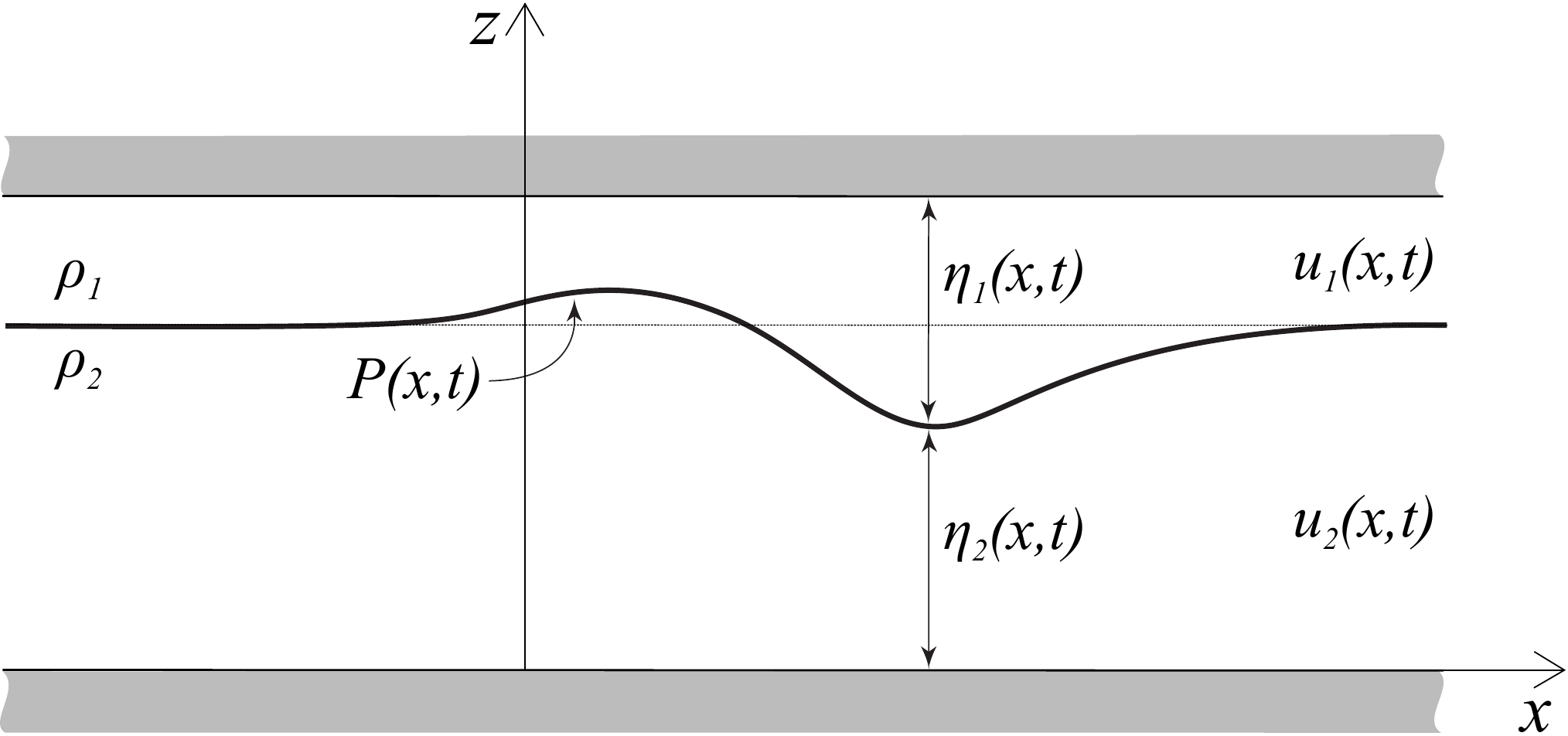}}
\caption{Schematic of two-layer fluid in a horizontal channel.}
\label{two_layers-fluid-fig}
\end{figure}

A simplification of system~(\ref{Eceq}) which retains the essential properties of stratification can be obtained by considering a system  of two fluids of homogeneous densities whose interface is described by a smooth function $z=\eta(x,t)=\eta_2(x,t)$, see figure \ref{two_layers-fluid-fig}.  As well known, taking layer-averages in this case offers a convenient starting point to derive simplified models. This results in the (non-closed) system 
\begin{equation}
\label{2layer}
 \begin{split}
 & {\eta_i}_t+(\ou_i\eta_i)_x=0 \, , \qquad i=1,2 \\ &
 {\ou_1}_t+{\ou_1}{\ou_1}_x -g {\eta_1}_x + \frac{P_x}{\rho_1} + D_1 =0 \, , \\
 &  {\ou_2}_t+{\ou_2}{\ou_2}_x +g {\eta_2}_x + \frac{P_x}{\rho_2} + D_2 =0 \, ,
 \\ &\eta_1+\eta_2=h, \qquad (\eta_1 \ou_1 + \eta_2 \ou_2)_x=0 \, . 
 \end{split}
\end{equation}
Here $\ou_{1,2}$ are the layer-mean velocities, defined as
\[
\ou_1(x,t)=\frac{1}{\eta_1(x,t)}\int_{h-\eta_1(x,t)}^{h} u_1(x,z,t)\d z, \quad \ou_2(x,t)=\frac1{\eta_2(x,t)}\int_0^{\eta_2(x,t)} u_2(x,z,t) \d z\, ,
\]
 and $P(x,t)$ is the interfacial pressure. 
The terms $D_1$, $D_2$ at the right hand side of system~(\ref{2layer}) are 
\begin{equation} 
D_i= \frac{1}{3 \eta_i} \partial_x [\eta_i^3 ({\ou_i}_{xt} +{\ou_i}{\ou_i}_{xx}-({\ou_i}_x)^2)] + \dots, \qquad{i=1,2}\, , 
\label{disperterm}
\end{equation}
where dots represent terms with nonlocal dependence on the averaged velocities. These terms collect the non-hydrostatic correction to the pressure field, and
make the evolution of system~(\ref{2layer}) dispersive, as can be readily seen by linearizing the equations around a flat interface with just the explicit terms in~(\ref{disperterm}).
When an asymptotic expansion  based on the long-wave assumption, e.g., 
\begin{equation}
\varepsilon \equiv \max_{(x,t)}[\eta_i/L] \ll 1 \, , \qquad i=1,2 \,  
\label{vareps}
\end{equation}
(with $L$ denoting a typical horizontal scale of the initial data), is carried out, expressions~(\ref{disperterm}) explicitly define the leading order dispersive terms  in the small parameter $\varepsilon$; truncating  equations~(\ref{2layer}) at this order results in the strongly nonlinear system studied in~\cite{miyata,CC99}, which is a two-layer version of the so-called Serre-Green-Naghdi model (see, e.g., \cite{serre,su,green,lannes}) for a single layer with a free surface.

The constraints in the last line of (\ref{2layer}) can be used to reduce this model to a system of only two unknowns (see, e.g.,~\cite{CFO}), which simplifies considerably if the dispersive terms $D_1$ and $D_2$ are neglected altogether. These non-dispersive equations of motion are most naturally expressed in terms of the thickness 
$\eta(x,t)\equiv \eta_2(x,t)$ of the lower layer and the layer-averaged weighted vorticity
\begin{equation}
\sigma(x,t) \equiv \frac1{h}\left(\rho_2 \ou_2-\rho_1 \ou_1\right)\,. 
\label{sigma+def}
\end{equation}
In these variables the resulting equations read (see e.g.,~\cite{CFO}):
\begin{equation}
\begin{array}{l}
\dsl{ \eta_t= }\, \dsl{-\partial_x\left( {
{ h \eta (\eta-h) \sigma \over 
 (\rho_2-\rho_1) \eta -\rho_2 h }} \right)} \, , 
 \vspace*{0.3cm}
 \\
\dsl{{\sigma}_t}=
\dsl{-\partial_x\left({h\over 2}\, 
{\rho_2(h-\eta)^2 -\rho_1 \eta^2
\over 
( (\rho_2-\rho_1) \eta -\rho_2 h)^2}\, \sigma^2
+\frac{g(\rho_2-\rho_1)\eta}{h} \right)} \, .
\end{array}
\label{e-m-full}
\end{equation} 
When the limit of vanishing upper fluid density is further considered, together with the nondispersive limit obtained by dropping the dispersive terms $D_2$, system~(\ref{2layer}) reduces to the well known shallow water Airy's model, as $\rho_1=0$ implies $P_x=0$ (consistently with the condition of constant pressure at the free surface in the classical water wave problem) and the lower layer $i=2$ equations read simply 
\begin{equation}
\begin{array}{l}
\eta_t=-\partial_x(\eta \, \ou_2)  \, , 
\vspace*{0.3cm}
\\
{\ou_2}_t=-\partial_x \left( {1\over 2}\, {\ou_2}^2 +g \eta \right) \, . 
\end{array}
\label{c_airy}
\end{equation}

In general, the nondispersive approximations above can be expected to lose accuracy as time progresses when steep gradients in the dependent variables become relevant. As we shall see, for the classes of initial data and time scales we study the relative effects may typically be considered negligible when tested against numerical solutions of the parent Euler equations. An exception occurs when quantitatively accurate predictions of thresholds for critical phenomena become necessary, as seen later in~\S\ref{crthgt}.

Having defined the physical setup and motion equations, as well as their models in the appropriate limits, we next look at what information can be extracted in general from the parent system by simple kinematic considerations.  

\subsection{Kinematic effects of the boundary-interface contact}
We consider the full Euler equations in a two-layer setup. The case of free surface under constant pressure can be obtained as the appropriate limiting case from this, as $\rho_1\to 0$. Suppose that the interface, assumed to be a graph of class $C^1$,  initially touches the bottom at one point $x_0$, that is, $\eta(x_0,0)=0$; then there is a moving contact point $x=x_\kappa(t)$, 
with $x_\kappa(0)=x_0$, at the bottom such that $\eta(x_\kappa(t),t)=0$ for all times $t$. 
It is not difficult to see that 
$x_\kappa(t)$  
is the solution of the Cauchy problem
\begin{equation}
\label{cauchy-pb}
{\dot x}(t)=u_2(x(t),\eta(x(t),t),t)\, , \qquad 
x(0)=x_0\, , 
\end{equation}
where $u_2$ is the horizontal component of the velocity field in the lower layer.
Indeed, let $Z_1(t)=\eta(x_\kappa(t),t)$. Then, using 
the interface velocity condition
$\eta_t+\eta_x u_2=w_2$, we obtain
\begin{equation}
\begin{split}
{\dot Z}_1(t)&=\eta_x(x_\kappa(t),t)\dot x_\kappa(t)+\eta_t(x_\kappa(t),t)=\eta_x(x_\kappa(t),t)u_2(x_\kappa(t),\eta(x_\kappa(t),t),t)+\eta_t(x_\kappa(t),t)\\
&=w_2(x_\kappa(t),\eta(x_\kappa(t),t),t)=w_2(x_\kappa(t),Z_1(t),t).
\end{split}
\end{equation}
Hence $Z_1(t)$ is a solution of the Cauchy problem
\begin{equation}
{\dot z}(t)=w_2(x_\kappa(t),z(t),t)\, , \qquad
z(0)=0 \, . 
\end{equation}
Due to the boundary condition $w_2(x,0,t)=0$, the identically zero function is also a solution of the same Cauchy problem. If $(t,z)\mapsto w_2(x_\kappa(t),z,t)$ 
is a Lipschitz function with respect to 
$z$ and uniformly continuous with respect to $t$, then the solution is unique and we can conclude that $Z_1(t)=0$ for all $t$, as claimed. 

Suppose now that $\eta(x_0,0)=0$ for all $x_0$ in an interval $I_0$. Then we can apply the previous result to the end points of the interval 
(respectively for their left and right neighborhoods) as this evolves with time into $I_t$, and conclude that the end points never detach from the boundary. 
Hence, no lower layer fluid can come into contact with points within the interval $I_t$ as long as the interface remains a regular, single-valued function of the form $z=\eta(x,t)$. A similar argument holds for the contact of lower, denser fluid with the top boundary plate. In particular, when fluid velocities are initially zero in this case, 
contact is maintained during the slumping that can be expected in the evolution of the interface for as long as this remains a regular graph. Hence, to break the contact a different mechanism, one that relies on loss of regularity, is needed to detach the lower fluid from the top plate. 

In order to shed some light on these issues, we now turn our attention to {\it models} of Euler fluid dynamics, which under appropriate assumptions can isolate relevant dynamics in simplified settings more amenable to analysis.

\section{A single fluid over a flat surface: Airy's model} 
\label{sezione:Airy}

We now  consider the simplest model of 
a homogeneous fluid with free surface under a constant pressure, with a horizontal flat bottom placed at $z=0$. 
As summarized in~\S~\ref{Euler-2D}, such a model is provided by the long-wave approximation 
at leading order, which yields the classic Airy system~(\ref{c_airy}). Written in non-dimensional form by the rescaling 
\begin{equation}
 \eta\to\eta/h \, , \qquad \ou_2\to {\ou_2  \over \sqrt{g h}}\,, \qquad x\to {x \over h} \, , \qquad t\to t \, \sqrt{g \over h}\,, 
\label{rescairy}
\end{equation}
system~(\ref{c_airy}) becomes 
\begin{equation}
\eta_t+(\eta \, u)_x=0 \, ,
 \qquad u_t+u\, u_x+\eta_x=0 \, ,
 \label{airyeq}
\end{equation}
where we have abused notation a little by not differentiating between scaled and unscaled variables; further, unless explicitly indicated, by $u=u(x,t)$ we will denote the layer averaged-horizontal velocity instead of the corresponding variable in the Euler equations, 
and from here on we will drop bars and layer subscripts, restoring the full notation of~\S~\ref{Euler-2D} whenever necessary to avoid confusion. 
(Note that in this notation the gravity acceleration is normalized to have unit magnitude, which coincides with the convention of~\cite{Sto}, section~2.3).

Our first aim is to show that the persistence of contact points observed 
above for a two-fluid system governed by the full Euler equations continues to hold for this greatly simplified  model, as long as the regularity of the initial data is maintained by its evolution. 
We take $(\eta(x,t),u(x,t))$ to be a solution of 
(\ref{airyeq}) such that $\eta(x_0,0)=0$. This obviously means that initially there is (at least) one dry point on the floor. 
Our claim is that the bottom  will remain dry for some time and will not get immediately wet; more precisely, once again we have to find a (moving) point $x_\kappa(t)$ such that  
$\eta(x_\kappa(t),t)=0$ for all $t$ (for as long as this function is sufficiently regular before shocks form --- this caveat will often be left implicit in the following). 
It is easy to show that the function $x_\kappa(t)$ we are looking for is the solution of the Cauchy problem $\dot x(t)=u(x(t),t)$, $x(0)=x_0$. Indeed, 
\begin{equation}
\frac{\d}{\d t} \eta(x(t),t) = \eta_x\frac{\d x}{\d t} +\eta_t = \eta_x u -(\eta u)_x=-u_x \eta,
\end{equation}
so that the function $\eta_\kappa(t)\equiv \eta(x_\kappa(t),t)$ is a solution of the Cauchy problem 
\begin{equation}
\label{pC-N1}
{\dot \eta_\kappa}(t)=-\upsilon_\kappa(t)\eta_\kappa(t)\, , \qquad
\eta_k(0)=0 \, , 
\end{equation}
where the velocity gradient at the contact point, $\upsilon_\kappa(t)\equiv u_x(x_\kappa(t),t)$, can  be viewed as a given function of time. Loss of regularity in finite times of the function $u$ typically occurs in the form of a shock, and hence leads to unbounded $\upsilon_\kappa$'s. Due to uniqueness, 
we can conclude that $\eta_\kappa(t)=0$ for all times before a shock forms. If the 
interface is tangent to the bottom somewhere, then it remains tangent, even if the tangency point can move along the bottom. 
Indeed, if we assume  that initially $\eta_x(x_0,0)=0$, we can introduce the function $\gamma_\kappa(t)\equiv\eta_x(x_\kappa(t),t)$, the slope of the surface at the contact point,  and easily show that $(\eta_\kappa(t),\gamma_\kappa(t))$ is a solution of a Cauchy 
problem for a homogeneous first-order linear system, with initial data $(\eta_\kappa(0),\gamma_\kappa(0))=(0,0)$. Again from uniqueness, we can conclude that $\eta(x_\kappa(t),t)=0$ and $\eta_x(x_\kappa(t),t)=0$
for all times before the possible shock.

We remark that
\begin{equation}
\label{x(t)speed}
\frac{\d}{\d t}u(x_\kappa(t),t)=-\eta_x(x_\kappa(t),t)=0 \, ,
\end{equation}
meaning that the velocity of the dry point is constant. Hence, with the additional hypothesis $u(x_0,0)=0$ we see that $x_\kappa(t)=x_0$, that is, a dry point does not move if the  initial velocity is zero at that point. (This fact was already noticed in \cite{MT}.) In particular, if the velocity field of the fluid initially vanishes, we have two 
consequences: (i) suppose that there is only one dry point $x_0$, then this point does not move (even if the fluid configuration is not symmetric with respect to the line $x=x_0$, which would imply this as a consequence of preservation of symmetry); (ii) 
suppose that there is a whole interval of dry points, then these points will remain dry at all times so long as a shock does not develop. 
We remark that working with the ``sound speed'' variable (see e.g.,~\cite{Sto}, Chapter 10) $c_s=\sqrt{\eta}$, both consequences can be seen as corollaries of the vanishing of this speed at the contact point.
(As will be seen in section \ref{sezione:parab-solut}, consequence (ii) is in fact false if the interface does not touch the bottom with sufficient regularity: an example is given by the initial data $\eta(x,0)=(1-x^2)\chi_I(x)$, $u(x,0)=0$,
where $\chi$ is the characteristic function of the interval $I=(-1,1)$.)
Viceversa, if the interface does not touch the bottom at time $t=0$, then the uniqueness  hypothesis for solutions of equation~(\ref{pC-N1}) in backward times shows that  
the interface will never reach the bottom, as long as only classical solutions of the Airy system are considered (an explicit example will be shown in section \ref{sezione:parab-solut}). 
A qualitative description of the role of shocks can be also given: 
suppose a smooth interface has a small-amplitude minimum at a given time $t_0$, say 
$\eta(x_0,t_0)=\epsilon\ll 1$ and $\eta_x(x_0,t_0)=0$. Then 
\begin{equation}
  \eta_t(x_0,t_0) = -(u_x \eta +u\eta_x)\Big{|}_{x=x_0,t=t_0}=-u_x(x_0,t_0)\epsilon= O(\epsilon) 
\end{equation}
for as long as the velocity gradient $u_x(x_0,t_0)$ remains bounded. Thus, the argument would break down if a large velocity gradient were to develop; this would imply a fast change of the interface which could eventually 
make contact of the interface with the bottom possible.

It is worthwhile to add a comment on the curve $t\mapsto (x_\kappa(t),t)$. Since $\eta(x_\kappa(t),t)=0$, a posteriori we can say that it  belongs to both solution families (the characteristics) of
\begin{equation}
 \frac{\d x}{\d t} = \lambda_\pm = u \pm \sqrt{\eta} \, , \qquad x(0)=x_0\, , 
 \label{chars_1}
\end{equation}
where $\lambda_\pm=u \pm \sqrt{\eta}$ are the eigenvalues 
of the matrix 
\begin{equation}
\left(\begin{array}{cc} 
u & \eta\\
1 & u
\end{array}
\right)
\end{equation}
of the quasilinear system~(\ref{airyeq}). Thus, another interpretation of the curve $(x_\kappa(t),t)$ is that the point $(\eta(x_\kappa(t),t),u(x_\kappa(t),t))$ stays on the ``sonic" line $\eta=0$ 
where the eigenvalues coincide.

\subsection{Airy's system shocks for simple waves}
\label{sezione:simpler-catastrophe}
When the free surface  contacts the bottom, the process of shock formation for Airy's system~(\ref{airyeq}) solutions deserves a closer look: as we have seen,  wetting can only happen with loss of smoothness, i.e., the free surface cannot detach from the bottom before a shock forms. For a class of special solutions of system~(\ref{airyeq}), the so-called ``simple waves,"  we will prove 
that the vertical location of the shock position depends on the initial steepness of the interface near the bottom. 

As well known, 
along curves defined by~(\ref{chars_1}) the quantities $R_\pm$ (the Riemann invariants) 
\begin{equation}
R_\pm=u\pm2\sqrt{\eta}
\label{riemannpm}
\end{equation}
are 
constant; when such constant is independent of the initial condition $x_0 \in {\cal{I}}$ for some open interval $\cal{I}\subset \barr$, the corresponding solution $(\eta(x,t),u(x,t))$ is commonly referred to as  simple wave (see, e.g, ~\cite{Whitham}, p.164). 
Let us focus on the $R_-$ invariant,  
\begin{equation}
\label{simplewave}
R_-=u - 2 \sqrt{\eta}\, , 
\end{equation}
and let  $c$ denote its constant value. Thus, since $u(x,t)= 2 \sqrt{\eta(x,t)} + c$, the Cauchy problem for the Airy system reduces to
\begin{equation}
\label{SWNLSeqn}
\eta_t +(3 \sqrt{\eta}+c) \eta_x =0 \, , \qquad
\eta(x,0)=\eta_0(x)   \, . 
\end{equation}
We briefly review the shock formation in this case. The generic (hodograph) solution is 
\begin{equation}
 x-(3\sqrt{\eta}+c)t=F(\eta),
\end{equation}
where $F$ is the (local) inverse function of the initial condition $\eta_0$. 
The shock point $(x_c,t_c)$ is the lowest point, in the $(x,t)$ plane, of the shock curve 
\begin{equation}
 x-(3\sqrt{\eta}+c)t=F(\eta), \qquad -\frac{3}{2\sqrt{\eta}}t=F'(\eta),
\end{equation}
where the second equation comes from ${\d x}/{\d \eta} = 0$. Therefore, 
the parametric equations of the shock curve are
\begin{equation}
 \label{catas-curve}
\begin{array}{l}
  x(\eta)= F(\eta)-2\, \eta \, F'(\eta)-{\displaystyle \frac{2c}3} \sqrt{\eta}  \,\,  F'(\eta)  \, ,\\ 
  \noalign{\smallskip}
  t(\eta)= -{\displaystyle \frac23} \sqrt{\eta} \, \, F'(\eta) \, . 
\end{array}
\end{equation}
Thus, the shock value $\eta_c$ of the field $\eta$ is the minimum of the function $t(\eta)$, where 
${\d t}/{\d\eta}=0$. Hence we have
\begin{equation}
 F'(\eta_c)+2 F''(\eta_c)\eta_c=0
\end{equation}
and 
\begin{equation}
 x_c= F(\eta_c)- 2\eta_c F'(\eta_c) -\frac{2c}{3}\sqrt{\eta_c}F'(\eta_c), \qquad t_c=-\frac23 \sqrt{\eta_c}F'(\eta_c).
\end{equation}

\begin{figure}[h!]
\centering
{\includegraphics[width=10cm]{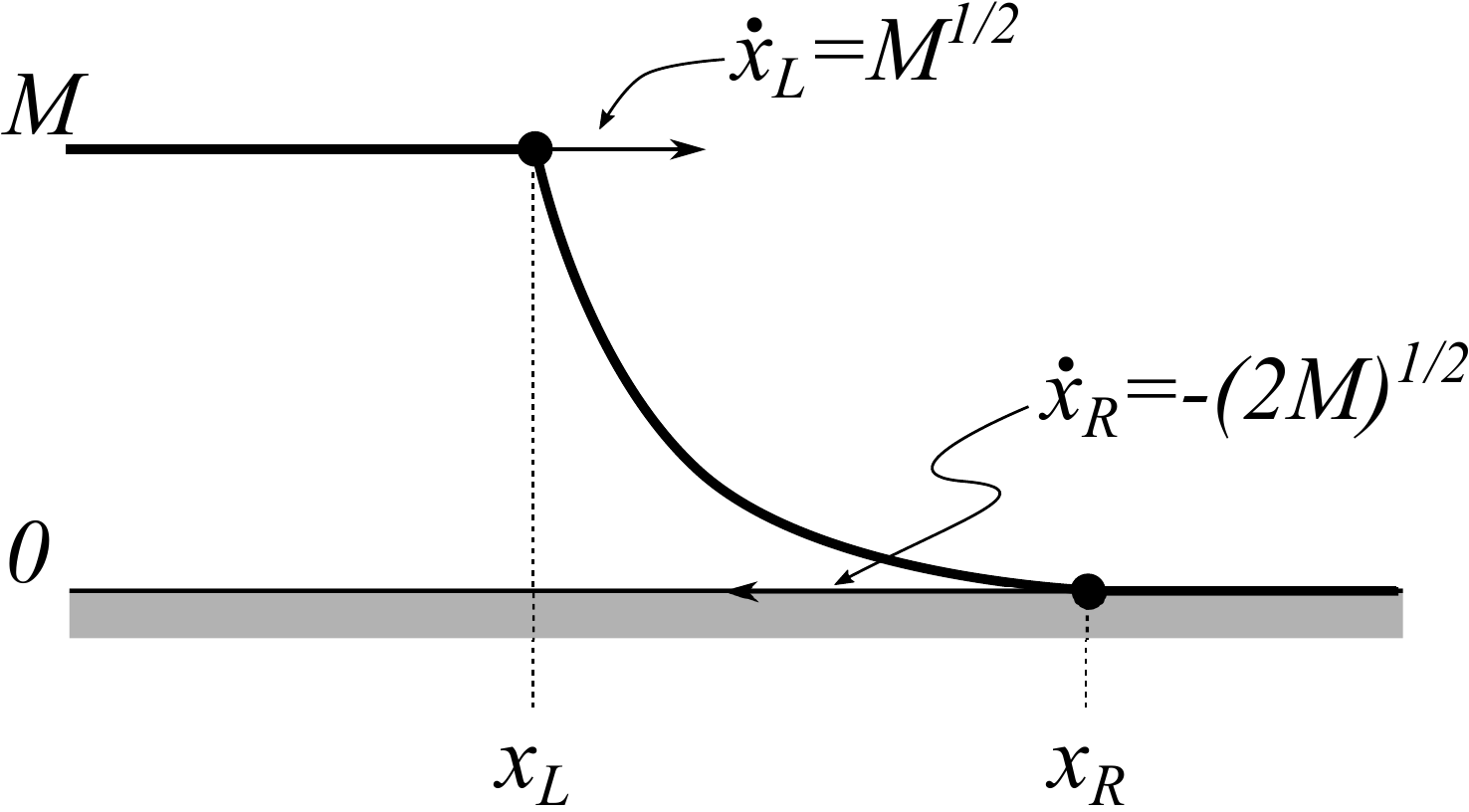}}
\caption{Qualitative behavior of the interface for simple-wave solutions related to (\ref{initialdataxa}) before the shock. 
The velocity is $u=2\sqrt{\eta}-2\sqrt{M}$.}
\label{shoshock-fig}
\end{figure}

Next, we examine the shock's position dependence on the regularity of the interface function 
$\eta$  at the intersection (dry) point at the bottom boundary. We consider the family of initial interfaces 
\begin{equation}
\label{initialdataxa}
 \eta(x,0)= \left\{  \begin{array}{cccl}
                         M  & & &  x<-M^{1/\alpha} \\   
                         (-x)^\alpha &&& -M^{1/\alpha} \leq x <0 \\
                         0 &&& x \geq 0
                     \end{array}
\right. \, 
\end{equation}
with $M> 0$ and $\alpha > 1$. We choose the constant value of the Riemann invariant $R_-$ 
to be~$c=-2\sqrt{M}$, so that $u=2\sqrt{\eta} -2 \sqrt{M}$ and the fluid is at rest at infinity.
The hodograph solution is 
\begin{equation}
\label{hodo-sol}
 x-(3\sqrt \eta -2 \sqrt{M})t=-\eta^{1/\alpha}.
\end{equation}
For these initial data, we can see that a shock occurs at finite times. Indeed, at the points $(x_L(t),M)$ and $(x_R(t),0)$, depicted in figure
\ref{shoshock-fig}, from (\ref{hodo-sol}) it follows  that  ${\dot x}_L=\sqrt{M}$ and ${\dot x}_R=-\sqrt{2M}$.
The shock curve is given by
\begin{equation}
 (x(\eta),t(\eta)) = 
 \left(\frac{2}{3\alpha}\eta^{(2-\alpha)/(2 \alpha)}(3\sqrt \eta -2 \sqrt{M}) - \eta^{1/\alpha} 
 ,\frac{2}{3\alpha}\eta^{(2-\alpha)/(2 \alpha)}\right).
\end{equation}
When $\alpha\ne 2$ the second component $t(\eta)$ is a monotonic function  of $\eta$, and
the shock can only appear at the limiting values of the interface transition region,  
\begin{equation}
 \eta_c= \left\{  \begin{array}{cccl}
                          0 &&& \mbox{if $1<{\alpha}<2$}\\   
                        M  & & &  \mbox{if ${\alpha}>2$}
                     \end{array}
\right.
\end{equation}
and the shock time is
\begin{equation}
t_c= \left\{  \begin{array}{cccl}
                          0 &&& \mbox{if $1<{\alpha}<2$}\\
                       \displaystyle{ \frac{2}{3\alpha}}M^{(2-\alpha)/(2 \alpha)}  & & &  \mbox{if ${\alpha}>2$}
                     \end{array}
\right. \, . 
\end{equation}
\begin{figure}[h!]
\centering
{\includegraphics[width=10cm]{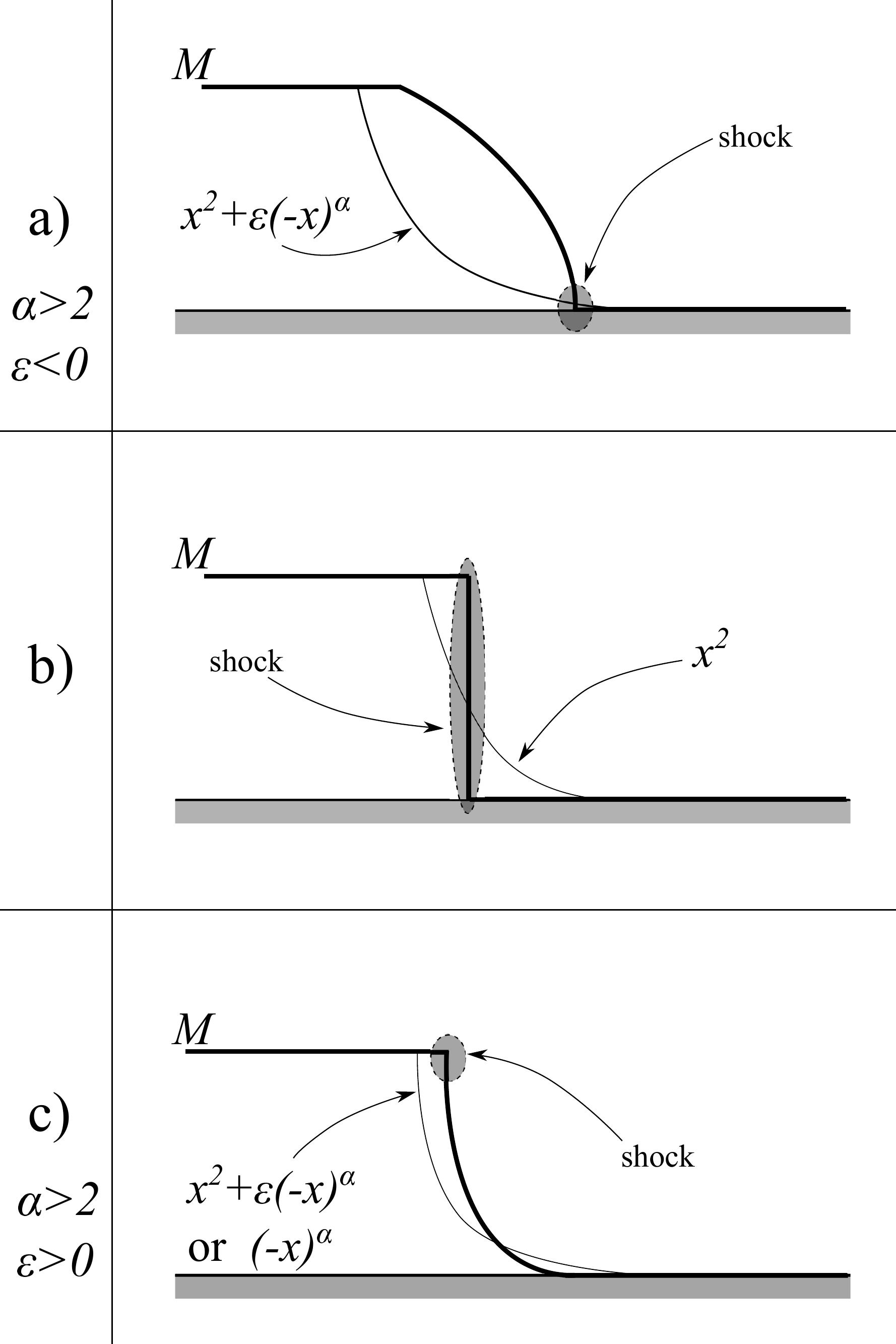}}
\caption{Qualitative behavior of the interface for simple-wave solutions related to (\ref{PbC-sw-x2xa}). The thin line represents the initial 
interface condition while the thick line represents the interface at the shock time formation. We choose the initial velocity as $u=2\sqrt{\eta}-2\sqrt{M}$.}
\label{CatastrophePoint-sw-xa-fig}
\end{figure}
Thus, the interface shock formation depends on whether the exponent $\alpha$ is smaller or greater than the threshold value $\alpha =2$. For sufficiently steep 
interfaces the shock appears far from the bottom, while for flatter interfaces the shock appears right  at the bottom  at $t=0^+$.
For $\alpha=2$, the shock time is $t_c=1/3$, and the shock starts as a finite amplitude jump from $\eta=0$ to $\eta=M$, since the evolution is nothing but the time-reversal version of the classical dam-break 
problem (see, e.g., \cite{Sto}).
At this critical case, higher-order terms in the expansion of the initial datum near the bottom determine the evolution. 
Hence, while still in the simple-wave approximation, it is instructive to study an initial interface of the form 
\begin{equation}
\label{PbC-sw-x2xa}
 \eta(x,0)= \left\{  \begin{array}{cccl}
                         M  & & &  x<x_M \\   
                         x^2+ \epsilon \, (-x)^\alpha &&& x_M \leq x <0 \\
                         0 &&& x \geq 0
                     \end{array}
\right.
\end{equation}
where now $\alpha>2$ and $x_M^2+ \epsilon \,  x_M^\alpha=M$. 
For sake of simplicity we let  $|\epsilon| \ll 1$ and 
consider the leading order  terms in this small parameter.  For the branch that corresponds to $x<0$,  the asymptotic expansion for $\epsilon \to 0$ of the inverse of the initial condition is 
\begin{equation}
 x \sim -\sqrt{\eta} + \frac{\epsilon}{2} {\eta}^{(\alpha-1)/2} 
\end{equation}
 with an error $o(\epsilon)$. The hodograph solution is therefore approximately given by
\begin{equation}
 x-(3 \sqrt{\eta}-2\sqrt{M})t \sim -\sqrt{\eta} + \frac{\epsilon}{2} {\eta}^{(\alpha-1)/2},
\end{equation}
so that (assuming the asymptotic relation holds for derivatives as well) 
\begin{equation}
 \frac{3}{2 \sqrt{\eta}} t_c \sim \frac{1}{2 \sqrt{\eta_c}} - \epsilon\frac{\alpha-1}{4} {\eta_c}^{(\alpha-3)/2}
\end{equation}
or
\begin{equation}
\label{critcurv2n}
  t_c \sim \frac{1}{3} -  \epsilon\frac{\alpha-1}{6} {\eta_c}^{(\alpha-2)/2}.
\end{equation} 
Thus, the  critical time $t_c$ continues to be a monotonic function of the critical interface height $\eta_c$. The shock position depends uniquely on the sign of $\epsilon$ 
and not on the exponent $\alpha$ (see figure \ref{CatastrophePoint-sw-xa-fig}). 
If $\epsilon$ is positive the minimum of (\ref{critcurv2n}) is attained when $\eta_c=M$, meaning that the shock appears far 
from the bottom. When $\epsilon$ is negative the shock appears  at the contact point with the fluid's bottom boundary. 
The main difference with respect to the qualitatively similar case analyzed above with the local interface $\eta(x,0)=(-x)^\alpha$ for $1<\alpha<2$ is given by the 
critical time which is now $O(\epsilon)$-close to $t=1/3$ and does not develop immediately at $t=0^+$ as in the previous case.

The fact that the interface touches the bottom boundary plays a fundamental role in the shock position. To illustrate this, let us consider a 
vertically translated version of the local initial data (\ref{initialdataxa}),
\begin{equation}
 \label{initialdataxawet}
 \eta(x,0)= \left\{  \begin{array}{cccl}
                         M  & & &  x<-M^{1/\alpha} \\   
                         (-x)^\alpha+\delta &&& -(M-\delta)^{1/\alpha} \leq x <0 \\
                         \delta &&& x \geq 0
                     \end{array}
\right.
\end{equation}
where $\delta<M$ and $1<\alpha <2$. 
The relation between the velocity and $\eta$ is taken as previously as in the simple-wave case, i.e.,  $u=2\sqrt{\eta}-2\sqrt{M}$. 
The hodograph solution 
is
\begin{equation}
 x-(3\sqrt \eta -2 \sqrt{M})t=-(\eta-\delta)^{1/\alpha},
\end{equation}
which implies that the time component of the shock curve is
\begin{equation}
\label{critcurvfarwet}
t =\frac{2}{3 \alpha} {\sqrt{\eta}} (\eta-\delta)^{-(\alpha-1)/ \alpha}.
\end{equation}
The minimum of this function is given by $\eta_c=\alpha\delta/(2-\alpha)$, which is strictly greater than $\delta$.

While we do not provide here a proof, numerical simulations show that the above results may apply to more general classes of initial data, whose local behavior is illustrated by the special cases considered here. In particular, in Appendix B  
we will provide an example of how the smoothing of the interface at $\eta=M$ does not affect the
qualitative shock formation displayed in this section.


\subsection{Self-similar parabolic solutions}
 \label{sezione:parab-solut}
All solutions of the Airy system~(\ref{airyeq}) we have considered so far belong to the simple wave class, that is, one of the two Riemann invariants is constant across characteristics. This gave rise to the examples we have considered for shock formation when the free surface initially contacts the bottom boundary with $\eta(x,0)=0$ on the half-line~$x\geq0$. 

Inspired by the well-known linear scaling solution of the Hopf equation $u_t+uu_x=0$ with $u=x/t$, 
we now study a particular class of local similarity 
solutions, already briefly mentioned in~\cite{Ovs}, 
for which the simple wave restriction can be relaxed, while maintaining continuity (in most cases) for finite times. These similarity solutions  will also allow us to study setups where the free surface can come arbitrarily close to the bottom, but does not contact it. Furthermore, this class of solutions can be used as an advantageous starting point to move into two-fluid setups, and illustrate the difference between evolution from initial conditions where the interface touches the upper (as opposed to the 
bottom) boundary in the air-water case of two-fluid dynamics. 
We remark that self-similar solutions of polynomial form for free surface flows are known for the parent Euler equations themselves 
(as opposed to models)~\cite{John,Longuet1,Longuet2}, albeit for fluids of infinite depth.


\subsubsection{Local behavior of parabolic interfaces}
\label{appendice:loc-behav-parab}

 We consider solutions of the Airy system evolving from the class of initial data locally defined  by  
\begin{equation}
 \eta(x,0)= \gamma_0 x^2 + \mu_0
 , \qquad u(x,0)= \nu_0 x
 ,
\end{equation}
with
  $\mu_0>  0$ and 
  $\gamma_0>0$.
Contact of the free surface with the bottom is clearly achieved at $x=0$ if $\mu_0=0$. It is immediate to see that the second order polynomial dependence of the right hand side of the Airy system~(\ref{airyeq}) is maintained by these data, and similarly to 
the polynomial balance argument for the linear scaling solution of the Hopf equation $u=x/t$, whereby nonlinear terms balance with space and time derivative to yield cancellations,  we make the ansatz
\begin{equation}
  \eta(x,t)= \gamma(t)\, x^2+\mu(t)
  , \qquad u(x,t)= \nu(t)\, x
  ,
  \label{numuga}
\end{equation}
which we assume holds locally for some finite time $0\leq t <T$. Substitution into the Airy's system~(\ref{airyeq}) then yields the following ODE's  for the scaling parameters,
\begin{equation}
 \dot{\nu}+\nu^2 +2\gamma =0
 , \qquad  \dot{\mu}+\nu \mu=0
 , \qquad  \dot{\gamma}+3 \nu\gamma=0
 ,
\label{coeffODEs}
\end{equation}
with 
\begin{equation}
 \nu(0)=\nu_0
 , \qquad  \mu(0)=\mu_0
 , \qquad  \gamma(0)=\gamma_0
 .
\end{equation}
Being nonlinear, these ODE's will in general support a movable singular time $T>0$ depending on the initial data. Since the first equation is of Riccati type, it can be linearized and the system solved by quadratures (note that the $\nu$ and $\gamma$ equations are uncoupled from the 
$\mu$ evolution). The general solution in parametric form follows from
\begin{equation}
 \nu^2=K_2 \gamma^{2/3} +4 \gamma\, , \qquad \mu= K_1\gamma^{1/3}\, ,
\end{equation}
where the constant $K_1$ and $K_2$ are determined by the initial conditions 
\begin{equation}
 K_1= \mu_0 \gamma_0^{-1/3} \, , \qquad K_2= \nu_0^2\gamma_0^{-2/3} -4 \gamma_0^{1/3}  \, .  
\end{equation}
The minimum $\eta(0,t)=\mu(t)$ of the parabola displays the persistence of the contact point with the bottom, as the second equation in system~(\ref{coeffODEs}) shows that $\mu=0$ is an  invariant manifold for these ODE's. 
The $\mu$-evolution by eliminating $\nu$ and $\gamma$ can be written as 
\begin{equation}
 \dot{\mu}^2= \frac{\mu^4}{K_1^2} \left(K_2+4 \frac{\mu}{K_1}\right)
 =\frac{\mu^4}{\mu_0^2} \left(\nu_0^2+4 \gamma_0 \left(\frac{\mu}{\mu_0}-1\right)\right)\, .
 \label{parab-vert}
\end{equation}
Thus, if $\mu_0>0$, $\nu_0 \neq 0$ and the negative branch of the square root is taken so that $\dot\mu(0)<0$, then $\mu(t)$ is a decreasing function of time. Nonetheless, by solution uniqueness the bottom can never be reached, unless a singularity develops and Lipschitz continuity of~(\ref{coeffODEs}) or~(\ref{parab-vert}) is lost. 
Indeed, if there exists a time $\bar t$ such that
$\mu(\bar t)=0$, then from (\ref{parab-vert}) the quadrature solution leads to a divergent integral
\begin{equation}
\label{divt}
 \bar t=\int_0^{\mu_0} {\mu_0 \over \mu^2} \left/ \sqrt{\nu_0^2+4 \gamma_0 \left(\frac{\mu}{\mu_0}-1 \right)}\right.\, d\mu  .
 \end{equation}

Elimination of $\mu$ and $\nu$ shows that the curvature parameter $\gamma(t)$ satisfies the differential equation
\begin{equation}
 \dot{\gamma}^2=9 K_2 \gamma^{8/3}+36 \gamma^3\, .
 \label{ODEt}
\end{equation}
 The auxiliary variable $\tau=(\gamma/\gamma_0)^{1/3}$ simplifies this evolution equation to 
 \begin{equation}
  \dot{\tau}^2= 4\gamma_0(\tau^5-\tau^4)\, , 
 \end{equation}
 and is useful for the parametrization of the quadrature solution. 
(The variable $\mu=K_1 \gamma^{1/3}$ would play a similar role, however it is inconvenient  in general because $\mu$ should be allowed to vanish,  which leads to $K_1=0$ and thus to $\mu$ vanishing for all times.)
The sign of $\dot{\gamma}$ in the solution of (\ref{ODEt}) is determined by the initial data
through the third equation in~(\ref{coeffODEs}), as $\dot{\gamma}(0)=-3\nu_0 \gamma_0$. The explicit solution of (\ref{ODEt}) splits into different regimes, according to the sign of $\nu_0$ and $\gamma_0$. 

For the purpose of comparing model predictions with the dynamics of the full Euler fluids,  we now focus on  initial conditions for which the fluid is initially at rest, i.e., $\nu_0=0$ and the parabolic surface has positive concavity, $\gamma_0>0$. 
Therefore $\dot{\gamma}(0)=0$, and the correct sign in (\ref{ODEt}) is chosen, when $\nu_0=0$, by
observing that  
 \begin{equation}
  \ddot{\gamma}(0)=6 \gamma_0^2\, , 
 \end{equation}
so that the curvature parameter $\gamma$ is monotonically increasing at time $t=0^+$. We have
\begin{equation}
 \dot{\gamma}= 6 \sqrt{ \gamma^3 - \gamma_0^{1/3} \gamma^{8/3}} \, , 
\end{equation}
whose implicit solution is
 \begin{equation}
  t={1 \over 2 \sqrt{\gamma_0}} \left({\sqrt{\tau-1}\over \tau}+ \mathrm{arctan} 
  \sqrt{\tau-1} \right)\,,  \quad \tau \geq 1\,\, ,
    \label{curvt}
 \end{equation}
with respect to the auxiliary time-like variable $\tau$. 
As functions of this variable, the coefficients in~(\ref{numuga}) are, explicitly, 
\begin{equation}
  \gamma(\tau)=\gamma_0 \tau^3 \, , \quad \mu(\tau)=\mu_0 \tau\, , \quad \nu(\tau)=-\sqrt{4 \gamma_0 \tau^3 +(\nu_0^2-4 \gamma_0)\tau^2} \, . 
\label{summry}
\end{equation}
The curvature diverges in finite time $t=t_s$, where
\begin{equation}
t_s\equiv\frac{\pi}{4 \sqrt{\gamma_0 }}.
 \label{shocktime}
\end{equation}
At this time  the velocity coefficient $\nu$ diverges to negative infinity and, if the fluid surface is initially away from the channel bottom boundary, or $\mu_0>0$, then the height coefficient $\mu$ diverges as well. However, if initially the surface is in contact with the channel bottom, or $\mu_0=0$, 
then the minimum of the parabolic surface will remain at zero, $\mu(t) \equiv 0$, at all times before the singularity of infinite curvature develops at $t=t_s$.

We remark that when $\gamma_0<0$, i.e., the parabola has negative concavity initially, the formulae above must be modified to account for the correct branches of the square roots. We report the appropriate, explicit relations here for completeness:  
\begin{equation}
\begin{split}
t=&{1 \over 2 \sqrt{|\gamma_0|}} \left({\sqrt{1-\tau}\over \tau}+ \mathrm{arctanh} 
  \sqrt{1-\tau} \right)\,,  \quad 0<\tau \le 1\, , \\
\gamma(t)=&\gamma_0 \tau^3 \, , \quad \mu(t)=\mu_0 \tau\, , \quad \nu(t)=\sqrt{4 \gamma_0 \tau^3 +(\nu_0^2-4 \gamma_0)\tau^2} \, . 
\end{split}
\label{summrym}
\end{equation}

\subsubsection{Local parabolic solutions continuously connected to uniform depth state}
\label{bckgrd_stte}
The local parabolic solutions can be globally extended by ``gluing" them together with a constant background $\eta(x,t)=M>0$. We will report the full details of the analysis of the time evolution of such solutions elsewhere, as these deserve a separate study to analyze all relevant scenarios. Here, we simply report results for the initial condition 
\begin{equation}
\eta(x,0)=\big(\mu_0+\gamma_0 x^2\big)\chi_{{\cal A}_0}(x)+ M \chi_{{\cal A}_0^c}(x)\, , \qquad u(x,0)=0\, ,
\label{ic_paraairy}
\end{equation}
where $\chi_{\cal I}(x)$ is the characteristic function of the interval 
${\cal I}=(a,b) \subset\barr$, that is  
$$
\chi_{\cal I}(x) = H(x-a)-H(x-b)
$$ 
with $H$ the Heaviside step function, and ${\cal I}^c$ is the complement set in $\barr$; here ${\cal A}_0=(-a_0,a_0)$ is a symmetric interval about the origin $x=0$.
If we choose $M>\mu_0>0$ and $a_0=\sqrt{(M-\mu_0)/\gamma_0}$ then the initial data are continuous and piecewise differentiable.
The evolution by system~(\ref{airyeq}) preserves continuity and piecewise differentiability, as long as shocks do not form.  The number of nondifferentiable points can 
change in time, and possibly in nature, due to their collisions. While a full analytical representation of the solution can be provided, for the present scope it is enough to simply derive equations for the motion of these nondifferentiable points, by describing them as shocks for the spatial derivatives of the dependent variables, using distributional calculus. 
We make the ansatz that the global behavior of the solution before a singularity time $t=t_s$ is
\begin{equation}
\begin{split}
\eta(x,t)=&(\mu(t)+\gamma(t) x^2)\chi_{\cal A}(x)+W(x,t)\chi_{{\cal I}_+}(x)+W(-x,t)\chi_{{\cal I}_-}(x)  + 
M\chi_{{\cal B}^c}(x) \\
u(x,t)=&\nu(t) x \chi_{\cal A}+V(x,t)\chi_{\cal I_+}-V(-x,t)\chi_{{\cal I}_-}\, ,  
\label{Anssol}
\end{split}
\end{equation}
where $W(x,t)$ and $V(x,t)$ are, respectively, local  surface and velocity solutions of the Airy system~(\ref{airyeq}), and $a=a(t)$, $b=b(t)$ are  
suitable time-dependent end points of the intervals ${\cal A } =(-a,a)$, ${\cal B} =(-b,b)$, ${\cal I}_+=(a,b)$, ${\cal I}_-=(-b,-a)$ which keep track of the nondifferentiable points of the global solution. Because spatial
symmetry of solutions is preserved in time under Airy evolution, it is sufficient to follow  positive half-line evolution only. 
Differentiating the Airy system with respect to $x$ yields 
\begin{equation}
  u_{xt}+\left(\frac{u^2}{2}+\eta\right)_{xx}=0\, , \qquad \eta_{xt}+(u \, \eta)_{xx}=0\, .
 \label{Airymodelcsq}
\end{equation}
When the solution ansatz~(\ref{Anssol}) is substituted in these equations, three different kinds of distributions (Heaviside $H$, Dirac$-\delta$ 
and Dirac$-\delta'$) appear and need to be separately balanced.
The terms involving Heaviside functions cancel out because (\ref{Anssol}) is a piecewise solution of the Airy system, and the same happens with
the terms  involving Dirac$-\delta'$ because of the continuity of~(\ref{Anssol}). Therefore the only terms that give the evolution of the points $a(t)$ and $b(t)$
come from the Dirac$-\delta$'s. 
Using the relation
\begin{equation}
 \chi'_{\cal I}=\delta(x-a)-\delta(x-b)\, ,
\end{equation}
yields the following set of ordinary differential equations for the evolution of the end points: 
\begin{equation}
\begin{split}
-\dot{a} \xi + (\gamma a^2 +\mu) \upsilon + a \nu \xi =0\, ,\qquad -\dot{a} \upsilon + a \nu \upsilon +\xi=0\, ,\\
\mathrm{where} \quad  \upsilon\equiv V_x(a)-\nu\, \qquad {\rm and} \qquad \xi\equiv W_x(a)-2 \gamma a\, ,
\end{split}
\end{equation}
for the terms supported at $x=a(t)$ 
and 
\begin{equation}
W_x(b) \dot{b} -M V_x(b)=0 \, \,, \qquad  V_x(b) \dot{b} - W_x(b)=0 
\end{equation}
for the terms supported at $b(t)$. The (explicit) evolution of the singular 
points $a$ and $b$ follows from the  ratio of the $W$ and $V$-field derivative jumps  at the singular points, 
\begin{equation}
 \frac{\xi}{\upsilon}=-\sqrt{\gamma a^2+\mu} \, , \qquad \frac{W_x(b)}{V_x(b)}=\sqrt{M}\, .
\end{equation}
(The negative sign in the first equation is necessary since, for all times, $\upsilon>0$ and $\xi<0$.)
Hence, the evolution of the location of the singular points $a$ and $b$ is simply
\begin{equation}
\dot{a}=\nu a -\sqrt{\gamma a^2+\mu}  \, , \qquad \dot{b} =\sqrt{M}\,, \qquad {\rm with }\quad a(0)=b(0)=\sqrt{(M-\mu_0)/\gamma_0}\,.
\end{equation}
The second equation is easily solved,  
\begin{equation}
 b(t)= \sqrt{M}\,  t +\sqrt{(M-\mu_0)/\gamma_0} \, .
\end{equation}
The first equation requires the solution of the self-similar parabola parameters ($\nu,\mu,\gamma$).
Since the auxiliary variable $\tau=(\gamma/\gamma_0)^{1/3}$ is increasing in time (as shown in (\ref{curvt})), 
the $\tau$ evolution
\begin{equation}
 \dot{\tau}=2\sqrt{\gamma_0} \sqrt{\tau^5-\tau^4}\, , \qquad \tau(0)=1\, ,
\end{equation}
allows the parametric solution of this equation,  via
\begin{equation}
 \frac{\d}{\d t} = \dot{\tau} \frac{\d}{\d \tau}=2 \sqrt{\gamma_0} \sqrt{\tau^5-\tau^4}  \frac{\d}{\d \gamma}.
\end{equation}
 The evolution equation for $a$ becomes
 \begin{equation}
 \frac{\d a}{\d \tau}= -\frac{a}{\tau}-\frac{1}{2}\sqrt{\frac{\tau^2 a^2 +(\mu_0/\gamma_0)}{\tau^4-\tau^3}}  \, , 
 \qquad  {\rm with} \quad a(1)=\sqrt{\frac{M-\mu_0}{\gamma_0}}\, , 
 \end{equation}
 whose solution is
\begin{equation}
 a(\tau)= \frac{\sqrt{\tau(M-\mu_0)}-\sqrt{M(\tau-1)}}{\tau\sqrt{\gamma_0}} \, .
 \label{aevog}
\end{equation}
When the point $x=a(t)$ reaches the origin  it collides with its symmetric twin at $x=-a(t)$. The value $\tau_c$ of $\tau$ at the collision time $t_c$ 
can be obtained from (\ref{aevog}) by imposing $a(\tau_c)=0$, thus yielding
\begin{equation}
 \tau_c=\frac{M}{\mu_0}.
\end{equation}
From (\ref{curvt}) one can explicitly obtain the collision  time in terms of the initial data,
\begin{equation}
 t_c={1\over 2 \sqrt{\gamma_0 } }
 \left(
 \sqrt{\frac{\mu_0}{M}\left(1-\frac{\mu_0}{M}\right)}
 +\mathrm{arctan} \left(\sqrt{\frac{M}{\mu_0}-1}\right)
 \right)\, .
\end{equation}
In the limit $\mu_0 \to 0$ the collision time coincides with the shock time, defined by (\ref{shocktime}), when the parabola collapses (closes) onto its axis, 
\begin{equation}
 \lim_{\mu_0 \to 0} t_c=t_s 
 =\frac{\pi}{4 \sqrt{\gamma_0 }}
 \, .
 \label{globalshockt}
\end{equation}
This equality is confirmed by the fact that in the same limit $\tau$ diverges and so does the curvature $\gamma$ of the parabolic part 
of the interface. 
\begin{figure}[t]
	\centering
	\includegraphics[scale=1]{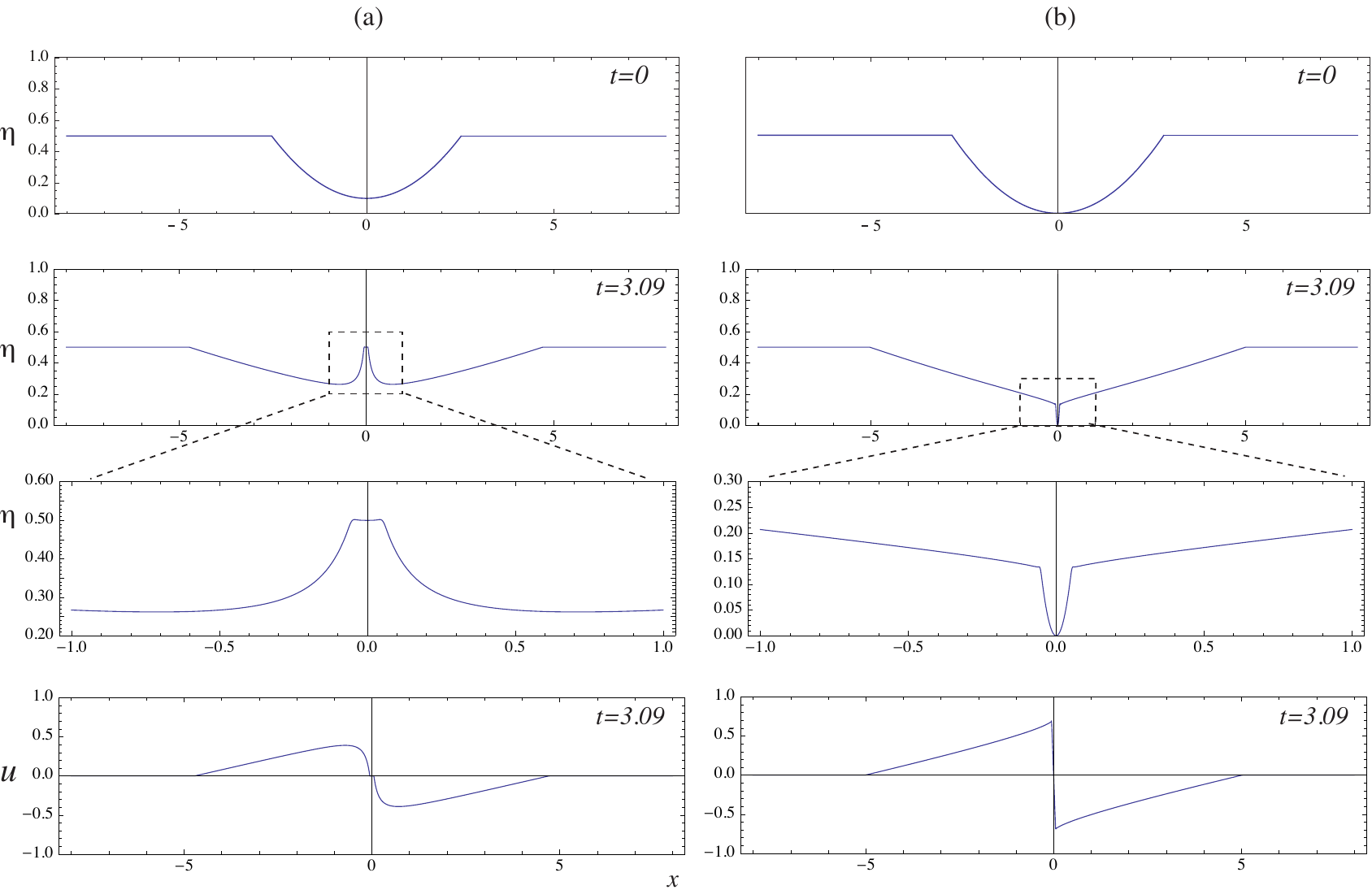}
	\caption{Snapshots of the evolution governed by system~(\ref{airyeq}) for the class of initial data~(\ref{ic_paraairy}) with $\gamma_0=1/16$ and $\nu_0=0$: (a) $\mu_0=0.1$; (b) $\mu_0=0$. At the  time $t=3.09$, the flat region around $x=0$ immediately following the parabola's disappearance can be noticed in the $\eta$-blowup (bottom panel) for case (a), while the central region connected to 
$\eta=0$ shrinks in the limit $t\to t_s$ (for this case $t_s=\pi$) to a finite height segment $\eta(t_s,0)=M/4$ but maintains its parabolic shape, as seen from the blowups  (bottom panels) for case (b). All snapshots from numerical simulations, see \S\ref{para_airy} for details.}
	\label{para_airy}
\end{figure}
Two interesting quantities are the values of the interface and the velocity at $x=a$. By continuity, these are given  by
\begin{equation}
\begin{split}
& \eta(a(\tau),\tau)=\gamma a^2 +\mu=  (a^2 + \mu_0 \gamma_0^{1/3}) \gamma = \tau\left(\sqrt{\tau(M-\mu_0)}-\sqrt{M(\tau-1)}\right)^2+\mu_0 \tau \, , \\
& u(a(\tau),\tau)= \nu a =-2a \sqrt{\gamma - \gamma_0^{1/3}\gamma^{2/3} }= -2 \left(\sqrt{(\tau^2-\tau)(M-\mu_0)}-\sqrt{M}(\tau-1) \right)\, . 
\end{split}
 \end{equation}
At the collision  time, for strictly positive $\mu_0$, such quantities are
\begin{equation}
 \eta(a(t_c),t_c)= M\, , \qquad u(a(t_c),t_c)=0\, .
\end{equation}
As shown in figure \ref{para_airy}(a) this implies
that at the collision time the bottom $\mu$ has reached the background level $M$ and the parabola shrinks to a point and disappears,  with the point expanding to flat region at level $M$. When $\mu_0=0$ the fields value at 
$t_c=t_s$ are
\begin{equation}
 \eta(a(t_c),t_c)= \frac{M}{4}\, , \qquad u(a(t_c),t_c)=-\sqrt{M}\, .
\end{equation}
As typical in these models,  we also see that spatial and time limits do not commute,
\begin{equation}
\lim_{\mu \to \mu_0} \lim_{t \to t_c} \eta(a(t),t) \neq \lim_{t \to t_c} \lim_{\mu \to \mu_0}  \eta(a(t),t)\, .
\end{equation}
With  $\mu_0>0$ the parabola disappears before a shock forms, with only two ``(anti-)bumps" remaining. These bumps can be shown to be simple wave solutions moving leftward and rightward, respectively for the right and left bump. These simple waves eventually develop 
a standard shock, whose time of formation can be computed explicitly.
When there is a dry point, i.e., $\mu_0=0$, the parabola persists, becoming increasingly narrow
as 
shown in figure~\ref{para_airy}(b), to finally collapse to a vertical segment 
\begin{equation}
x=0 \qquad {\rm and} \qquad  0\leq \eta(0,t)\leq M/4 \, , 
\label{vertseg}
\end{equation}
reminiscent of the schematic depicted by figure~\ref{aw-vs-2layer-fig}(a).  Having lost regularity, this offers a mechanism for the ``detachment" of the interface from the bottom (see figure~\ref{aw-vs-2layer-fig}(a)), since the interface at the origin has reduced to the 
segment~(\ref{vertseg}), which can be eliminated to redefine the fluid domain with a new interface at the elevation 
$\eta(0,t_s)=M/4$ away from the bottom. Of course, how the dynamics of the Airy system can be continued past this point is not immediately clear,  as at  the collision time a shock in the velocities also develops on a whole segment. 

\subsubsection{Compact-support self-similar solutions connecting to dry state}
\label{parcompact} 
A variant of the parabolic self-similar solutions of the Airy's system~(\ref{airyeq}) which corresponds to a compactly supported fluid region is of particular interest as it illustrates, outside of the special class of simple waves and within a closed form expression, the evolution of the interface and its wet domain boundary. 
For this purpose, it is useful to change notation
slightly to focus on the support along the floor, and rewrite the corresponding similarity equations in the new variables. Thus, we now seek solutions in the form
 \begin{equation}
 \label{parAirysol}
  \eta(x,t)=  \gamma(t)\left[(\beta(t))^2-x^2\right]\,\chi_{\cal B}(x)\, , \qquad   u(x,t)= \nu(t) \, x\, \chi_{\cal B}(x)\, , 
\end{equation}
 where $\chi_{\cal B}(x)$ is the characteristic function of the interval ${\cal B}=(-\beta,\beta)$, and the coefficients $\nu$, $\beta$ and $\gamma$ depend only on time.
If $\eta$ describes a parabola with negative concavity, i.e., 
 \begin{equation}
 \gamma(t)>0\, , \qquad \beta(t) \neq 0,
 \end{equation}
 then the free surface $\eta(x,t)$ connects continuously to $\eta(x,t)=0$ for $x\notin {\cal B}$, while the velocity $u(x,t)$ connects through shocks at $x=\pm \beta(t)$ to 
 the rest state $u=0$ for $x\notin {\cal B}$, for as long as $\nu(t)\neq 0$. 
 In order for the fields (\ref{parAirysol}) to be solutions of the Airy system (\ref{airyeq}),  the coefficients $(\nu, \beta, \gamma)$ must evolve 
 according to  the ODE system
 \begin{equation}
 \label{par-par}
  \dot{\nu}+\nu^2-2 \gamma=0\, , \qquad \dot{\beta}-\nu \beta=0\, , \qquad \dot{\gamma}+3 \nu \gamma = 0,
 \end{equation}
with the three parameter family of initial conditions 
\begin{equation} 
\nu(0)=\nu_0, \qquad \beta(0)=\beta_0, \qquad \gamma(0)=\gamma_0.
\label{icalbega}
\end{equation}
When $\gamma_0>0$ and $\beta_0\neq 0$ ($\beta_0 >0$, say), the fields $\eta(x,t)$ and $u(x,t)$ always asymptote to zero at 
long times. This can be seen by direct quadrature of the ODE system~(\ref{par-par}), which again follows from the Riccati-like structure of the $\nu$ equation, 
and the conservation law implied by the second and third equation. It is easy to see that 
\begin{equation}
 \beta^3\gamma=\hbox{const.}\equiv K_1,
\label{massb}
\end{equation}
and so 
\begin{equation}
\dot\beta=\pm\sqrt{2K_2-4K_1/\beta},
\label{massbb}
\end{equation}
for some constant $K_2>2K_1/\beta_0$. The first quadrature~(\ref{massb}) shows that 
$\gamma$ retains its initial sign for all times if $\beta$ does not go through zero. 
The functions $\nu(t)$ and $\gamma(t)$ are therefore determined by $\beta(t)$ through
\begin{equation}
\nu(t)= \pm {\sqrt{2 K_2}\over \beta(t)}\sqrt{1-{2K_1\over K_2\beta(t)}}\,, \qquad \gamma(t)={K_1\over (\beta(t))^3}, 
\label{massbbb}
\end{equation}
while $\beta(t)$ itself is determined implicitly by the last quadrature~(\ref{massbb}).
This corresponds to the one-dimensional dynamics along the $\beta$-axis of a charged particle in the repulsive field generated by a same-signed charge at the origin.  
Thus, if $\dot\beta(0)<0$, 
which is implied by  $\nu<0$ initially, $\beta(t)$ is a decreasing function of time down to the minimum  
\begin{equation}
\beta_{\rm{min}}\equiv 2K_1/K_2, 
\label{top}
\end{equation}
 attained at $t=t_{0}$,  corresponding to the time when $\dot\beta=0$; for times $t>t_0$, the function $\beta(t)$ is monotonically increasing. Conversely, if 
$\nu_0>0$, then $\dot{\beta}>0$ initially and 
 $\beta(t)$ is always monotonically growing for all times $t>0$. These properties imply $\eta(x,t)\to 0$ and 
 $u(x,t)\to 0$  as $t \to \infty$, since $\gamma(t)$ and $\nu(t)$ are both proportional to inverse powers of $\beta(t)$. The time $t=t_0$ gives the maximum amplitude 
 reached by $\eta(x,t)$, as 
\begin{equation}
\max_{t>0}\max_{x\in \small{\sbarr}}\, \eta(x,t)= \max_{t>0}\, (\beta^2 \gamma) =\max_{t>0}\, {K_1 \over \beta}= {K_1 \over \beta_{\rm{min}}}= {K_2 \over 2}. 
\label{maxtop}
\end{equation}

Note that depending on the initial values $\nu_0$ and $\gamma_0$ 
 the solution $\nu(t)$ may have a stationary point $\dot{\nu}=0$ at some intermediate time. For instance, if $\dot{\nu}(0)<0$ (i.e., $\nu_0^2-2\gamma_0>0$) 
 and $\nu_0<0$, then $\nu^2$ increases until its time derivative changes sign, since in terms of the quadrature constants $\dot\nu=2(K_2-3K_1/\beta)/\beta^2$, 
 which crosses zero when 
 $\beta(t)=3K_1/K_2>\beta_{\rm{min}}$. This can be given a simple physical interpretation: the initial fluid velocity gradient parametrized by $\nu_0<0$ can be 
 chosen in such a way that the parabolic ``lump" shrinks, and if its initial peak amplitude $\gamma_0$ is sufficiently large the fluid will possess sufficient inertia 
 to further steepen the velocity gradient as the lump's amplitude grows.  However, regardless of the initial sign of $\dot\nu$, even for the initial condition 
 leading to initially growing lump's amplitudes
 $\nu_0<0$, gravity eventually always dominates, causing the relaxation of the fluid velocity $u$ to zero as the peak amplitude 
 $\eta=K_1/\beta_{\rm{min}}$ is attained at $x=0$ and $t=t_{0}$. From then on $\nu(t)>0$ and the fluid  slumps, spreading indefinitely over ever larger intervals 
 $x\in (-\beta, \beta)$. (Of course, this is also the only possibility for the case 
 $\nu_0>0$ --- i.e., $u$ with the same sign of $x$ --- which leads to monotonic decrease of $\nu$ and $\gamma$ as $\beta$ grows to infinity.)  
 
It is instructive to interpret the dynamics of this class of solutions from the perspective of the conservation laws of the Airy system~(\ref{airyeq}). By connecting 
at $x=\pm \beta$ to a state with $\eta=0$, it is clear that all the balance (transport) laws 
 $$
 {\partial F[\eta,u]\over \partial t}=-{\partial G[\eta,u]\over \partial x}
 $$ 
 whose fluxes $G[\eta,u]$ depend on $\eta$ and its derivatives effectively satisfy insulating boundary conditions at $x=\pm \beta$, which lead to conservation of 
 the corresponding quantity $\int F[\eta,u] dx $. Thus, for instance, the mass density $F[\eta,u]=\eta$, with flux 
 $G[\eta,u]=\eta u$, is conserved; the total mass of the fluid is given in terms of the similarity coefficients by
 \begin{equation}
  \int_{-\beta}^\beta \eta\, \d x = \frac{4}{3} \beta^3 \gamma \equiv \frac{4}{3} K_1. 
 \end{equation}
Similarly, both the momentum 
\begin{equation}
\int_{-\beta}^\beta \eta \, u \, \d x 
\label{mass}
 \end{equation}
(which by antisymmetry is always identically null) and the  
energy 
\begin{equation}
  \int_{-\beta}^\beta \left(\frac{u^2 \eta}{2}+\frac{\eta^2}{2}\right)\, \d x=  
  \frac{ 2 \gamma \beta^5}{15} \left( \nu^2+4 \gamma\right)= {4 \over 15} K_1 K_2
  \label{energy}
\end{equation}
are conserved, as can be easily checked directly from system~(\ref{par-par}). 
It is worth remarking that energy conservation survives despite the presence of shocks in the form of the solution~(\ref{parAirysol}), 
unlike the generic setup for weak solutions of the Airy system,  whereby conservation of mass and momentum can be enforced, but energy 
and other conservation laws (with higher order polynomial densities) are not conserved and decay according to the appropriate Rankine-Hugoniot conditions. 
  
The quadrature~(\ref{massbb}) provides an exact functional form for $\beta(t)$ (expressed implicitly through its inverse $t(\beta)$). 
The detailed expression does not seem to be important, but the quadrature can be further analyzed to extract the asymptotic behavior of 
the solution at long times, to yield at leading order
\begin{equation}
 \nu \sim t^{-1}\, , \qquad \beta \sim  \sqrt{2K_2} \,  t \, , \qquad \gamma \sim {K_1 \over  (2K_2)^{3 / 2}} \, {1\over t^3}. 
\end{equation}
As expected, the asymptotic behavior of $\nu$ is independent of the initial data and 
the interface $\eta$ flattens asymptotically as $t^{-1}$, at any fixed location 
$x$, independently of the initial velocity.

This class of solutions can serve as a starting point of a perturbation analysis for a two-fluid system, which will be our focus from now, in the limit of vanishing density of the upper fluid,  in particular for modeling the splashing process when a top plate is positioned at the maximum height~(\ref{maxtop}).


\section{Two-fluid system in a horizontal channel}
\label{sezione:two-fluid}
We are now ready to move on to the stratified case of two fluids divided by an interface, constrained to move in a horizontal 
channel of height $h$ (see figure \ref{two_layers-fluid-fig}). As reviewed in~\S~\ref{Euler-2D}, 
a dispersionless model for such a physical setup is system~(\ref{e-m-full}). 
Such a system can be given a Hamiltonian structure, with 
Hamiltonian density (see, e.g., \cite{CFOP-proc,CFO})
\begin{equation}
\label{hami-disp}
{\cal H}=\frac12
\left( \frac{h^2\eta(h-\eta)\sigma^2}{\rho_2 h-(\rho_2-\rho_1)\eta} + g (\rho_2-\rho_1)\eta^2 \right)
\, , 
\end{equation}
and Hamiltonian operator
\begin{equation}
{\cal J}\equiv -\frac1{h}\left(\begin{array}{cc} 
0 & \partial_x\\
\partial_x & 0
\end{array}\right) \, , 
\end{equation} 
which leads to the equations of motion in the form 
\begin{equation}
\label{eq-disp}
\left( \begin{array}{c} \eta_t \\ \sigma_t \end{array} \right)  +
\left( \begin{array}{cc} A & B \\ C & A  \end{array} \right) 
\left( \begin{array}{c} \eta_x \\ \sigma_x \end{array} \right)=
 \left( \begin{array}{c} 0 \\ 0 \end{array} \right),
\end{equation}
where the functions $A(\eta,\sigma)$, $B(\eta,\sigma)$, and $C(\eta,\sigma)$ are
\begin{equation}
A 
= \frac{h\left(\rho _2 (h-\eta )^2- \rho _1\eta ^2\right)}{\left(\rho_2 h-(\rho_2-\rho_1)\eta\right)^2}\sigma,\quad
B=\frac{h\eta  (h-\eta )}{\rho_2 h-(\rho_2-\rho_1)\eta},\quad 
C 
= -\frac{\rho_1\rho_2 h^3 \sigma^2}{\left(\rho_2 h-(\rho_2-\rho_1)\eta\right)^3}+\frac{(\rho_2-\rho _1)g}{h}.
\end{equation}
System~(\ref{eq-disp}) becomes elliptic for large values of $|\sigma|$ (see, e.g., \cite{Chumakova00,tabak-2004} for the fluid-dynamics relevance of this feature).

The structure of the two-fluid system~(\ref{eq-disp}) is best analyzed in terms of non-dimensional variables, by introducing the parameter 
$r=(\rho _2-\rho _1)/{\rho _2}$ and the rescaled variables
\begin{equation}
 \eta\to\frac{\eta}{h},\qquad \sigma\to\sigma\sqrt{\frac{h}{{\rho_2}^2 gr}},\qquad x\to\frac{x}{h},\qquad t\to t\sqrt{\frac{gr}{h}},\qquad
{\cal H}\to \frac{{\cal H}}{\rho_2 gr h^2}. 
\label{scales}
\end{equation}
Once again abusing notation by not distinguishing between dimensional and scaled variables, as long as this does not generate confusion, the Hamiltonian density (\ref{hami-disp}) takes the form
\begin{equation}
\label{hami-disp-adim}
{\cal H}=\frac12
\left( \frac{\eta(1-\eta)\sigma^2}{1-r\eta} + \eta^2 \right)
,
\end{equation}
and equations~(\ref{eq-disp}) retain the same form with rescaled matrix elements
\begin{equation}
\label{ABC-riscalati}
A 
= \frac{1-2\eta +r\eta^2}{\left(1-r\eta\right)^2}\sigma,\qquad
B=\frac{\eta  (1-\eta )}{1-r\eta},\qquad 
C 
= -\frac{(1-r) \sigma^2}{\left(1-r\eta\right)^3}+1.
\end{equation}
Explicitly, in conservation form, system~(\ref{eq-disp}) (the nondimensional version of system~(\ref{e-m-full})) reads 
\begin{equation}
\begin{array}{l}
\dsl{ \eta_t= }\, \dsl{-\partial_x\left( {
{ \eta (1-\eta) \sigma \over 
1-r \eta  }} \right)} \, , 
\vspace{0.3cm}
 \\
\dsl{{\sigma}_t}=
\dsl{-\partial_x\left({1\over 2}\, 
{1-2\eta +r\eta^2
\over 
( 1-r\eta)^2}\, \sigma^2
+\eta\right)} \, .
\end{array}
\label{es_nondim}
\end{equation}

The limit $r\to 1^{-}$ when the upper fluid's density becomes negligible is related to the Airy system as long  as the interface 
does not get into contact with the upper lid (see Appendix B
).
The opposite limiting case $r \to 0^+$ gives the so-called Boussinesq approximation of negligible density effects on the fluids' inertia. 
It is remarkable that in this limit the equations of motion can be mapped into the Airy's model of a single layer fluid~\cite{Chumakova}. In fact, the map 
\begin{equation}
\tilde{u}  \equiv (1-2\eta) \sigma \qquad \tilde{\eta} \equiv(1-\sigma^2)(\eta-\eta^2)
\label{airy_B_map}
\end{equation}
sends solutions of the $(\eta,\sigma)$-system~(\ref{es_nondim}) into solutions of the classical Airy system~(\ref{airyeq}), with dependent variables now adorned by tildes,
\begin{equation}
 \tilde{\eta}_t+(\tilde{\eta} \, \tilde{u})_x=0, \qquad \tilde{u}_t+\tilde{u}\, \tilde{u}_x+\tilde{\eta}_x=0 \, .
\label{airyeq_tilde}
\end{equation}
However, care must be paid when interpreting solutions $(\tilde{\eta},\tilde{u})$ of the Airy's system~(\ref{airyeq_tilde}) with respect to the original variables~$(\eta,\sigma)$ since the 
map~(\ref{airy_B_map}) is not one-to-one~\cite{EP}. Thus, under the map~(\ref{airy_B_map}), the same initial data $\tilde{\eta}(x,0)$ and $\tilde{u}(x,0)$ are the image, in general, of four different initial conditions in the original variables $(\eta,\sigma)$. Some consequences of this multivaluedness in the context of the boundary confinement effects will be briefly explored below in 
\S\ref{ss_bousi}.

The results proved for the Airy system can be extended to the generic form~(\ref{eq-disp})
which includes all the Hamiltonian systems with the ``canonical structure" 
\begin{equation}
{\cal J}\equiv -\left(\begin{array}{cc} 
0 & \partial_x\\
\partial_x & 0
\end{array}\right)
\end{equation} 
and whose Hamiltonian density ${\cal H}(\eta,\sigma)$ does not depend on $x$-derivatives. In this case, the matrix entries in (\ref{eq-disp}) have the general form 
\begin{equation}
A={\cal H}_{\sigma\eta},\qquad B={\cal H}_{\sigma\sigma},\qquad C={\cal H}_{\eta\eta}. 
\end{equation}
The characteristic velocities of (\ref{eq-disp}) are 
\begin{equation}
\lambda_\pm=A\pm\sqrt{BC} 
\end{equation}
so that the sonic curve (i.e., the region in the $(\eta,\sigma)$-plane where $\lambda_+=\lambda_-$) has 
two components, 
\begin{equation}
B(\eta,\sigma)=0\quad \mbox{or}\quad C(\eta,\sigma)=0. 
\end{equation}
Given the symmetry between these two cases, it suffices to analyze one of them, with the results easily extended to the other case. In the following we will be mainly interested 
in the first component of system~(\ref{eq-disp}), hence we focus on the first choice, $B(\eta,\sigma)=0$. Note that in the Airy case this amounts to $\eta=0$. We assume that $(\eta(x,t),\sigma(x,t))$ 
is a solution of system~(\ref{eq-disp}) such that $B(\eta(x_0,0),\sigma(x_0,0))=0$ for some $x_0$. Under suitable hypotheses, we want to show that there exists a 
moving point $x_\kappa(t)$ such that 
\begin{equation}
x_\kappa(0)=x_0\quad \mbox{and}\quad B(\eta(x_\kappa(t),t),\sigma(x_\kappa(t),t))=0
\end{equation}
for all $t$ before shocks emerge. This can be done by considering the solution $x_\kappa(t)$ of the Cauchy problem
\begin{equation}
\dot x(t)=A(\eta(x(t),t),\sigma(x(t),t))\, , \qquad x(0)=x_0 \,.
\end{equation}
Indeed, define $\beta_\kappa(t)= B(\eta(x_\kappa(t),t),\sigma(x_\kappa(t),t))$ and compute
\begin{equation}
\label{evol-N1}
{\dot \beta}_\kappa = B_\eta(\eta_x\dot x+\eta_t)+B_\sigma(\sigma_x\dot x+\sigma_t)=\alpha_1 \beta_\kappa + \alpha_2 \gamma_\kappa,
\end{equation}
where $\alpha_1(t)=-\left(B_\eta \sigma_x\right)\big|_{x=x_\kappa(t)}$, $\alpha_2(t)=-\left(B_\sigma C\right)\big|_{x=x_\kappa(t)}$, and 
$\gamma_\kappa(t)\equiv \eta_x(x_\kappa(t),t)$.
If the hypothesis $B_\sigma=0$ is satisfied (i.e., if $B$ depends only on the variable $\eta$), then the previous equation is a (linear) homogeneous equation 
for $\beta_\kappa(t)$. Since $\beta_\kappa(0)=0$, we obtain our thesis. 

Now let us remove the hypothesis $B_\sigma=0$ and let us suppose that $\eta_x(x_0,0)=0$. 
From the definition of $\gamma_\kappa(t)$, it is easily found that 
\begin{equation}
\label{evol-N2}
{\dot \gamma}_\kappa = \alpha_3 \beta_\kappa+\alpha_4 \gamma_\kappa+\alpha_5 \gamma_\kappa^2,
\end{equation}
where $\alpha_3(t)=-\sigma_{xx}(x_\kappa(t),t)$, $\alpha_4(t)=-\left((A_\sigma+B_\eta)\sigma_x\right)\big|_{x=x_\kappa(t)}$, 
and $\alpha_5(t)=-A_\eta\big|_{x=x_\kappa(t)}$. By uniqueness of solutions of system~(\ref{evol-N1}-\ref{evol-N2}) we can conclude that 
\begin{equation}
 B(\eta(x_\kappa(t),t),\sigma(x_\kappa(t),t))=0\quad \mbox{and}\quad \eta_x(x_\kappa(t),t)=0\quad\mbox{for all $t$.} 
\end{equation}
As a consequence, one can easily check that $\frac{\d}{\d t}A(\eta(x_\kappa(t),t),\sigma(x_\kappa(t),t))=0$, so that the ``contact point'' $x_\kappa(t)$ moves at the constant 
speed 
\begin{equation}
 \dot x_\kappa(t)=A(\eta(x_0,0),\sigma(x_0,0)). 
\end{equation}
In particular, if $A(\eta(x_0,0),\sigma(x_0,0))=0$ then $x_\kappa(t)=x_0$, 
so that $B(\eta(x_0,t),\sigma(x_0,t))=0$ and $\eta_x(x_0,t)=0$ for all $t$.

Analogous conclusions can be drawn for the case $C=0$, under either of the conditions $C_\eta=0$ or $\sigma_x(x_0,0)=0$.

We can apply these general results to the 
two-fluid system~(\ref{es_nondim}), where $B$ does not depend on $\sigma$.
Hence, if $(\eta(x,t),\sigma(x,t))$ is a solution such that $B=0$ at $x=x_0$ and $t=0$, then, for as long as this solution maintains regularity, 
$B(\eta(x_\kappa(t),t),\sigma(x_\kappa(t),t))=0$, where $x=x_\kappa(t)$ is the solution of the Cauchy problem 
\begin{equation}
 \dot x(t)=\frac{\left(1-2\eta(x(t),t) +r\eta(x(t),t)^2\right)\sigma(x(t),t)}{\left(1-r\eta(x(t),t)\right)^2},\qquad\qquad
 x(0)=x_0.
\end{equation}
Since $B=0$ if and only if $\eta=0$ or $\eta=1$, we have the following conclusions, 
for as long as the solution $(\eta(x,t),\sigma(x,t))$ exists and is regular:
(i) if the upper (lower) fluid initially touches the bottom (top) plate at a point $x_0$, then it will stay in contact 
with the bottom (top) at the moving point $x_\kappa(t)$;
(ii) $\eta_x(x_0,0)=0$ implies that $\eta_x(x_\kappa(t),t)=0$ and that $\dot x_\kappa(t)=A(\eta(x_\kappa(t),t),\sigma(x_\kappa(t),t))$ is constant; 
(iii) if the velocity field of the fluid is zero initially,  then 
$\sigma(x_0,0)=0$ and $\dot x_\kappa(t)=\dot x_\kappa(0)=0$, so that $x_\kappa(t)=x_0$, and the remarkable behavior noticed in points (i) and (ii)  
after equation (\ref{x(t)speed}) above can be observed in this setting as well
(however, note that, as we have seen in the previous section, the contact point can certainly move if the interface has a discontinuous derivative at the contact point).

\subsection{Behavior of simple waves near hyperbolic-elliptic transition curves}
\label{sezione:behav-sw}
We recall that an alternative, more general  definition of a {simple wave\/} for system~(\ref{eq-disp}) is a particular solution of the form 
\begin{equation}
\label{sw-def}
 \eta(x,t)=N(\theta(x,t)),\qquad  \sigma(x,t)=S(\theta(x,t))\, . 
\end{equation}
According to this definition, we  will refer to the map $\theta\mapsto (N(\theta),S(\theta))$ in the $(\eta,\sigma)$-plane 
as the {\em simple wave curve\/}.
It can be easily checked that~(\ref{sw-def}) is a solution of~(\ref{eq-disp}) if 
\begin{equation}
\label{sw-eq}
\frac{N'}{S'}=\pm\sqrt{\frac{B}{C}},
\end{equation}
that is, the simple wave curves are solutions of the equation
\begin{equation}
\label{sw-eq2}
\frac{\d\eta}{\d\sigma}=\pm\sqrt{\frac{B}{C}}.
\end{equation}
In order to give a geometrical interpretation of the condition $B_\sigma=0$ encountered in the previous section, we consider the curve $B(\eta,\sigma)=0$. 
We observe that equation (\ref{sw-eq2}) implies that any simple wave curve has a horizontal tangent at a point $(\eta_0,\sigma_0)$ where $B$ vanishes. 
In the generic case $B_\sigma\ne 0$, the behavior of simple waves near to the sonic line is~\cite{KO}
\begin{equation}
\eta=\eta_0+\sqrt{\frac{\left(B_\sigma\right)_0}{4C_0}}(\sigma-\sigma_0)^{3/2}+\cdots,
\end{equation}
where the subscript 0 stands for evaluation at $(\eta_0,\sigma_0)$.
If additionally this curve has vanishing derivative $B_\sigma(\eta_0,\sigma_0)=0$, then the tangent to the curve $B(\eta,\sigma)=0$ at $(\eta_0,\sigma_0)$ is also horizontal, so that the simple wave curve~(\ref{sw-def}) and the curve defined implicitly by $B=0$ are tangent at $(\eta_0,\sigma_0)$.

The behavior of a simple wave near a generic point of intersection with the curve $B(\eta,\sigma)=0$, under the assumption $B_\sigma=0$ at this point, can now be analyzed. The asymptotic expansion for $\sigma \to \sigma_0$ \begin{equation}
 \eta(\sigma)=\eta_0+\Omega (\sigma-\sigma_0)^\alpha+\cdots
\end{equation}
in equation~(\ref{sw-eq2})
 yields at the leading order 
\begin{equation}
 \Omega\alpha(\sigma-\sigma_0)^{\alpha-1}=\pm\sqrt{\frac{B_0+\left(B_\sigma\right)_0(\sigma-\sigma_0)+\left(B_\eta\right)_0\Omega(\sigma-\sigma_0)^\alpha}{C_0}}.
\end{equation}
Since by assumption $B_0=\left(B_\sigma\right)_0=0$, this asymptotic relation is satisfied for $\alpha=2$ and 
$\Omega=\left(B_\eta\right)_0/(4C_0)$, so that 
the generic behavior of simple waves  near this class of intersection points is parabolic, independently of the explicit system (\ref{eq-disp}):
\begin{equation}
\label{sw-approx}
 \eta(\sigma)=\eta_0+\frac{\left(B_\eta\right)_0}{4C_0}(\sigma-\sigma_0)^2+\cdots.
\end{equation}
The value $\sigma_0$
is by construction approximately constant along the simple wave curve (\ref{sw-approx}). 
We now recall that the Riemann invariants $R_\pm$ are the transported quantities along the solutions of (\ref{eq-disp}) with characteristic velocities 
$\lambda_\pm$, i.e.,  $\left(R_\pm\right)_t +\lambda_\pm \left(R_\pm\right)_x=0$. In the $(\eta,\sigma)$-plane, this means that the level curves of $R_+$ or $R_-$ are   
simple wave curves. This implies that $\sigma_0$ gives the local behavior of the Riemann invariants.

These results should be compared with the Airy's case, analyzed in section~\ref{sezione:Airy}, where the previous computations are exact. 
Indeed, we have $(\eta_0,\sigma_0)=(0,u_0)$, $B(\eta,u)=\eta$ and $C(\eta,u)=1$, so that (\ref{sw-approx}) takes the form
\begin{equation}
\label{sw-airy}
 \eta(\sigma)=\frac14(\sigma-\sigma_0)^2.
\end{equation}
Hence, the Airy case gives the local generic behavior of the more general system~(\ref{eq-disp}).

We remark that for simple wave curves getting into contact with the sonic lines, generically the matrix in system~(\ref{eq-disp}) will not be diagonalizable (being of Jordan form with an off-diagonal nonzero entry and corresponding eigenvalue of geometric multiplicity one). The special case $B=C=0$, i.e, when the matrix is proportional to the identity, is of interest. Simple wave curves behave  linearly in a neighborhood of these points, or $\eta \propto \sigma$ (for generic analytic  functions $B$ and $C$). For our systems, this occurs at the corners of the hyperbolicity domain (discussed in Appendix A, figure~\ref{aaa}), which coincide with the locations of the physical domain boundary. A general study, when these points occur away from confining boundaries, is possibly more relevant, and a substantial step in this direction can be found in~\cite{CT10}. We leave this to future studies, and move on to examining the shock structure at the fluid domain boundaries next.

\subsection{Two-fluid system's shocks for simple-waves}
\label{airy_like_bottom_sw}

We want to extend the study of \S\ref{sezione:simpler-catastrophe} for the Airy system of solution behaviors  near the bottom of the channel at the shock time, 
by adapting it to the results of \S\ref{sezione:behav-sw}. 

\begin{figure}
\centering{
\includegraphics[width=7cm]{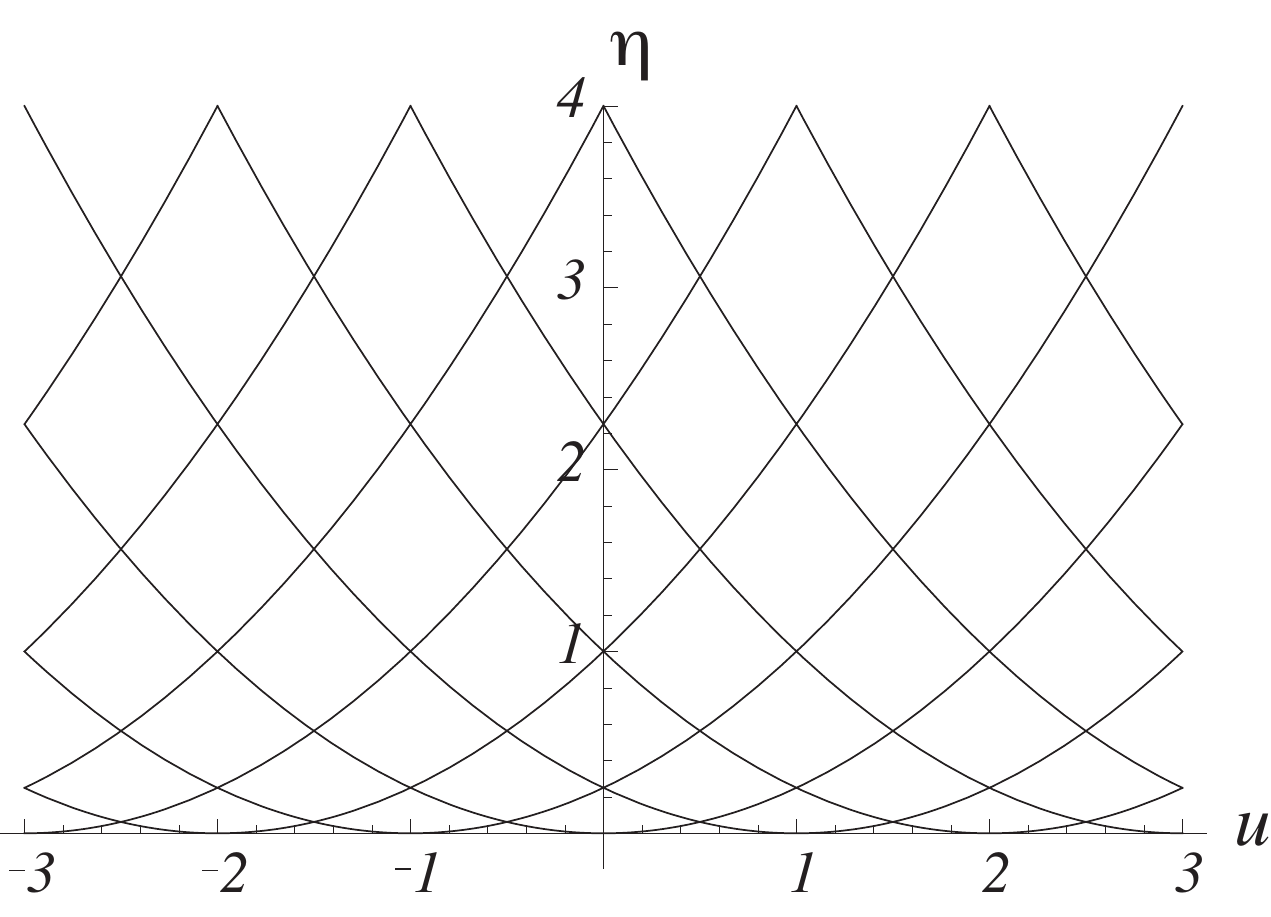}\qquad\qquad
\includegraphics[width=7cm]{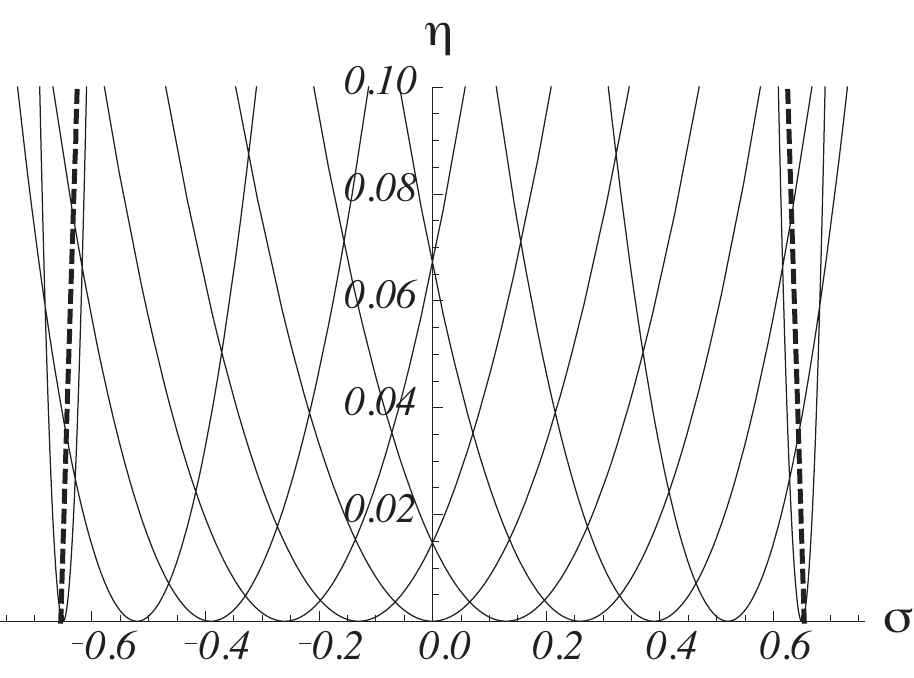}}
\caption{Left: Class of simple-wave curves for the Airy system ($r=1$) in the $(u,\eta)$-plane. 
Right: Class of simple-wave curves for the 2-layer system with $r=0.3$ in the $(\sigma,\eta)$-plane, near to $\eta=0$. 
The dashed lines are boundaries of the hyperbolicity domain.}
\label{2l-approx-sw-fig}
\end{figure}

As seen in that section, 
simple waves are found by solving the differential equation
 \begin{equation}
 \label{swODEh}
  \frac{\d \sigma}{\d \eta}= \sqrt{\frac{C}{B}}\,,
 \end{equation}
whose general solutions are not explicitly known in the two-fluid case. However, it is sufficient for the present purposes to consider approximations close to the channel boundary $\eta=0$, the other case $\eta=1$ being totally analogous. We look for a 
solution in a neighborhood of the point $(0,\sigma_0)$: from (\ref{sw-approx}) we have that
\begin{equation}
\label{approx-sw-2fluids}
 \sigma=\sigma_0+\sqrt{\frac{4C_0}{\left(B_\eta\right)_0}}\sqrt{\eta}+\cdots=\sigma_0+2\sqrt{1-(1-r){\sigma_0}^2}\sqrt{\eta}+\cdots,
\end{equation}
see figure \ref{2l-approx-sw-fig} for a comparison with the Airy case.

Moreover, from $\left(R_+\right)_t +\lambda_+ \left(R_+\right)_x=0$ it follows that $\eta(\sigma(x,t))$ is a trasported quantity, 
\begin{equation}
\label{sw-eta}
\eta_t +\lambda_+ (\eta,\sigma(\eta))\eta_x=0.
\end{equation}
The characteristic velocity $\lambda_+(\eta,\sigma(\eta))$ can be approximated by (\ref{approx-sw-2fluids}),
\begin{equation}
\label{charvelgen-sw}
 \lambda_+=A+\sqrt{BC}=\sigma_0+3\sqrt{1-(1-r){\sigma_0}^2}\sqrt{\eta}+\cdots,
\end{equation}
so that an approximation of (\ref{sw-eta}) is given by
\begin{equation}
\label{charvelH-sw}
 \eta_t+\left(\sigma_0+3\sqrt{1-(1-r){\sigma_0}^2}\sqrt{\eta}\right)\eta_x=0.
\end{equation}
After rescaling $\eta\to \left(1-(1-r){\sigma_0}^2\right)\eta$, this reduces to the same equation as in~(\ref{SWNLSeqn}). Therefore, the results of section~\ref{sezione:simpler-catastrophe} apply to the two-fluid system as well.
In particular, the degree of regularity of the initial interface in contact with the 
bottom affects the shock position, and the critical exponent is still $\alpha =2$ (both for the initial conditions 
(\ref{initialdataxa}) and (\ref{PbC-sw-x2xa})).

Higher order terms in the approximation of the characteristic velocity $\lambda_+$ could correct this behavior. 
However, 
at the next order in $\eta^{1/2}$ (see (\ref{approx-sw-2fluids})), the 
dependence of  $\sigma$ on $\eta$ is given by
\begin{equation}
 \sigma
 =\sigma_0+ 2  \sqrt{1-(1-r) {\sigma_0}^2} \sqrt{\eta } -2  (1-r) {\sigma_0} {\eta } + O(\eta^{3/2})
\end{equation}
and the characteristic velocity $\lambda_+$ in (\ref{charvelgen-sw}) becomes
\begin{equation}
 \lambda_+= \sigma_0+ 3  \sqrt{1-(1-r) {\sigma_0}^2} \sqrt{\eta } -6  (1-r) {\sigma_0} {\eta} + O(\eta^{3/2}).
\end{equation}
If we consider the initial condition (\ref{initialdataxa}), the study of the shock curve associated with the hodograph solution
\begin{equation}
x - \lambda_+ t = -\eta^{1/\alpha}
\end{equation}
leads to the same qualitative behavior as in the single layer Airy's case.

\subsection{Self-similar solutions for two-layer fluids: the high and low density contrasts }
\label{twofluidpar}
The study of the self-similar solutions of the Airy system in section~\ref{sezione:parab-solut} is directly relevant to the case of two-fluid flows in two (opposite) limits mentioned in section~\ref{sezione:two-fluid}: when $r \to 0$, i.e., the densities of the two fluids are close,  the Boussinesq approximation can be invoked, so that under mapping~(\ref{airy_B_map}) the two-layer system reduces to the Airy's model; in contrast, when $r\to 1^{-}$, i.e., the upper layer fluid density becomes negligible, the two-fluid system becomes effectively single layer and the Airy's model applies. However, asymptotic corrections to these limits introduce subtle issues, which we will briefly examine next. 
\begin{figure}[t]
	\centering
	\includegraphics[scale=.3]{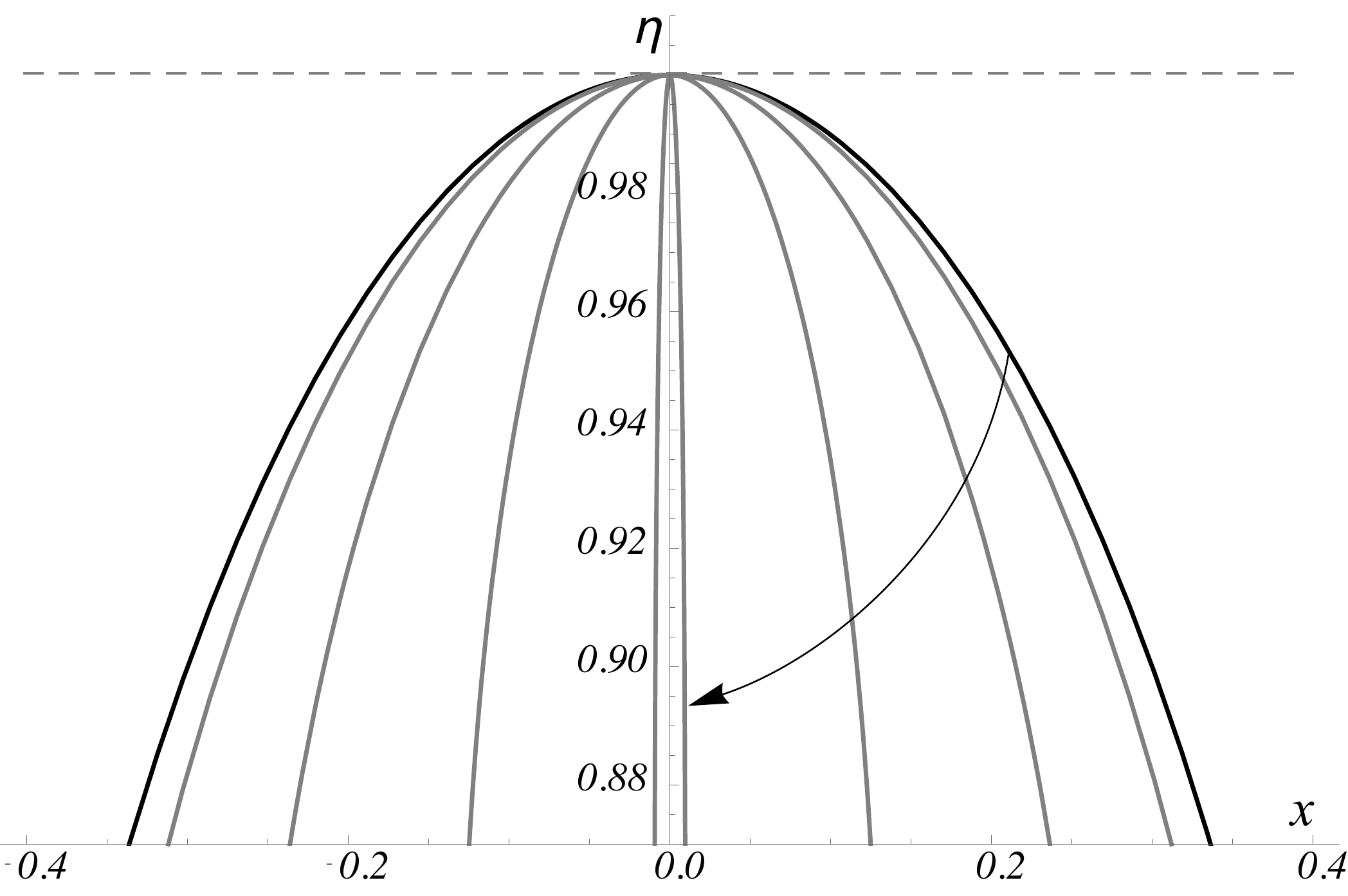} \hspace{0.5cm}
	\includegraphics[scale=.3]{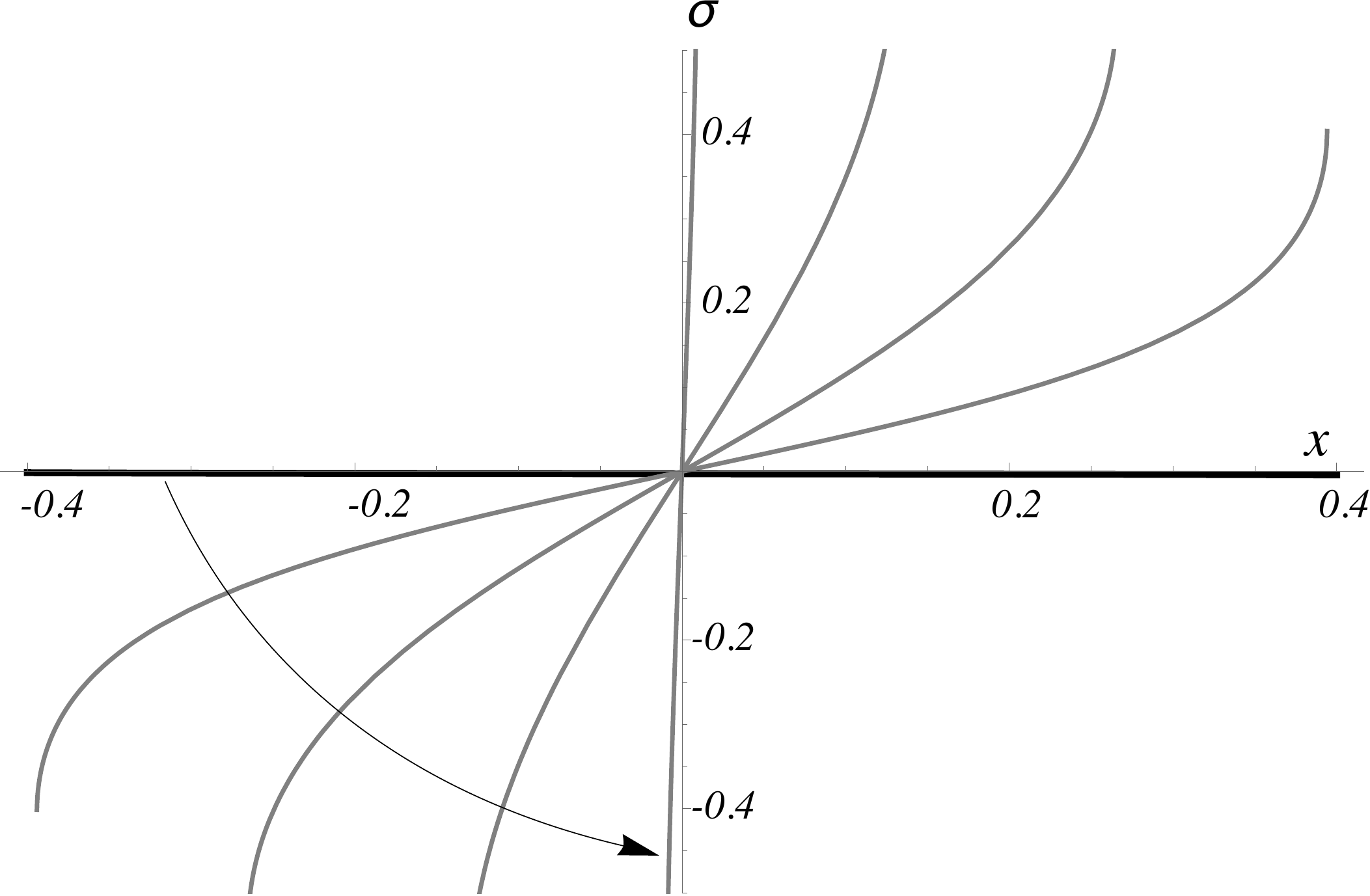}
	\caption{Local behavior of solutions to  system~(\ref{es_nondim}) obtained through map~(\ref{maptop}) from the class~(\ref{numuga}), when the interface contacts the upper plate, and the initial velocities are zero (with $\gamma_0=1$ and $\nu_0=\mu_0=0$).  Arrows indicate the evolution of dependent variables $\eta(x,t)$ and $\sigma(x,t)$ with increasing time. Snapshots taken at times $t=0,0.2,0.4,0.6,0.78$ (this last time is close to the global shock formation $t_s=\pi/4$, when the parabola closes onto its axis). 
	}
	\label{paraup}
\end{figure}

\subsubsection{The Boussinesq limit $\rho_1 \sim \rho_2$}
\label{ss_bousi}
As mentioned at the beginning of this section, 
the Airy's system can play a significant role for its two-fluid counterpart in the Boussinesq limit, as solutions of~(\ref{airyeq_tilde}) under the inverse of map~(\ref{airy_B_map}) yield solutions of system~(\ref{es_nondim}) in this limit. 

Explicitly, the (multivalued, one-to-four) inverse of~(\ref{airy_B_map}) is
\begin{equation}
\left\{
\begin{array}{l}
\eta= {1 \over 2} \left( 1+ \sqrt{{Y_\pm \over 2}}\right)\\
\noalign{\smallskip}
\sigma=-\sgn(\tilde{u})\,
\sqrt{{Y_\mp}\over2}
\end{array}
\right.
\qquad\mbox{ or }\qquad
\left\{
\begin{array}{l}
\eta= {1 \over 2} \left( 1- \sqrt{{Y_\pm \over 2}}\right)\\
\noalign{\smallskip}
\sigma=\sgn(\tilde{u})\,
\sqrt{{Y_\mp}\over2}
\end{array}
\right.
\label{quattroauno}
\end{equation}
where 
\begin{equation}
Y_\pm(\tilde{\eta},\tilde{u})=(\tilde{u}^2 - 4 \,\tilde{\eta}+1 ) \pm \sqrt{(\tilde{u}^2- 4 \,\tilde{\eta}-1 )^2 - 16 \,\tilde{\eta}}\, . 
\label{ypm}
\end{equation}
In particular, under this map the class of self-similar solutions~(\ref{numuga}) yields the local behavior of solutions of the two-fluid system~(\ref{es_nondim}) near the bottom or top plate. 
For instance, when the interface $\eta(x,t)$ is in contact with the top plate, or $\eta(0,t)=1$ for some time interval $t\in[0,t_s)$, the map is, explicitly,
\begin{equation}
\eta(x,t)= \frac12 \left(1+\sqrt{Y_+\big(\gamma(t) x^2,\nu(t) x\big)\over 2}\right)\, , \qquad \sigma(x,t)= \sgn\big(\nu(t)x\big)\sqrt{Y_-\big(\gamma(t) x^2,\nu(t) x\big)\over 2} \, , 
\label{maptop}
\end{equation}
 where the coefficients $\gamma$ and $\nu$ evolve in time according to~(\ref{coeffODEs}), with $\mu_0=\mu(t)=0$ until the global shock time $t_s$~(\ref{globalshockt}). 
The evolution of the local interface~$\eta$ (for this class of initial data,  this coincides at $t=0$ with half an ellipse with  vertical and horizontal axes) and shear~$\sigma$ given by this solution is depicted
in figure~\ref{paraup}. 
These local solutions can be connected to a background constant state   
$\eta=M\geq 1/2$ as in~\S~\ref{bckgrd_stte}. The evolution of the Boussinesq approximate solutions thus obtained can be compared with a direct numerical integration of the two-fluid system~(\ref{es_nondim}), when the densities are sufficiently close, with initial conditions in the class~(\ref{maptop}). This will be taken on again below in section~\ref{sezione:numerics}.

\subsubsection{``Air-water" limit $\rho_1\to 0$}
\label{ss_airwater}
Solutions of the Airy's system in the form~(\ref{parAirysol}), studied in~\S~\ref{parcompact}, can be useful to investigate the case of a two-fluid system in a channel in the limit of vanishing density of the upper fluid.  
For this, it is convenient to go back to the original formulation in dimensional form of the two-fluid system in long-wave asymptotics~(\ref{2layer}), sketched in section~\ref{Euler-2D}, which retains the interfacial pressure $P(x,t)$. 
Thus, $\eta_1(x,t)$ and $u_1(x,t)$ (respectively, $\eta_2(x,t)$ and $u_2(x,t)$) denote the thickness and the vertically averaged horizontal velocity of the upper (respectively, lower) fluid at $x$, with $\rho_1$, $\rho_2$ their constant densities. In the hydrostatic, nondispersive approximation, the equations of motion are obtained by dropping the $D_j$ terms
\begin{equation}
       {\eta_j}_t+(\eta_j u_j\big )_x =0\, ,\qquad
      \rho_j {u_j}_t + \rho_j u_j {u_j}_x +(-1)^j \rho_j g \, {\eta_j}_x+P_x=0\, , \qquad j=1,2 \, . 
\label{momj}
\end{equation}
If $h$ is the height of the channel and the velocities vanish at infinity, we have the constraints
\begin{equation}
\eta_1+\eta_2=h\, , \qquad \eta_1 u_1+ \eta_2 u_2=0 \, , 
\label{constraint1}
\end{equation}
the second one expressing volume conservation from incompressibility. These constraints link the upper fluid thickness $\eta_1$ and velocity $u_1$ to the ones ($\eta_2$ and $u_2$) of the lower fluid by 
\begin{equation}
\eta_1=h-\eta_2\, , \qquad u_1=-{\eta_2 u_2 \over h-\eta_2}, 
\label{upflu}
\end{equation}
\begin{figure}[t]
	\centering
	\includegraphics[scale=.99]{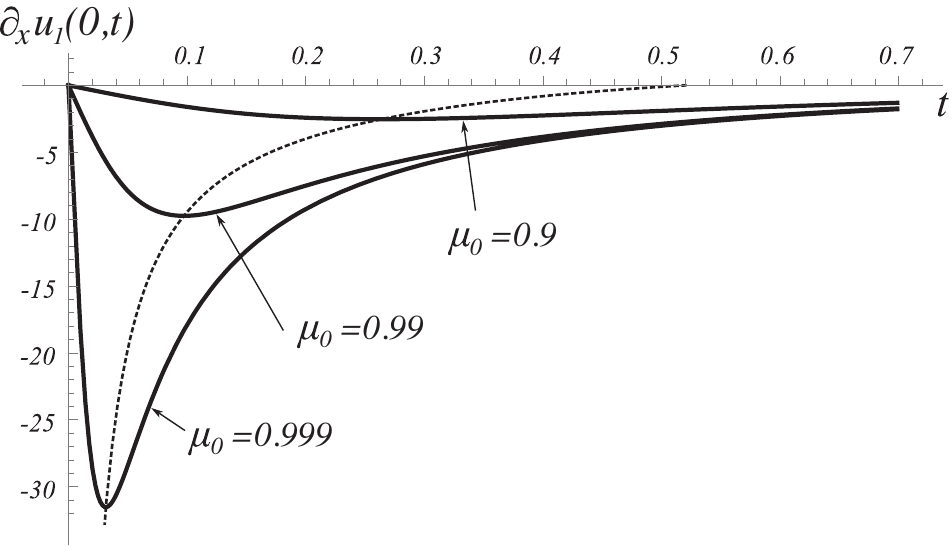}
	\caption{Time evolution of the leading order 
	contribution to the upper fluid velocity slope ${u_1}_x$ at $x=0$, corresponding to the initial data $\eta(x,0)=\mu_0-x^2$, $u_1(x,0)=u_2(x,0)=\sigma(x,0)=0$, for different $\mu_0$ values.  The dotted line tracks  the extrema of  
	${u_1}_x(0,t)$ and  marks the times $t=t_m(\mu_0)$ where these are attained: as $\mu_0 \to 1$, i.e., the initial position of the interface approaches the top plate, $t_m \to 0$.
 	}
	\label{garigagava}
\end{figure}
Use of the density weighted shear $\sigma(x,t)$ is suggested by the elimination of the interfacial pressure gradient $P_x$, and in fact this can be seen as the counterpart, for two-layer stratification, of the well-known step of eliminating the pressure gradient by taking the curl of the momentum equation in~(\ref{Eceq}). Just as in that case, if a solution $(\eta,\sigma)$ of system~(\ref{e-m-full}) is known, the interfacial pressure $P$ can then be found, in this specific case through relations~(\ref{sigma+def}),  (\ref{momj}) and (\ref{constraint1}). In particular, the layer mean-velocities can be expressed solely in terms of $\eta$ and $\sigma$,  e.g., 
\begin{equation}
u_2={h\over \rho_2}{1-\eta/h \over 1-r \, \eta/h}\, \sigma \, ,
\label{u2sigma}
\end{equation} 
so that, in turn,  the interfacial pressure gradient can be expressed in terms of $\eta$ and $\sigma$ through either one of the momentum equations~(\ref{momj}). 

The above relations allow to set up an asymptotic expansion of the solutions to system~(\ref{momj}), or, equivalently, system~(\ref{e-m-full}), in the `air-water' limit, that is, in the small parameter $\rho_r\equiv 1-r=\rho_1/\rho_2 \ll 1$, 
\begin{equation}
u_j=u_j^{(0)}+\rhor u_j^{(1)}+O(\rhor^2) \, , \qquad \eta=\eta^{(0)}+\rhor\eta^{(1)}+O(\rhor^2) \, , 
\qquad P_x=P_x^{(0)}+\rhor P_x^{(1)}+O(\rhor^2) \, , 
\label{asro} 
\end{equation}
starting from solutions of the Airy's system~(\ref{c_airy}). In fact, at leading order the upper layer momentum equation yields $P_x^{(0)}=0$ (i.e., the interfacial pressure is constant, as in the usual free-surface water wave problem), so that at this order the system for governing the evolution of $\eta^{(0)}(x,t)$ and $u_2^{(0)}(x,t)$ coincides with Airy's. At the next order, we have 
\begin{equation}
P_x^{(1)}=-\rho_2
\left(
\partial_t{u_1^{(0)}}+ \partial_x\left({{(u_1^{(0)})}^2\over 2} -g \,  \eta^{(0)}\right)
\right) 
\label{u1lead}
\end{equation}
where the leading order upper layer mean horizontal velocity $u_1^{(0)}$ is related to the leading order solution of the Airy's equation $u_2^{(0)}$ through~(\ref{upflu})
\begin{equation}
u_1^{(0)}=-{\eta^{(0)} u_2^{(0)} \over h-\eta^{(0)}}  \, . 
\label{upflu0}
\end{equation}
Substituting this into~(\ref{u1lead}) and using the Airy system for the time derivatives yields an expression for the interfacial pressure gradient in terms of $\eta^{(0)}$, $u_2^{(0)}$ and their $x$-derivatives, 
\begin{equation}
P_x^{(1)}=-\rho_2
\left(
{ h \, \eta^{(0)} \over {(h-\eta^{(0)})}^2}\, \partial_x {\big(u_2^{(0)}\big)}^2+
{h^2 {(u_2^{(0)})}^2\over {(h-\eta^{(0)})}^3} \, \partial_x \eta^{(0)}
+g\,  {2h-\eta^{(0)} \over {h-\eta^{(0)}}}\, \partial_x \eta^{(0)}
\right) \, .
\label{p1x}
\end{equation}
The next order correction to the leading Airy evolution is then obtained by solving the linearized (inhomogeneous) Airy system
\begin{equation}
\begin{array}{l}
\partial_t{\eta^{(1)}}+\partial_x
\left(
u_2^{(0)}\eta^{(1)}+u_2^{(1)}\eta^{(0)} 
\right)=0 \, , 
\vspace*{0.3cm}
\\
\partial_t u_2^{(1)}+\partial_x
\left(u_2^{(0)}u_2^{(1)}+g \, \eta^{(1)}
\right)= - P_x^{(1)} \, , 
\end{array}
\label{nextu2}
\end{equation}
with zero initial data. 
An analytic expression for the solution of this equation can in principle be computed by characteristics (which coincide with those of the leading order Airy system's solution), in particular for those emanating from the class of initial conditions~(\ref{ic_paraairy}). We will not attempt this here, however some information can be gleaned from the knowledge of the interfacial pressure gradient $P_x^{(1)}$ when such leading order solutions are used. For instance, when the initial data are such that the interface approaches the upper plate, the denominators in~(\ref{p1x}) vanish, 
signaling the possibility of a divergence of the interfacial pressure gradient (which can lead to the breakdown of the asymptotic expansion for $\rho_r\to0$, by driving the first order corrections $\eta^{(1)}$ and $u_2^{(1)}$ to become of magnitude comparable to $O(1/\rhor)$). In contrast, no such mechanism for divergence exists when the interface contacts the lower plate. 

More specifically, when the class of self-similar solutions~(\ref{numuga}) is used for the leading order fields, that is $\eta^{(0)}=\mu(t)+\gamma(t) x^2$ and $u_2^{(0)}=\nu(t) x$,  the asymptotic approximation of the upper fluid velocity as $\rho_r\to 0$ evaluates to, from~(\ref{upflu}) and~(\ref{upflu0}), 
 
\begin{equation}
 u_1(x,t)=- 2 \sqrt{g |\gamma_0|} \, \, \frac{ \tau \sqrt{1-\tau}  \, \,  x \, (\mu_0 \tau - \gamma_0 \tau^3 x^2) }{h- (\mu_0 \tau - \gamma_0 \tau^3 x^2) }+O(\rho_r) \,,  
\label{u1exp}
\end{equation}
where $ \tau$ is the function of time defined implicitly by~(\ref{curvt}).
The slope of the upper fluid velocity $u_1(x,t)$ at $x=0$ is therefore, at leading order, 
\begin{equation}
\partial_x{u_1}^{(0)}(0,t)=- 2 \mu_0 \sqrt{g|\gamma_0|}\, \, \frac{ \, \tau(t)^2  \,  \sqrt{1-\tau(t)}}{h- \mu_0 \, \tau(t) }\, .
\end{equation}
This expression has an overall minimum at some time $t>0$, $t_m$ say, corresponding to 
\begin{equation}
 \tau_m=\frac{5h+2 {\mu_0}-\sqrt{4 {\mu_0}^2-28h {\mu_0}+25h^2}}{6 {\mu_0}}\, ,
\end{equation}
since $\tau(t)$ is a monotonic (decreasing) function of time (with $\tau(0)=1$ and $\tau(\infty)=0$).
Thus, as $\mu_0 \to h$ the amplitude of this minimum grows unbounded, while the corresponding time $t_m(\mu_0)$ tends to zero, see figure~\ref{garigagava}. Of course, as noted above, the asymptotics may become invalid as $\eta^{(0)}(0,t) \to h$, and it is necessary to check whether the $O(\rho_r)$ terms in~(\ref{u1exp}) are negligible. However, the coefficient of $\rho_r$ in this expression multiplies
the next order field $\eta^{(1)}$ and $u_2^{(1)}$, which are zero initially, thereby providing a way to preserve the asymptotic ordering in~(\ref{u1exp}) initially in time. This is notably different from the case of the parabola touching the bottom plate, where, e.g., the slope of the velocity at the origin, $\partial_x u_2^{(0)}(0,t)=\nu(t)$ diverges only after a {\it finite} time $t=t_s$ has elapsed.

\subsubsection{Critical depth-ratio and short time evolution}
\label{crit}
We now return to  system~(\ref{eq-disp}), and consider a parabolic initial interface (the same as that studied for the Airy equation
in~\S\ref{sezione:parab-solut})  with vanishing initial velocities. More precisely, for this section we go back to dimensional variables and assume that  
\begin{equation}
 \eta(x,0)=\max_{x\in \sbarr}[\mu_0+\gamma_0 x^2,M],\quad \sigma(x,0)=0,\quad\mbox{with $\gamma_0<0$, $M\in (0,h)$ and $M<\mu_0\le h$.}
\label{ic_sigeta}
\end{equation}
We focus on the evolution of the concavity at $x=0$, with the aim of isolating, in the time dependence generated by such initial data, those critical phenomena that can be used for comparisons with numerical simulations of both the dispersionless system and the stratified incompressible Euler equations. We stress that the choice of initial conditions~(\ref{ic_sigeta}) is purely for convenience of comparison with the limiting cases we studied, where the evolution can be found analytically. The lack of explicit expressions for characteristics and Riemann invariants prevents the analog of such calculations to be carried out for system~(\ref{eq-disp}). We also stress that what follows is generic for any initial data with sufficient regularity and symmetry around the location $x=0$.

In fact, because of the symmetry of the initial interface, we have that $\eta_x(0,t)=0$ for all $t$. Using the equation of motion (\ref{eq-disp}-\ref{ABC-riscalati}), 
yields  $\eta_t(0,0)=0$ and 
\begin{equation}
\eta_{tt}(0,0)=-2\, g \, r \, |\gamma_0| \, \mu_0 \frac{h-\mu_0}{h-r\mu_0}.
\end{equation}
We first remark that $\eta_{tt}(0,0)=0$ when $\mu_0=h$ (i.e., if the interface initially is in contact with the top of the channel). 
This is in agreement with the general results obtained in the previous sections, that is, $\eta(0,t)=h$ for all times $t$ as long as the solution maintains regularity. 
Moreover, and again in agreement with the previous findings, if $r=1$, that is, when $\rho_1=0$, then $\eta_{tt}(0,0)=-2 g h\mu_0{\gamma_0}<0$, indicating that the interface detaches from the upper lid at $t=0^+$.

Continuing with the evolution of the concavity, it is easy to show that $\eta_{xxt}(0,0)=0$ and that 
\begin{equation}
\label{fourth-deriv}
\eta_{xxtt}(0,0)=12 \, g\, r\, {\gamma_0}^2 \,  \frac{h^2-2\mu_0 h+r{\mu_0}^2}{\left(h-r\mu_0\right)^2}.
\end{equation}
This is negative when the initial height $\mu_0$ is such that 
\begin{equation}
 \mu_0>\frac{1-\sqrt{1-r}}{r}h \, .
 \label{etacr}
\end{equation}
Hence, when $r<1$ is fixed, there is a critical height $\eta(0,0)=\eta_{\rm cr}\equiv {\mu_0}_{\rm cr}$ given 
by the right-hand-side of~(\ref{etacr}) beyond 
which the curvature of the interface is initially growing at the maximum point $x=0$, even though the height $\eta(0,t)$ might be decreasing. Conversely, when $r=1$ the interface flattens for all values of $\mu_0$,
in agreement with the results of~\S\ref{sezione:parab-solut}.
It is interesting to study how the dispersion effects neglected by system~(\ref{eq-disp}) affect these conclusions on the initial evolution of this class of data. While the calculation can be laborious, for the class of initial data under consideration the details can be worked out by going back to the dispersive system~(\ref{2layer}), and it is not too difficult to show that the threshold initial height separating the two curvature behaviors is determined by
the modified curvature acceleration
\begin{equation}
\eta_{xxtt}(0,0)=12 \, g\, r\, {\gamma_0}^2 \,  \frac{h^2-2\mu_0 h+r{\mu_0}^2}{\left(h-r\mu_0\right)^2}+D {\gamma_0}^3  \, , 
\label{fourth-derivD}
\end{equation}
where the dispersion contribution $D$ is
\begin{equation}
D={ 8 g r}  \, {D_1 r^3+D_2 r^2 +D_3 r + D_4 \over 
(h-r\mu_0 )^4} \, , 
\label{DD}
\end{equation}
where the coefficients $D_n$, $n=1,\dots, 4$, are 
\begin{equation}
\begin{split}
& 
D_1=6\mu_0^4(h-\mu_0), \quad 
D_2=\mu_0(18\mu_0^3h-22\mu_0^2 h^2-9\mu_0 h^3+10 h^4), \\
&
D_3=-14\mu_0^3h^2+57\mu_0^2h^3-40\mu_0h^4+3h^5 , \quad
D_4=-24 \mu_0^2 h^3 +24\mu_0 h^4 - 3h^5 \, . 
\end{split}
\end{equation}

The threshold initial height $\mu_0$ is obtained by setting the acceleration~(\ref{fourth-derivD}) to zero. To be asymptotically consistent with the remainder of the long-wave expansion  for model~(\ref{2layer}), which corresponds to the limit $\gamma_0\to 0$, the zeros of the resulting polynomial expression must be computed in this limit, which yields
\begin{equation}
{\mu_0}_{\rm cr,D}= h\left(\frac{1-\sqrt{1-r}}{r} 
+{2\over 3}\, {5r^2 +14r (\sqrt{1-r}-2)-28(\sqrt{1-r}-1) \over 
r^2} \gamma_0 h +o(\gamma_0 h) \right) \, . 
\label{mucritD}
\end{equation}
While not asymptotically consistent, it is nonetheless remarkable that by following the 
root~${\mu_0}_{\rm cr,D}(\gamma_0)$ of the {\it full} right-hand-side of~(\ref{fourth-derivD}) (that is, with only a subset of the higher order terms in the small parameter $\gamma_0 h$ accounted for in the dependence of ${\mu_0}_{\rm cr,D}$ on $\gamma_0$), the qualitative dependence of the resulting critical height on the curvature $\gamma_0$, and in particular the nearly constant shift upward for the ``large" curvature parameter range $0.2<\gamma_0 h<1$, is reflected by the dynamics of parent Euler system's evolution, as we shall see next.

As a final remark before leaving this section, notice that sending $r\to 1^-$ and then $\mu_0/h\to 1^-$ in (\ref{fourth-deriv}) yields (recall $|\gamma_0 h|\ll 1$)
\begin{equation}
\lim_{\mu_0/h\to 1^-}\lim_{r\to 1^-}\eta_{xxtt}(0,0)=12 g {\gamma_0}^2(1+O(\gamma_0h))>0 \, ,   
\end{equation}
whereas exchanging the two limits gives 
\begin{equation}
\lim_{r\to 1^-}\lim_{\mu_0\to 1^-}\eta_{xxtt}(0,0)=-\infty.  
\end{equation}
This phenomenon is yet another instance, as already seen in \S~\ref{ss_airwater}, of the fact that when the interface gets closer and closer to the channel boundaries the air-water limit $r\to 1^-$ 
is {\it not} in agreement with the single Airy system's behavior.
\begin{figure}[t]
	\centering
	\includegraphics[scale=1]{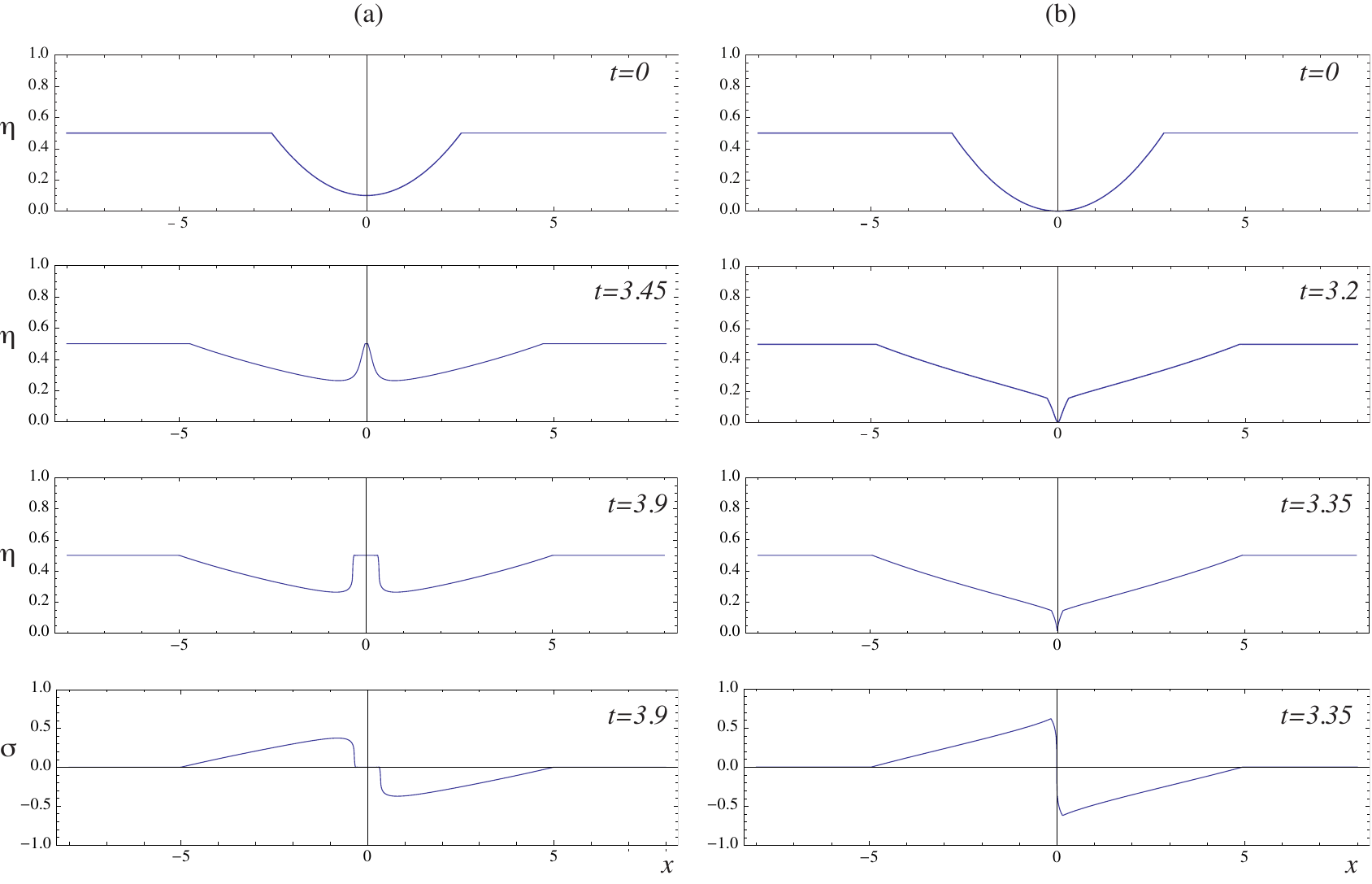}
	\caption{Snapshots of the evolution governed by system~(\ref{eq-disp}) with $r=0.89$, for the class of initial data~(\ref{ic_sigeta}) with $\gamma_0=1/16$: (a) $\mu_0=0.1$; (b) $\mu_0=0$. Periodic boundary conditions with domain $x\in[-16,16]$ are used throughout these simulations, with only half of the domain $x\in[-8,8]$ shown. Note the different times between the left and right columns: the flat region around $x=0$ has not yet developed at $t=3.2$ for case (a), while the central region connected to $\eta=0$ becomes too thin at the same spatial scales at $t=3.7$ for case (b). The intermediate snapshots at times $3.45$ (a) and $3.2$ (b) correspond to about $90\%$ of the times of shock formation, which occurs in both $\eta$ and $\sigma$ as depicted by the last frames of both left and right columns ($t=3.9$ and $t=3.4$, respectively). Note the shock forming away from the origin in the simple wave wings for case (a), as opposed at the origin in the `core' component for case (b).  
	}
	\label{quad_down}
\end{figure}

 \section{Numerical simulations}
  \label{sezione:numerics}
 The extent to which the two-fluid non-dispersive equations we have been studying so far, together with their density contrast limiting cases, manage to capture relevant elements of the dynamics of a real, stratified fluid is of course a question that asymptotics at a formal level cannot answer directly. To some extent, this question can be addressed by accurate numerical simulations of the model and parent Euler equations, and we now turn our attention to how predictions from asymptotic results 
are verified in this context. We choose to work in the context of specific examples where the initial velocity is taken to be zero, which eliminates the approximation errors associated with the reconstruction of the 2D velocity field from the layer-averaged horizontal component. The density field for the Euler simulations will be smooth, albeit with a sharp transition between two essentially homogeneous densities. We focus on critical phenomena, whereby the choice of initial data will lead to different evolutions that can be predicted by the models and discerned in the Euler direct numerical simulations. 
 
\subsection{Two-layer dispersionless model}
\label{2layer_numerics}
Continuous initial data with zero velocity, i.e. $\sigma=0$, and $M<\eta(x,0)\le 1$ (restoring nondimensional variables), with the constant $M>0$ chosen to be sufficiently large so as to avoid hyperbolic-elliptic transition for $t>0$ in the evolution governed by system~(\ref{eq-disp}), can be efficiently integrated numerically with the method of lines until shock forms. We use the version of this algorithm as implemented by the ``NDSolve" command environment of {\it Mathematica 8} software, with convergence tested against closed form solutions (when available) as well as by monitoring errors with the ``MaxStepSize" option ranging from $0.01$ (low resolution) to $0.0005$ (the maximum resolution for reasonable computation times -- in the order of tens of minutes -- on a MPB Air laptop).  
We choose to work with the (nondimensional) class of initial data~(\ref{ic_sigeta}), i.e., piecewise initial conditions formed  by arcs of parabolae centered at $x=0$ intersecting the constant elevation $\eta=M$ away from the origin. 
As the evolution out of these initial conditions bears strong similarities with their counterpart for Airy equations~(\ref{airyeq}), which can be computed analytically, it can serve as term of comparison to highlight the essential differences induced by the two layer model with generic density contrast. 
\begin{figure}[t]
	\centering
	\includegraphics[scale=1]{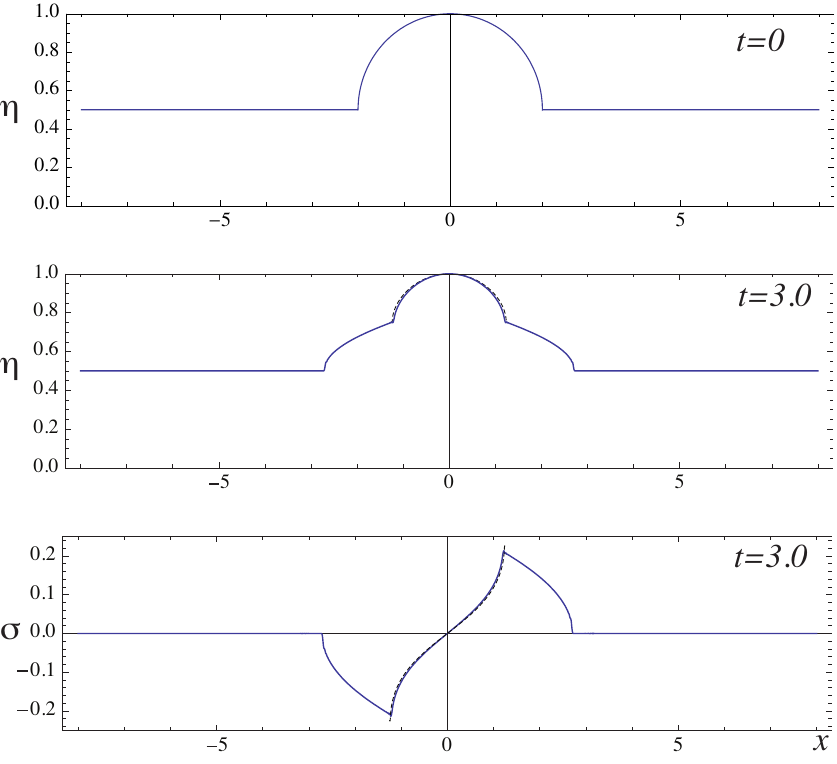}
	\caption{Snapshots of the evolution governed by system~(\ref{eq-disp}), computed numerically (continuous line) for the class of initial data obtained through the map~(\ref{quattroauno}) from~(\ref{ic_paraairy}) with $\gamma_0=1/16$, $\nu_0=0$ and $\mu_0=0$, and periodic boundary conditions at $x=\pm L$, $L=16$ (only half of the domain shown). The  dashed line depicts the analytical solution obtained by map~(\ref{maptop}).  
	The background constant state is at $\eta=\tilde{\eta}=M=1/2$. Here $\rho_1=0.8\, \rho_2$, i.e., $r=0.2$. The collapse time $t_s=\pi/(4 \sqrt{\gamma_0})$, scaled according to~(\ref{scales}), evaluates to $t\simeq7.02$, while for these initial data the simulation breakdown time is $t\simeq6.55$. 
	}
	\label{bousisu}
\end{figure}

\begin{figure}[t]
	\centering
	\includegraphics[scale=0.75]{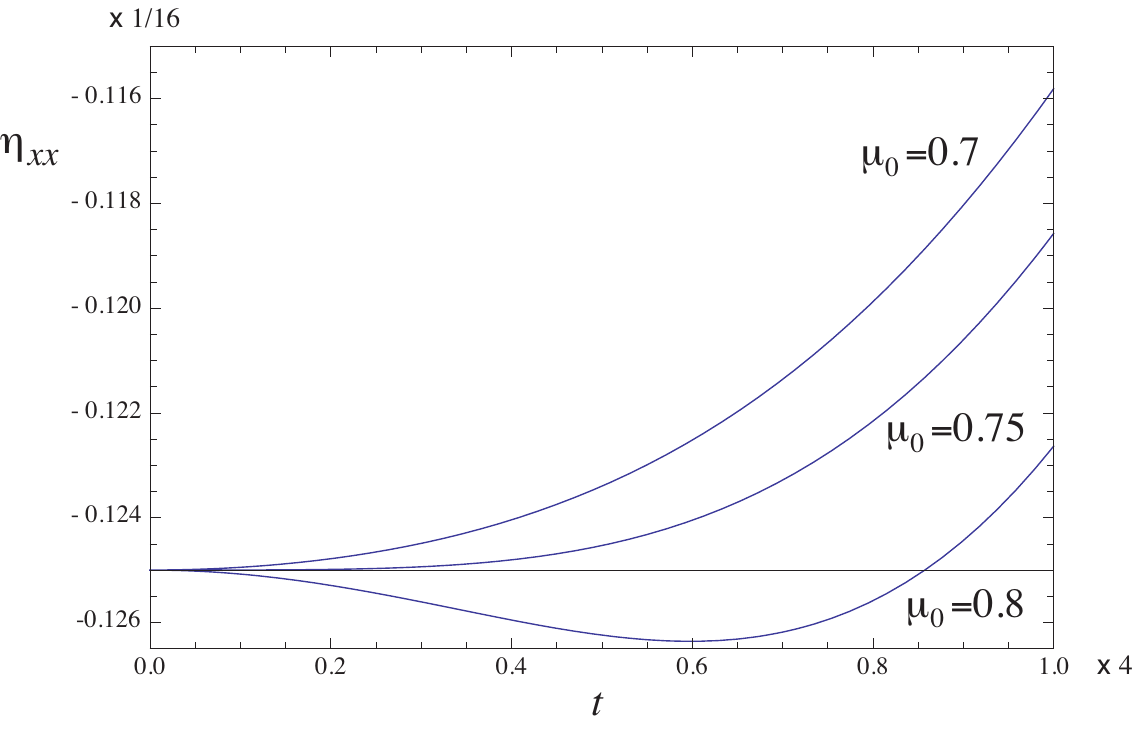}
	\caption{Time history of curvature $\eta_{xx}(0,t)$ of solutions of system~(\ref{eq-disp}) for three values of initial elevation 
	$\mu_0$ in the class of initial data~(\ref{ic_sigeta}) with $|\gamma_0|=1/16^2$ (scaled $\eta_{xx}$ and $t$ as per multipliers of axis labels) or $|\gamma_0|=1/16$ (unscaled variables).}
		\label{curvdisp}
\end{figure}
\begin{figure}[t]
	\centering
	\includegraphics[scale=0.75]{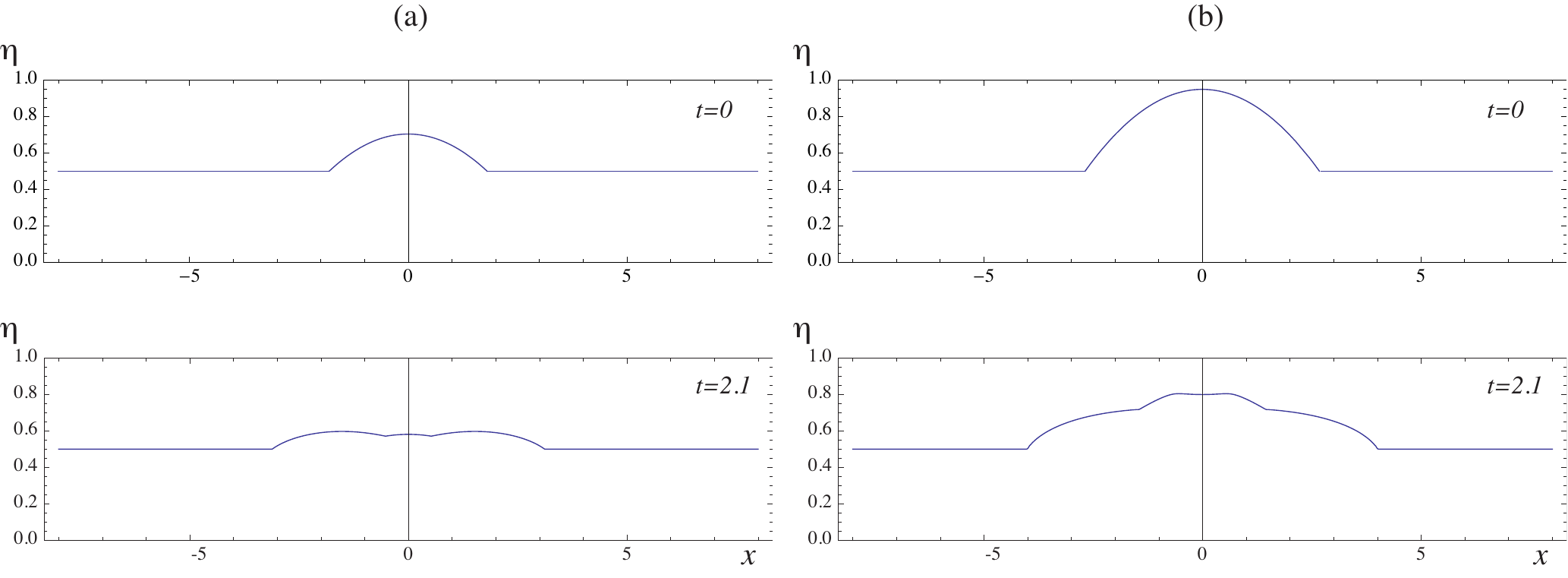}
	\caption{Snapshots of the evolution governed by system~(\ref{eq-disp}) with $r=0.89$: interface $\eta(x,t)$ at the initial time and at $t=2.1$ for initial data in the class~(\ref{ic_sigeta}) with (a) $\mu_0=0.7$ and (b) $\mu_0=0.95$, $|\gamma_0|=1/16$. Periodic boundary conditions in domain $x\in[-16,16]$, with only half of the domain shown. The curvature for the higher initial elevation case becomes positive at $x=0$, while the lower initial elevation leads to evolution maintaing the curvature sign at the same location.}
		\label{dimple_noD}
\end{figure}

\subsubsection{Positive interfacial curvature initial data: upper fluid extending into the lower one}
We first look at the parabolae  
with positive curvature, i.e., with the upper layer fluid closer to bottom, in the class
\begin{equation}
 \eta(x,0)=\min_{x\in \sbarr}[\mu_0+\gamma_0 x^2,M],\qquad \sigma(x,0)=0,\qquad\mbox{with $\gamma_0>0$, $M\in (0,h)$ and $0\leq\mu_0\le M$.}
\label{ic_sigetap}
\end{equation}
As the motion equation in divergent form~(\ref{e-m-full}) shows, the evolution for this class of data is invariant under the scaling 
$x\to x/\sqrt{\gamma_0}$, $t\to t/\sqrt{\gamma_0}$, so that different initial curvatures can simply be related to one single choice of $\gamma_0$. The densities will be taken to be $\rho_1=0.111$, $\rho_2=0.999$, corresponding to $r\simeq 0.888$, and the background constant $M=0.5$ throughout. 
As in the Airy's case in \S\ref{sezione:parab-solut}, generically when $\mu_0>0$ the elevation of the interface $\eta(0,t)$ grows till it recovers the background level $\eta(0,t)=M$ at some time $t=t_M$, after which the interface correspond to a flat, growing interval around $x=0$ where $\eta=M$, sandwiched between two symmetric simple waves (or ``wings").  Figure~\ref{quad_down} shows the initial conditions $\eta(x,0)$ and snapshots of their evolution $\eta(x,t)$ and $\sigma(x,t)$ at later times for the case $\mu_0=0.1$ and $\mu_0=0$, where the interface is close or contacts the bottom at $x=0$, respectively. Unlike their Airy's 
counterpart, for the former case the concavity of the rising interface at $x=0$ changes at some intermediate time before the flat interface region develops, while for the latter case the interface remains attached to the bottom with local positive curvature, similarly to the interface behaviour for the Airy system counterpart. In both cases, shocks develop at the ``back" of the simple waves joining  the local interface solution near $x=0$  at points where the derivative jumps.

\subsubsection{Negative interfacial curvature initial data: lower fluid extending into the upper one}
We now consider cases with {\it negative} initial curvature, where the lower layer, denser fluid approaches the top lid. We first look at an example where the Boussinesq approximation holds, and use 
this as a test for comparison with the exact solutions obtained by the map~(\ref{quattroauno}). 
Figure~\ref{bousisu} shows snapshots of the evolution from initial data obtained through this map from $\tilde{\eta}(x,0)$ and $\tilde{u}(x,0)$ in the class~(\ref{ic_paraairy}) (corresponding to the choice depicted in figure~\ref{para_airy} (b)). Here, to further test the limitations of the Boussinesq approximation, the density of the upper fluid is chosen to be $80\%$ that of the lower fluid, or $r=0.2$.  Note that at time $t=3.0$  the map of the exact solution~(\ref{numuga}) with respect to the variables ($\tilde{\eta},\tilde{u}$)
provides a close approximation to the ``core" (parabola's support in $(\tilde{\eta},\tilde{u})$ variables) region of the two-layer model, even with this relatively large departure from the Boussinesq limit $r=0$. 
\begin{figure}[t]
	\centering
	\includegraphics[scale=0.85]{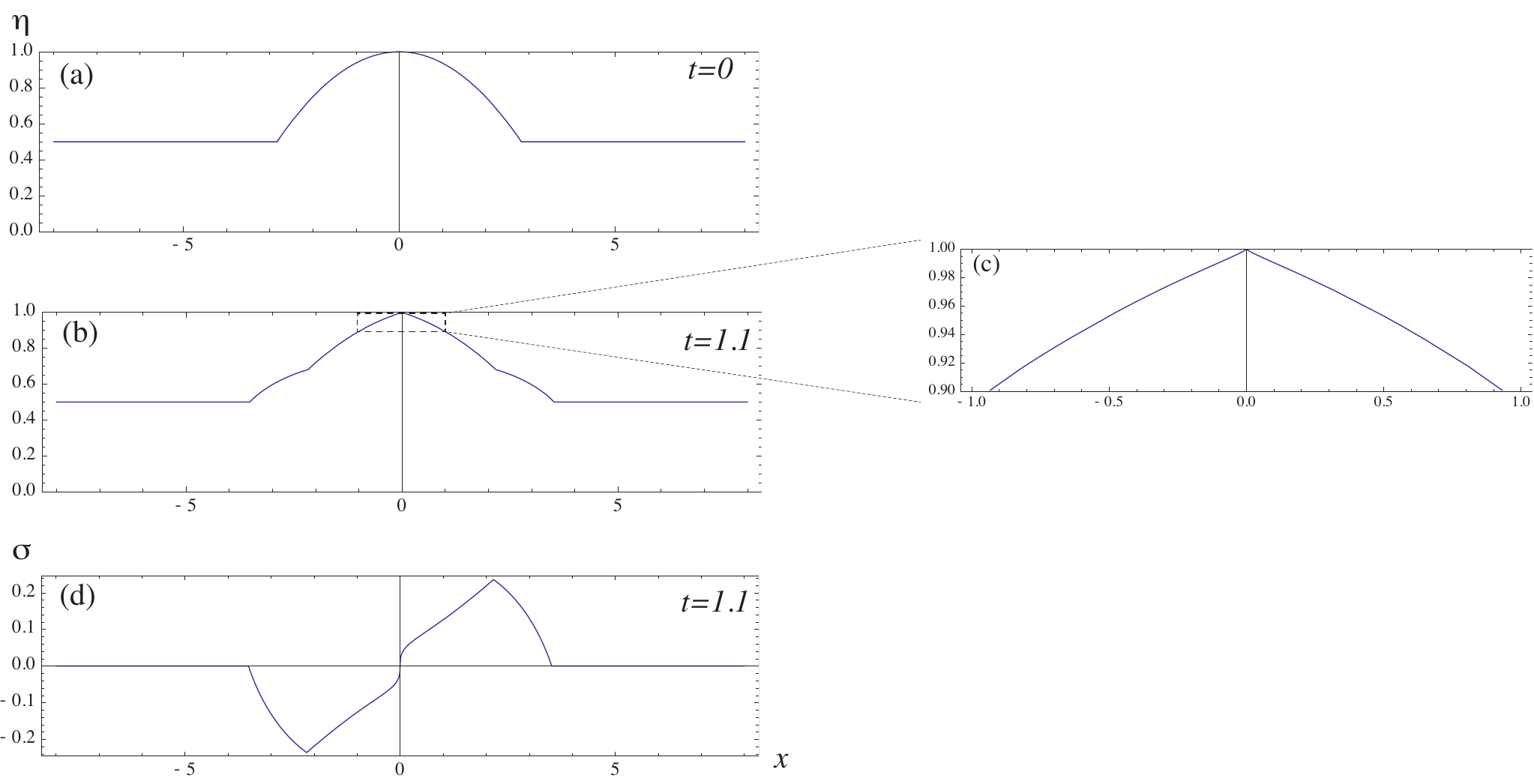}
	\caption{Snapshots of the evolution by system~(\ref{eq-disp}) with $r=0.89$: interface $\eta(x,t)$ and shear $\sigma(x,t)$ for the case of initial data~(\ref{ic_sigeta})  in contact with upper lid, $\mu_0=1$: (a) initial interface $\eta(x,0)$ ($\sigma(x,0)=0$); (b) interface at time $t=1.1$; (c) blow up of the region around the contact point at time $t=1.1$; (d) snapshot of $\sigma(x,t)$ at $t=1.1$, just before the shock forms at $x=0$. Here $|\gamma_0|=1/16$ and computation is performed with 
periodic boundary conditions in domain $x\in[-16,16]$), with only half of the domain shown.
}
		\label{stuck2layer}
\end{figure}
Even though the connection with the simple-wave wings may involve shocks, i.e., vertical (infinite) slopes at the connection points, the numerical solution tracks the 
analytical prediction until fairly close to the time $t_s$ of collapse of the parabola  (not shown here), and this time predicts the upper bound for the time of singularity formation detected numerically. 

Next, we focus on examples from the air-water limit counterpart. Unlike the Boussinesq limit, where there is up and down symmetry and the dynamics ensuing from the interface approach to the top lid mirrors the one to the bottom, 
the negative curvature cases can be tested for the critical phenomena discussed 
in~\S\ref{crit}. For the density ratio parameter $r=0.89$, the critical elevation predicted by the dispersionless model is ${\mu_0}_{\rm cr}=0.75$. We illustrate this by bracketing this value with  different choices of $\mu_0$ and tracking the time history of curvature in figure~\ref{curvdisp}.  

Simulations of the dispersionless system~(\ref{eq-disp}) further indicate that this critical phenomenon appears to manifest itself in a different form at later times in the evolution out of initial data in the class~(\ref{ic_sigeta}). In fact, when the apex of the parabola is sufficiently close to the top lid, the curvature at $x=0$ changes sign before the bump completely flattens, which makes a ``dimple" appear at the top of the bump during the slumping phase, while for initial elevations below the critical height ${\mu_0}_{\rm cr}$ the curvature maintains its (negative) sign throughout the slumping at $x=0$. This is illustrated by figure~\ref{dimple_noD}, where two snapshots at the same nondimensional time of the evolution from initial conditions $\mu_0=0.7$ and $\mu_0=0.95$ are shown side-by-side. We will see that such critical phenomena are capturing similar features in the evolution of the full Euler problem with continuous, nearly two-layer stratification. 

The last computation before leaving this section focuses on the extreme case of initial conditions~(\ref{ic_sigeta}) with negative curvature, whereby the interface is in contact with the upper plate at the parabola apex $\eta(0,0)=1$. Figure~\ref{stuck2layer} depicts the initial condition and the snapshots of the interface~$\eta(x,t)$ and shear~$\sigma(x,t)$ shortly before a shock develops in the latter field. The contact with the upper plate is not lost until this time, but, more importantly, the way the solution loses regularity is markedly different than the case when the contact point is at the bottom plate. The shock in $\sigma$ develops while the interface 
region over the lower fluid is still very broad with bounded slope; at the contact point there appears to be a change of concavity, with inflection points close to $x=0$ and a jump in derivative developing, in general agreement with the perturbative analysis of~\S\ref{ss_airwater}.

\subsection{Full Euler evolution from zero-velocity initial data}
We now present some numerical simulations of the full stratified, incompressible Euler 
equations~(\ref{Eceq}) performed using the numerical algorithm VarDen~\cite{Almgren-et-al}, with the aim of checking whether the critical phenomena isolated above can actually be detected in the evolution of the full Euler system. We typically use a square grid with $1024$ initial points along the vertical, although we have run cases with higher (up to 2048 points) and lower (down to 128 points) resolution
to assess convergence. We also used 6 levels for the adjustable mesh refinement limit. 
Lateral boundary conditions will be assumed to be periodic, while impermeable top and bottom plates (zero-vertical velocity) will be assumed. 
The code is optimized for simulations of stratified fluids and is second-order in both space and time using adaptive mesh refinement (see~\cite{Almgren-et-al} for further details about the algorithm). 
All the numerical integrations were performed on the University of North Carolina's KillDevil computing cluster.

\begin{figure}[t!]
	\centering
	\includegraphics[scale=.3]{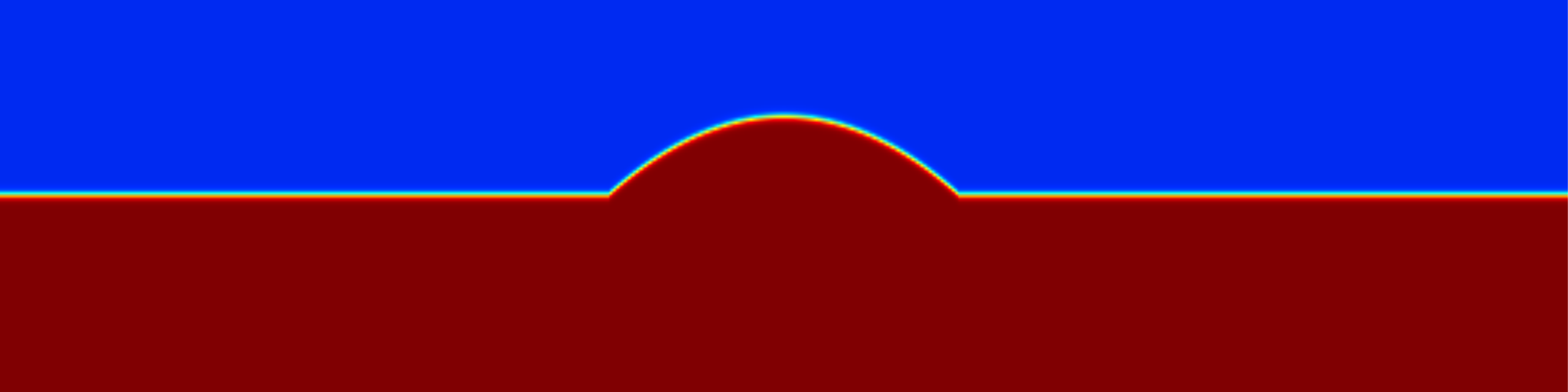}\\
	\vspace{0.5mm}
	\includegraphics[scale=.3]{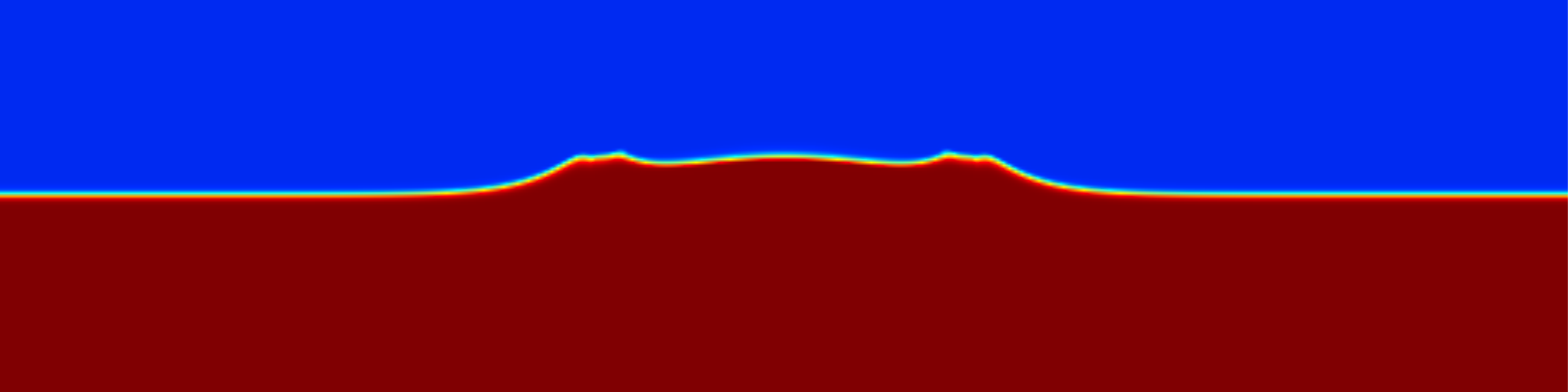}
		\caption{Initial condition and $t =2.1$ density field evolution governed by the Euler equations~(\ref{Eceq}), for the class of initial data~(\ref{inflect}) with $|\gamma_0| = 1/16$, $\mu_0 = .70$, limiting densities $\rho_1=0.111$ ($\rho_2=0.999$) for upper (lower) fluid. Image (scaled vertically by a factor $4$) restricted to the middle 16-unit wide window of the 32-unit domain, with $h=1$. The sign of the curvature of the interface (the thin transition region between two essentially uniform densities) at $x=0$ continues to be negative at this time.}
\label{zoom_nodimp}
\end{figure}

\begin{figure}[t!]
	\centering
	\includegraphics[scale=.3]{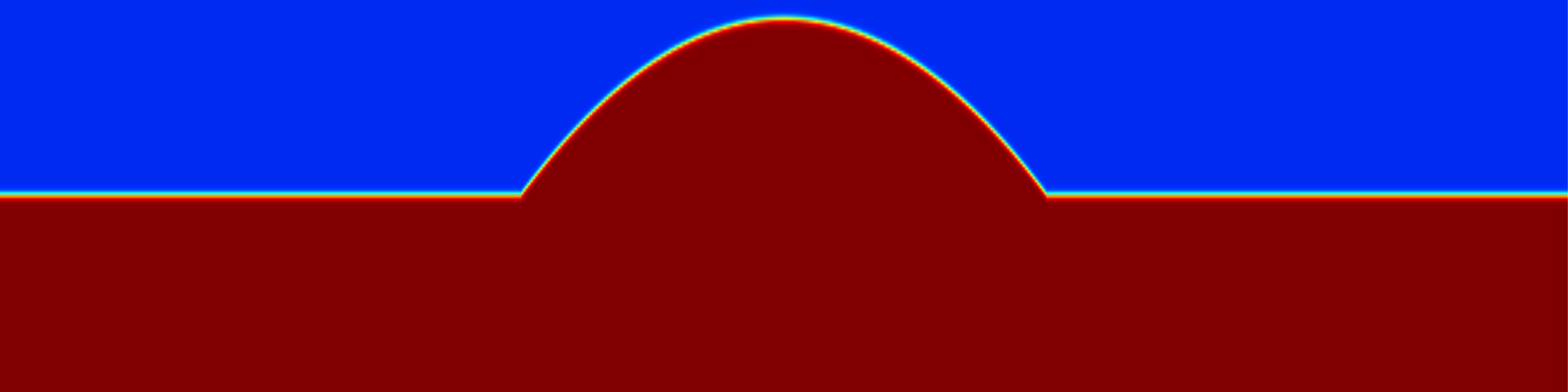}\\
	\vspace{0.5mm}
	\includegraphics[scale=.3]{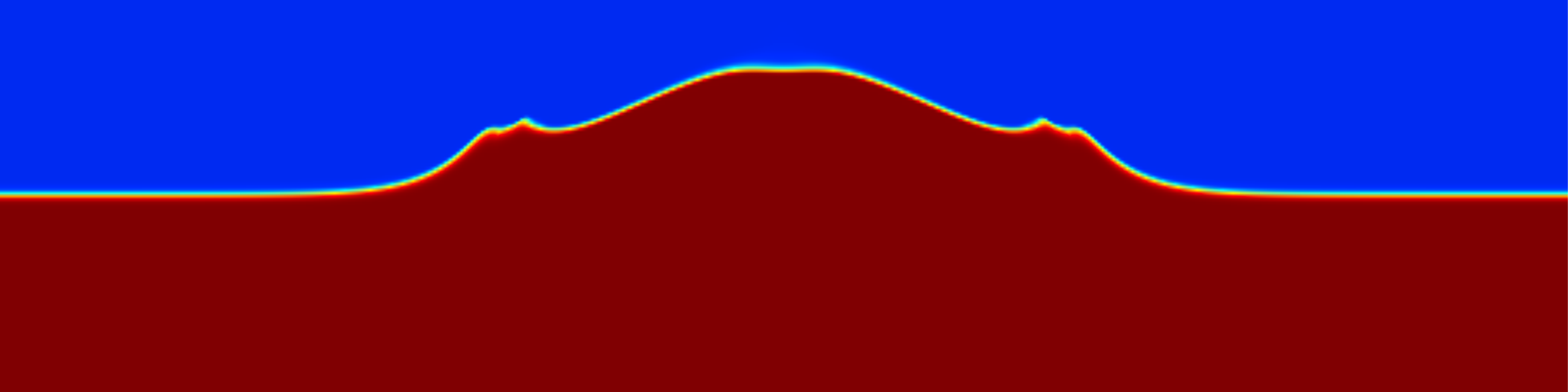}
		\caption{
		Same as figure~\ref{zoom_nodimp}, but with $\mu_0 = .95$. 
		The reversal of the sign of the curvature of the interface (the thin transition region between two essentially uniform densities) in the form of a ``dimple" in the region around $x=0$ is noticeable at this time.}
	\label{zoomed_dimp}
\end{figure}

For the simulations reported here, we again choose to work with a fairly high density contrast matching the one used for the dispersionless model,  i.e., $\rho_\text{min} = .111$ and $\rho_\text{max} = .999$. Likewise, the magnitude  of the acceleration of gravity is scaled to be unity. The initial conditions we consider are a close approximation to the class~(\ref{ic_sigeta}) for the two-layer limit, that is, the velocity is taken to be initially zero 
and the stratification is determined through the function
\begin{equation}
	y_{\text{mid}}(x) = \left\{\begin{array}{c}
	                     \max \\
	                     \min
	                     \end{array}\right\}_{x\in \sbarr}\left\{ \mu_0 + \gamma_0 x^2 , .5\right\} \, , 
	\label{inflect}
\end{equation}
(where the choice $\max$ or $\min$ is for negative or positive $\gamma_0$, respectively) 
which tracks the inflection point of a continuous density distribution
\begin{equation}
	\rho(x,y) = \rho_{\min} + \frac{\rho_{\max} - \rho_{\min}}{2}\left(1 + \tanh\left(\lambda^{-1} (y_{\text{mid}}(x) - y)\right)\right) \, . 
	\label{cont_dens}
\end{equation}
To approximate the two layer limit continuously, we typically work with the scale $\lambda=1/100$.  The evolution of the second derivative of the mean-density isoline at the center of the domain is computed by fitting a parabola centered at $x=0$, thus tracking the value of the best-fit second derivative over time. 
In all the figures reporting snapshots of direct numerical simulations, we limit ourselves to portraits of instantaneous density field,  color coded so that red corresponds to $\rho_2=0.999$ transitioning to blue at $\rho_1=0.111$. The computation domain is kept to rectangular box with aspect ratio $32\times 1$, with only the intermediate window $x\in[-8,8]$ visualized in the figures. Axis and color map are reported in the last figure,~\ref{eulerovrl}, where the overlay of the Euler density field with the interface position from the corresponding two-layer model at comparable times in the evolution is shown.

\subsubsection{Critical height and curvature evolution}
\label{crthgt}
As a first test of the Euler dynamics we look at the phenomenon identified by the two-layer model depicted in figure~\ref{dimple_noD} for the curvature evolution. Figures~\ref{zoom_nodimp} and~\ref{zoomed_dimp} depict the initial density configuration, together with a snapshot of 
the density evolution at time $t=2.1$. The cases considered correspond to the initial height 
$\mu_0=0.7$, figure~\ref{zoom_nodimp} and $\mu_0=0.95$, figure~\ref{zoomed_dimp}, respectively.  The initial curvature is kept constant at $\gamma_0=1/16$ for both cases. 
The formation of the ``dimple" for the higher initial elevation initial condition can be seen in the mean position of the density stratification, while the lower elevation case maintains the negative curvature sign around the origin. The general shape of the stratification is qualitatively similar to that of the dispersion model evolution at the same time depicted in figure~\ref{dimple_noD}, with the simple wave ``wings" being represented at this time by the full Euler dynamics, although of course some amount of smoothing by the dispersion, plus the effects of shears developing with the smooth stratification alter the fine scale structure of the solution with respect to the hyperbolic model. It is worth remarking that the full Euler system admits an upper bound on internal wave speeds for the stratification considered herein;  once this upper bound is estimated, it is easy to check  that no internal waves generated, for instance, by the discontinuous derivative of the density isolines at the parabola shoulders can reach the region around $x=0$ at the time of these snapshots, which further compels to attribute this behavior to the dispersionless response of the system.

\begin{figure}[h!]
	\centering
	\includegraphics[scale=.415]{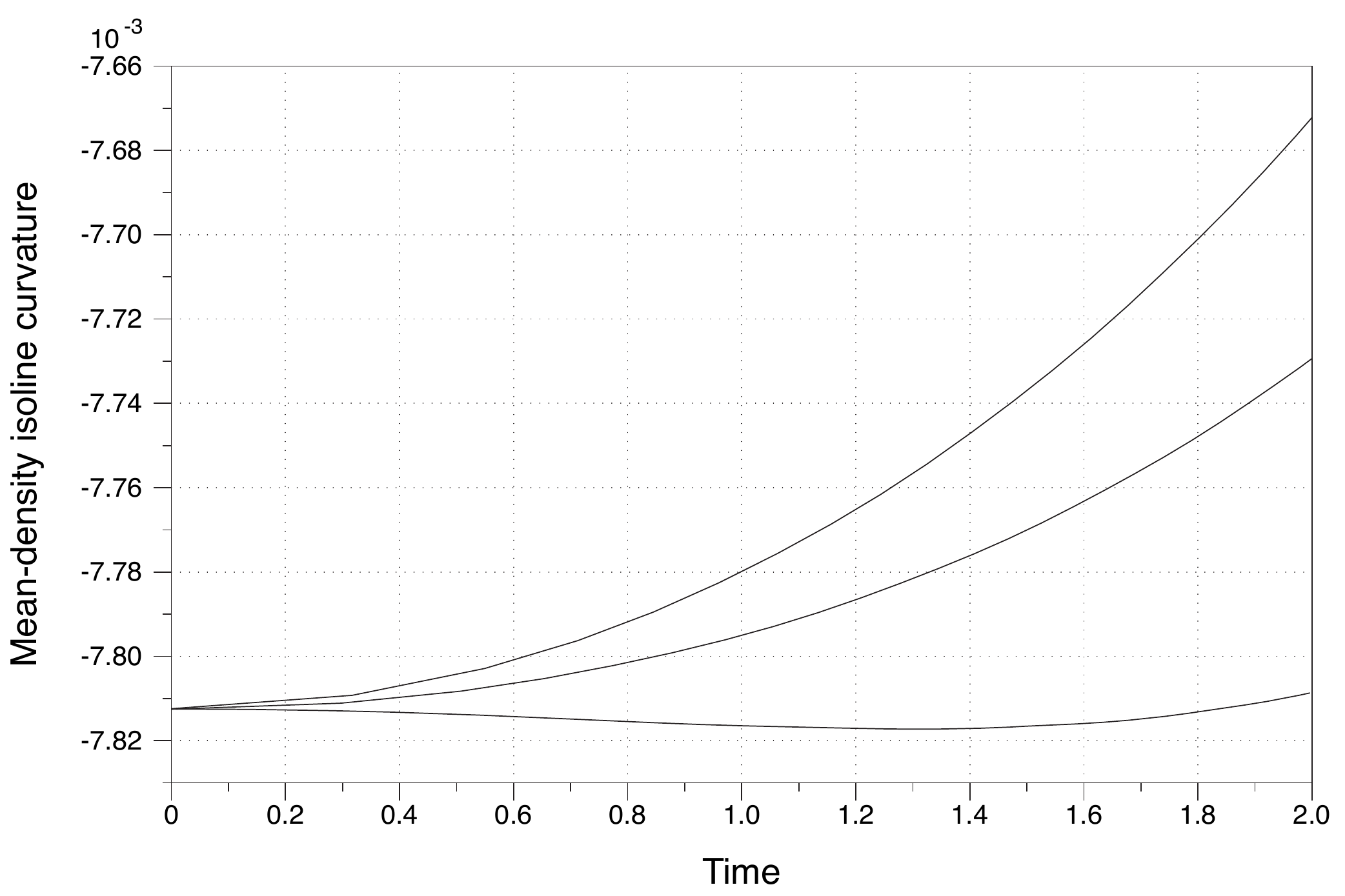} 
	\includegraphics[scale=.4]{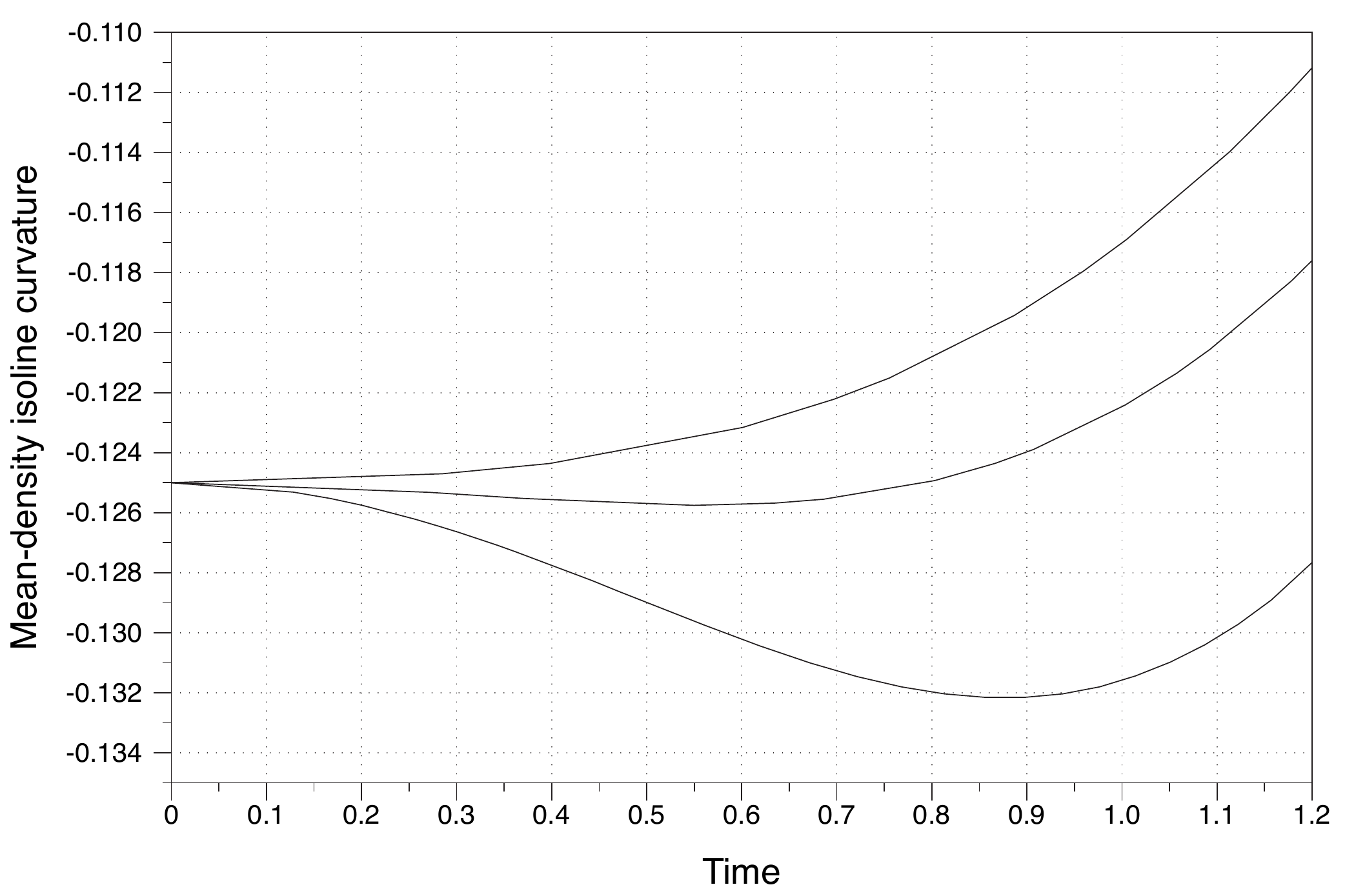}
	\caption{Time history of the mean density isoline curvature at $x=0$ corresponding to three different choices of the initial height bracketing 
	the critical elevation ${\mu_0}_{\rm cr,D}$~(\ref{mucritD}) for 
	the full Euler simulations with initial condition of class~(\ref{inflect}), with $\rho_1=0.111$, $\rho_2=.0.999$.  Left panel: $|\gamma_0| = 1/256$, and $\mu_0 = 0.7,\,  0.75, \, 0.8$; right panel: $|\gamma_0|= 1/16$, and $\mu_0 = 0.8, \, 0.85, \, 0.9$.}
	\label{zoomed}
\end{figure}

Next, we test how the Euler dynamics reflects the critical height of~\S~\ref{crit} for the curvature evolution from initial conditions in the class~(\ref{cont_dens}).   Figure~\ref{zoomed} shows the time history of curvature for two different choices of the curvature parameter 
$\gamma_0$,  $|\gamma_0|=1/256$ and 
$|\gamma_0|=1/16$ and three values of the initial heights $\mu_0$ which bracket the threshold elevation detected by monitoring how the initial curvature evolution switches from monotonic decreasing to increasing in time. Note the similarity, even at a quantitative level, with the hyperbolic two-layer model predictions of figure~\ref{curvdisp}.
\begin{figure}[h!]
	\centering
	\includegraphics[scale=.4]{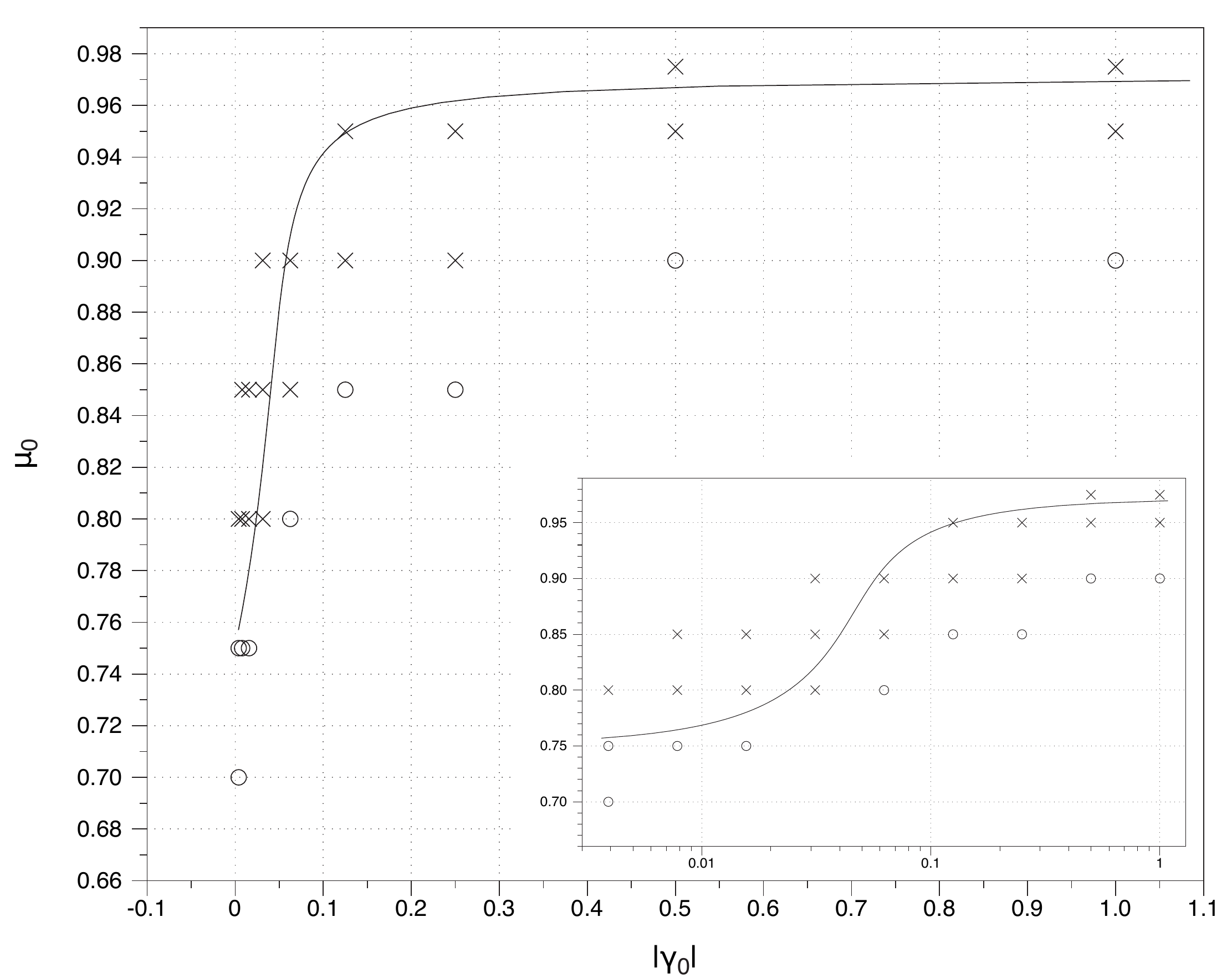}
	\caption{Critical height $\mu_0$ vs. curvature parameter $\gamma_0$, same setup as figure~\ref{zoom_nodimp}: sampling the two-parameter plane by evolving initial conditions in the class~(\ref{inflect}), with $\rho_1=0.111$, $\rho_2=0.999$. The solid curve is the dispersion correction of the strongly nonlinear model~(\ref{fourth-derivD}) fully included. This correction is asymptotically consistent with the strongly nonlinear model in the limit $|\gamma_0| h\ll1$, i.e., with the linear  relation~(\ref{mucritD}). Symbols represent the outcome of full Euler numerical simulations: ``$\small{\times}$"--curvature increase, ``o" curvature decrease. The curvature parameter $|\gamma_0|$ is set to be $2^{-n}$, $n=0,1\dots,6$. Inset: log-lin plot, to zoom in the region $|\gamma_0|$ small.
	}
	\label{bogus}
\end{figure}

\begin{figure}[h!]
	\centering
	\includegraphics[scale=.432]{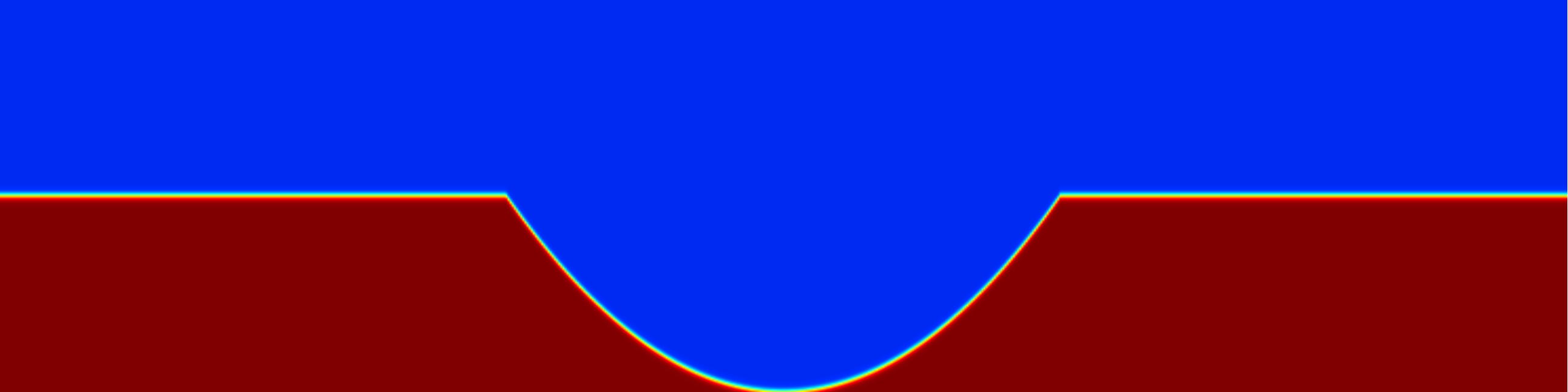}\\
	\vspace{0.2mm}
    \includegraphics[scale=.432]{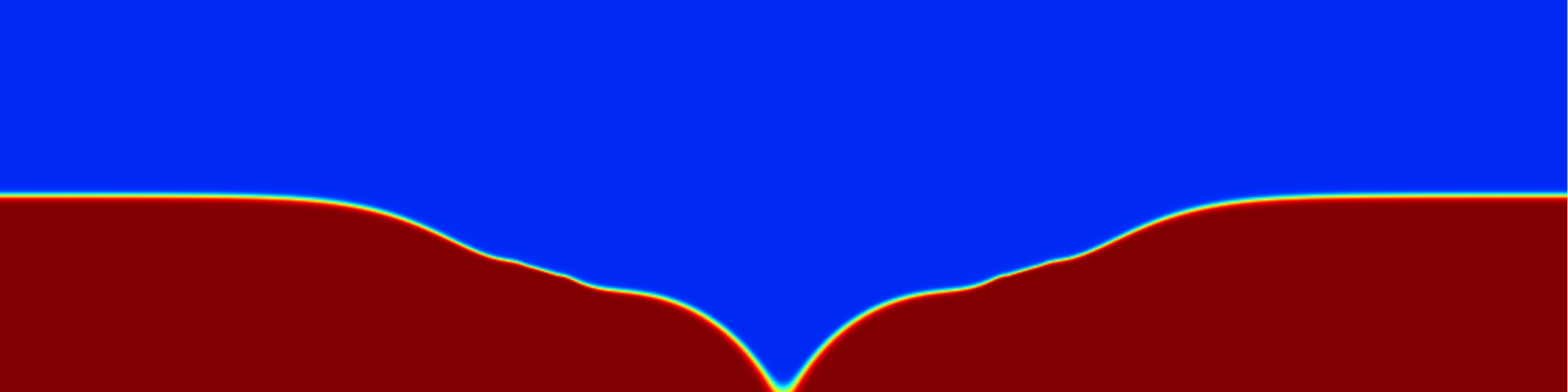}\\
	\vspace{0.2mm}
	\includegraphics[scale=.432]{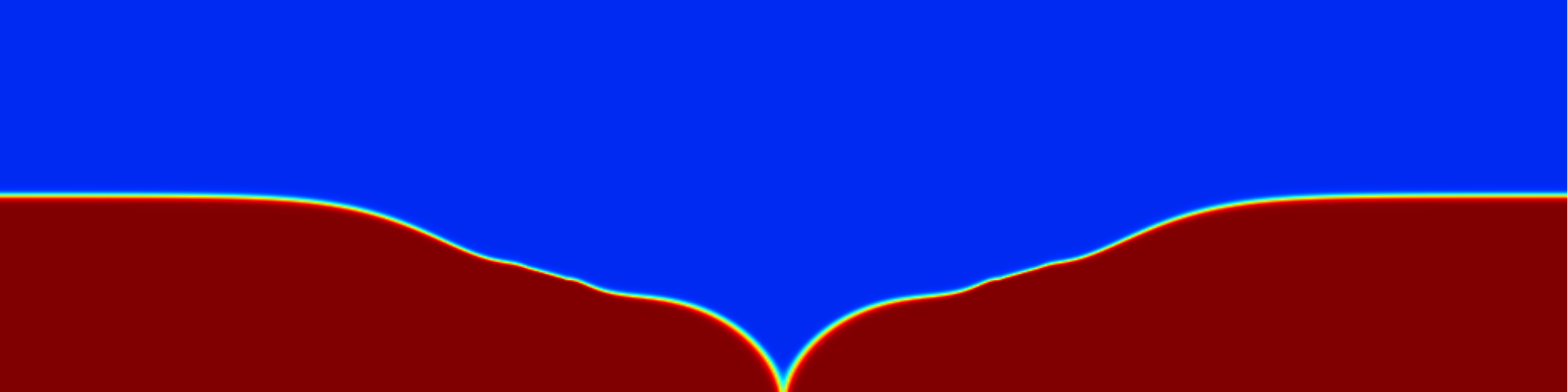}\\
	\vspace{0.2mm}
	\includegraphics[scale=.432]{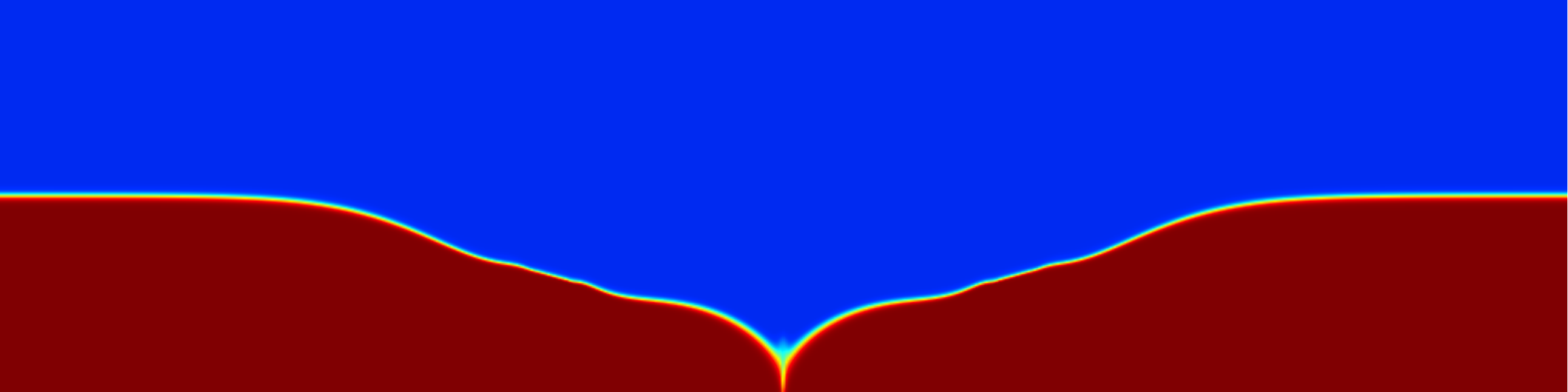}\\
	\vspace{0.2mm}
	\caption{Evolution of initial data of class~(\ref{inflect}) touching the lower plate, stratification same as in figure~\ref{zoom_nodimp}. Snapshots at times $t=0$ (top panel), $t=3.01$ (second), $t=3.25$ (third) and $t=3.41$ (fourth panel). Other parameters are the same as in figure~\ref{zoom_nodimp}. A filament of interfacial, intermediate density fluid appears to be forming and maintaing contact to the bottom at time $t=3.41$, in a manner reminiscent of the schematics depicted in figure~\ref{aw-vs-2layer-fig}(b).}
	\label{eulerpar}
\end{figure}

The data on the critical height with respect to height and curvature parameters $\mu_0$ and $\gamma_0$ from the simulations of the full Euler system are summarized in figure~\ref{bogus}. We consider an array of initial conditions, with heights $\mu_0$ and curvature parameters $\gamma_0$ bracketing the theoretical values identified in section~\ref{crit}. As can be seen from 
figure~\ref{bogus}, even when applied outside of its strictly asymptotic validity, the critical curve determined by equation~(\ref{fourth-derivD}) (lying outside and roughly paralleling  the convex hull of the circle symbols which mark the curvature-increase parametric region)  
can provide a fair estimate of the critical height even for large initial curvature parameter  $|\gamma_0| \simeq 1$. Of course, a more accurate quantitative agreement is beyond reach of this simple approximation: effects such as higher order dispersion in the model asymptotics as well as those originating from the full Euler system with smooth stratification must play a role; with this in mind, it is remarkable that the prediction is verified qualitatively at all.  We also stress that the relative steepness of the linear, asymptotically consistent estimate of the relation between 
 $\mu_{\rm cr,D}$ and $\gamma_0$ as $\gamma_0\to 0$ makes for a relatively sensitive dependence of the critical height on the curvature parameter and long-wave asymptotics, which can be seen with 
 the log-scale of the inset of 
 figure~\ref{bogus}. This shows that dispersive effects can become relatively important at short times, even before gradient catastrophes have begun to form, in this case in the smooth quadratic region near the parabolic apex. This early-time effects of dispersion for smooth initial data are reminiscent of those quantified by pressure imbalances in stratified fluids investigated in~\cite{CCFOP1,CCFOP2}. We remark that  each grid point in the two dimensional parametric search summarized by figure~\ref{bogus} represents a relatively expensive numerical computation  Euler equations, so that further refinement of the parametric search would require a more intensive investment of computation resources that is best left to a separate study.

\subsubsection{Singular behavior at the bottom and top plates}
In the same vein as the previous numerical simulations, it is interesting to see how the full Euler system evolves initial conditions for which the initial mean density isoline contacts the bottom or top plate, and observe the distinct behavior in the evolution of initial configurations brought forth by the two-layer model. In fact, as seen in figure~\ref{quad_down}(b), the bottom contact case evolves similarly to the Airy's solution,  with the interface remaining in touch with the plate, and the curvature remaining finite until a `global' shock develops in the interface profile at the contact point. In contrast, for the case of the interface touching the top plate, while the persistence of contact until loss of regularity is also observed, this occurs with the formation of a corner (jump in derivative) in the interface profile at the contact point  (accompanied by a standard shock -- derivative going to infinity at an inflection point --  at the origin for the corresponding velocity shear field~$\sigma$), see 
figure~\ref{stuck2layer}.

These trends can in fact be detected by the full Euler simulations.  
Figure~\ref{eulerpar} shows snapshots of the evolution of the density field for the bottom contact case. 
As can be seen, density isolines remain attached to the bottom plate, with curvature staying finite for the isolines near the contact region, for a finite time of the evolution, with the interfacial region bearing strong similarities with the two-layer counterpart. The Euler simulation is not limited by the shock formation of its two-layer counterpart, and offers a glimpse of how the evolution continues past the time of loss of regularity with the formation of an upward moving jet of higher density fluid.

\begin{figure}[t!]
	\centering
    \includegraphics[scale=.3]{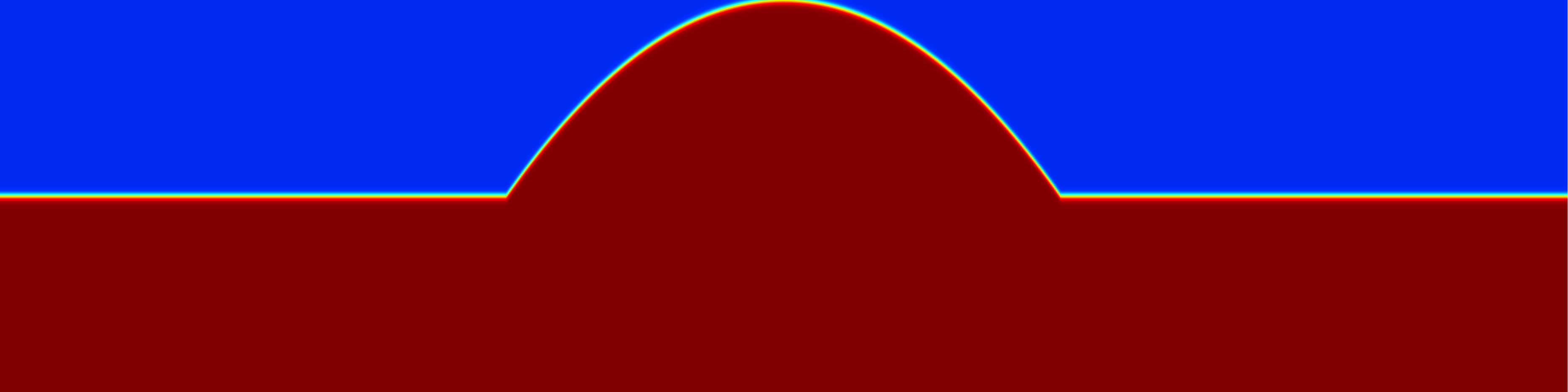}\\
	\vspace{0.2mm}
    \includegraphics[scale=.3]{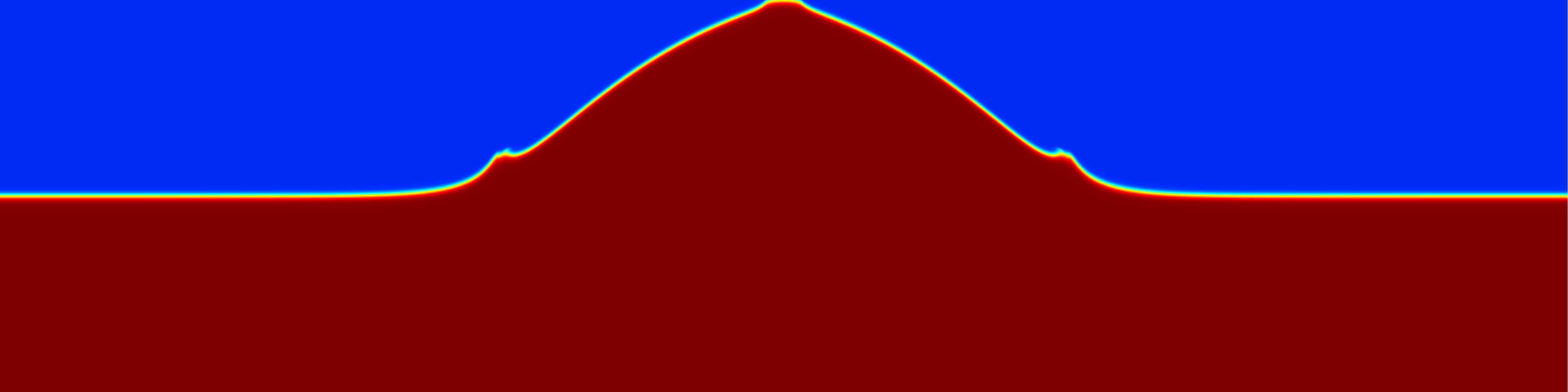}\\
	\vspace{0.2mm}
	\includegraphics[scale=.296]{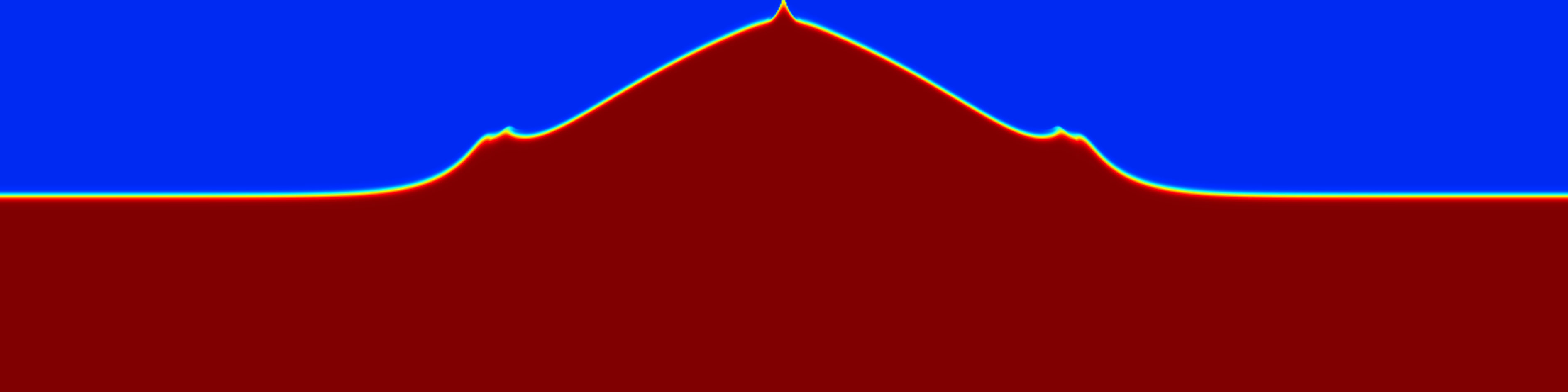}
	\caption{Evolution of initial data touching the upper plate. Snapshots at times $t=0$ (top), $t=1.1$ (middle) and $t=1.7$ (bottom). Other parameters are the same as in figure~\ref{zoom_nodimp}. The formation of the cusp-like shape of the interface near the top plate can be clearly seen at the time $t=1.7$ .}
	\label{eulercorn}
\end{figure}

Snapshots of the density field evolution for the case of top contact are depicted in figure~\ref{eulercorn}. The loss of regularity occurs much earlier than the bottom touching case. Density isolines remain attached to the top lid for times past that of the last panel in the figure, $t=1.7$, but eventually lack of resolution introduces artificial effects that can not truly represent the PDE solution of the Euler system. Shear instabilities may also develop at later times in the course of the evolution, however the ``stem" region attaching the lower layer fluid to the top plate seemingly persists up to those times.  

\begin{figure}[h!]
\hspace{3cm}
 \begin{minipage}{.5\textwidth}
	\begin{center}
	 \hspace{2.5cm}(a) \vspace*{-3mm}\includegraphics[scale=.280]{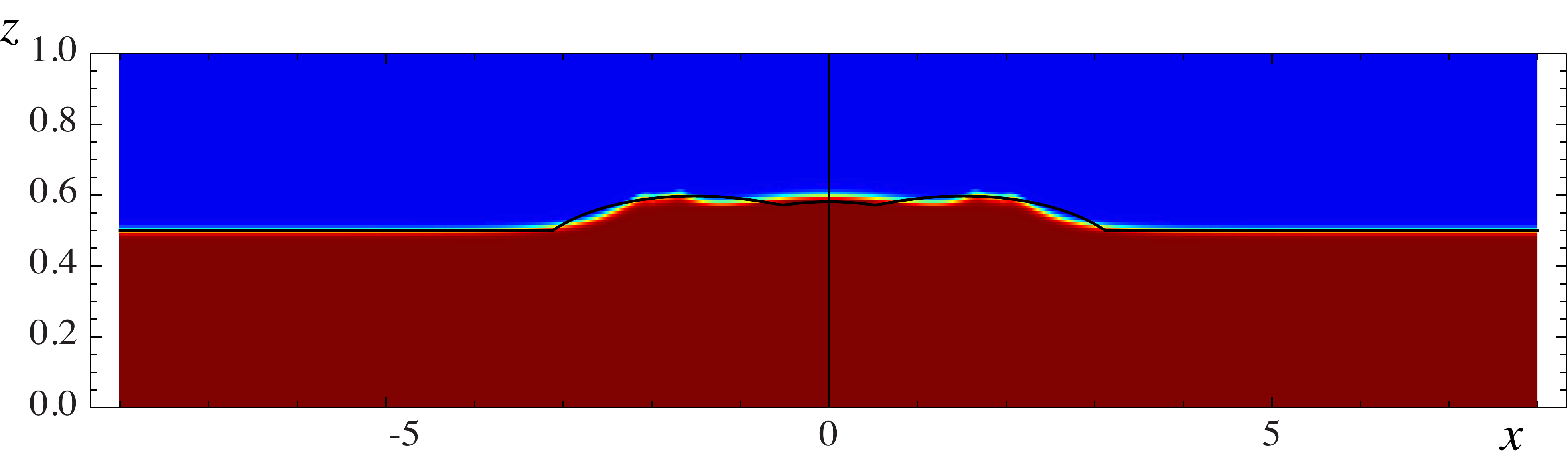}
	\end{center}
	
	\begin{center}
     \hspace{2.5cm}(b) \vspace*{-3mm}\includegraphics[scale=.280]{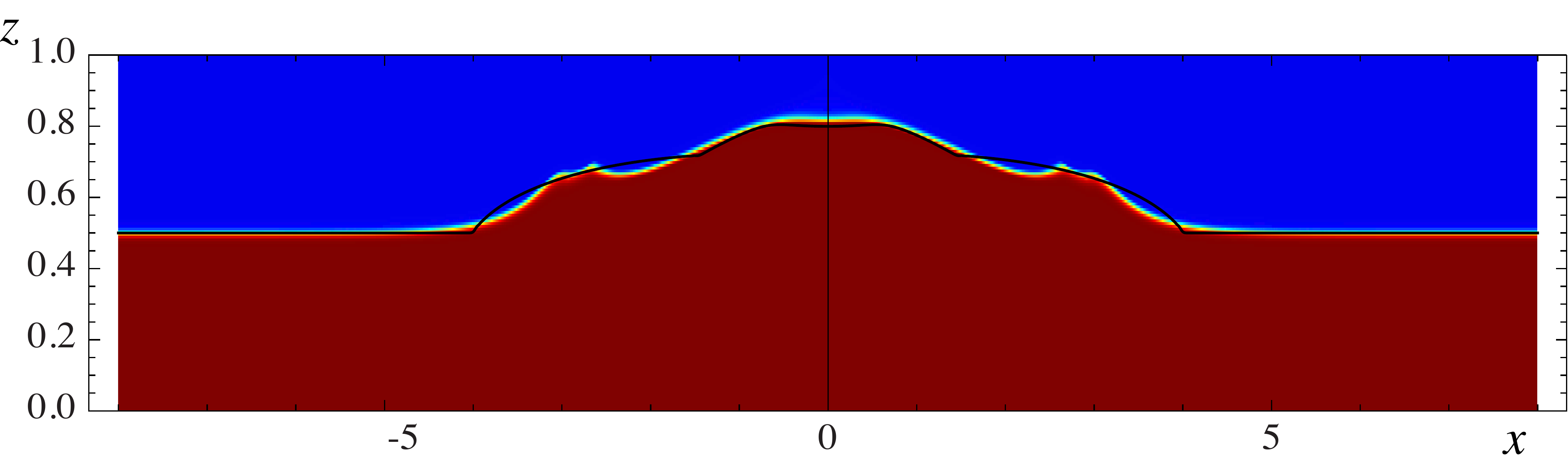} 
     \end{center}
     
     \begin{center}
	\hspace{2.5cm}(c) \vspace*{-3mm}\includegraphics[scale=.388]{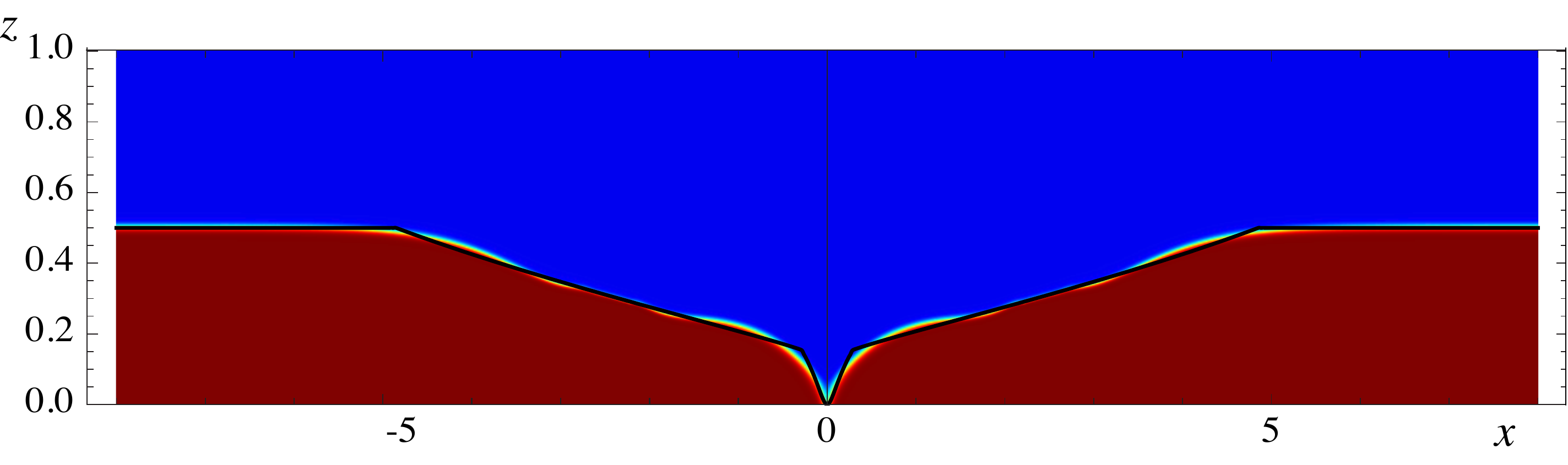}
	\end{center}

	\begin{center}
	\hspace{2.5cm}(d) \includegraphics[scale=.270]{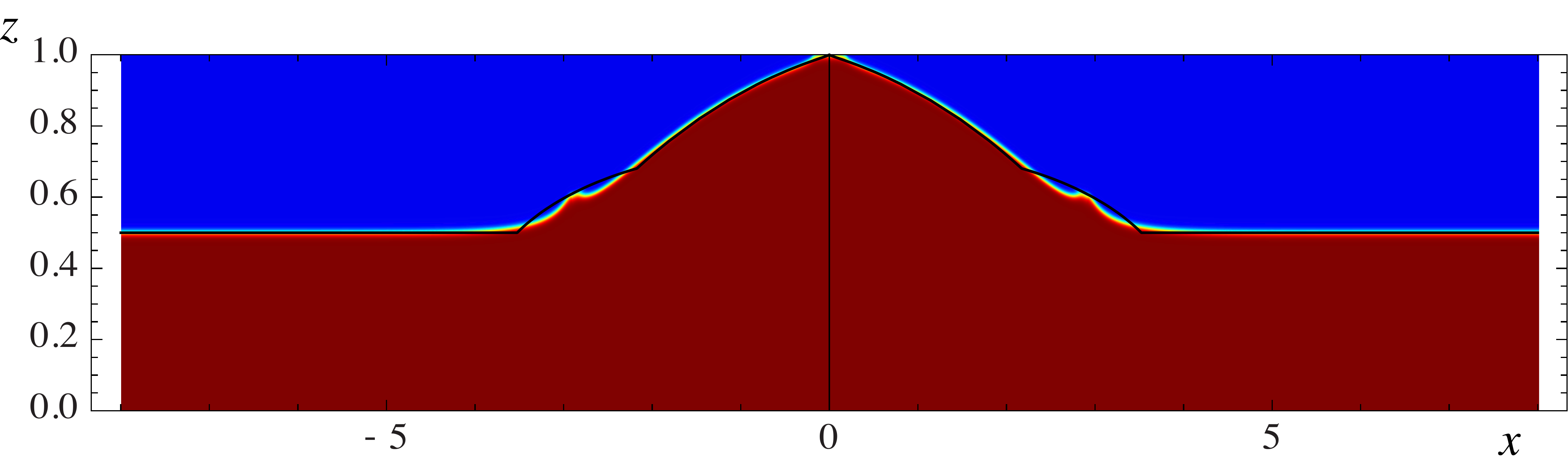}
	\end{center}
 \end{minipage}
 \begin{minipage}{.5\textwidth}
	 \hspace{3.cm}\includegraphics[scale=1]{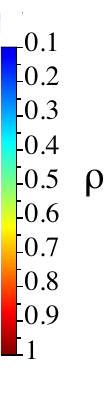} 
 \end{minipage}
 \caption{Comparison of solution profiles (black curves) from two-layer model overlaid onto Euler direct numerical simulations density-plots, for evolution of initial data of class~(\ref{inflect}). Stratification same as in figure~\ref{zoom_nodimp}, colormap depicted in sidebar. Each panel takes one snapshot of the evolution depicted 
in figures~\ref{zoom_nodimp}, \ref{zoomed_dimp}, \ref{eulerpar}, and \ref{eulercorn}, at times, respectively, (a) $t=2.1$, (b) $t=2.1$, (c) $t=3.25$ and (d) $t=1.1$, with superposed profiles from their two-layer model counterparts in, respectively, figures~\ref{quad_down}, \ref{dimple_noD}, and \ref{stuck2layer}. Other parameters as in these (latter) figures. }
	\label{eulerovrl}
\end{figure}

\section{Discussion and conclusions}

\label{sezione:lagrandefinale}
In this work we investigated the dynamic implications of interactions of stratified fluids with boundaries. These can be either of confining, rigid-wall type or of internal nature such as interfacial density jumps. We first looked at simple kinematic  information  that can be extracted from the full Euler equations, both for homogeneous fluids where a free surface contacts a bottom in 
``dry spot" configuration, and for a two-layer stratified fluid whose interface contacts either the bottom or top plate. In particular, we clarified in this context the interpretation of the typical characterization of boundary points as remaining on the boundary for all times. Next, working with limiting models, such as those that derive from two-fluid setups by nondispersive, long-wave asymptotics, including their limits to single homogeneous fluid with a free surface, we showed that detachment of an interface from smooth contacts with a rigid boundary can occur via a singularity development in the form of shocks. In particular, the limit of vanishing upper fluid density reveals substantial differences between the evolution of top and bottom contact points, whereby 
at top contact points shocks may develop immediately in this limit.  
Moreover, our study identified several critical phenomena with zero velocity initial data, and symmetric interfacial profiles, as well as the non-negligible role played even at short times by dispersive wave effects in some cases.  
One notable element that emerges from this analysis is a kind of upper lid ``suction," or tendency, above a critical threshold determined by the density ratio between the two fluids, for the peak region of a slumping bulge of heavier fluid to be delayed in its downward motion with respect to nearby lower fluid. This causes the curvature at the peak to increase temporarily, before reversing sign in its downward acceleration.

The critical cases, besides their inherent interest, can also serve as a testbed for numerical simulations, including direct ones for the parent Euler equations with smooth stratification (albeit with a sharp enough transition between two nearly uniform densities to approximate a two-fluid system). 
Performing these computations, with extensive testing of their numerical accuracy 
(see also~\cite{roxana,claudio}),  demonstrated a remarkable robustness of the nondispersive model predictions, at least for the class of initial conditions and time scales of evolution  we have focussed on, at both qualitative and, with few exceptions, quantitative levels. This is borne out of a comparison between interface profiles obtained through the numerical solutions of the two-layer model~(\ref{eq-disp}) and the direct numerical simulation of their Euler counterpart by overlaying the profile on the late time snapshots from figures~\ref{zoom_nodimp}, \ref{zoomed_dimp}, \ref{eulerpar}, \ref{eulercorn}, 
as reported in figure~\ref{eulerovrl}. A notable example that emerges from this comparison is the prediction of ``wing" structures (simple waves for the hyperbolic models) and their locations. These wings connect the core symmetric state centered at the origin with the constant background state: while only qualitatively similar, all the simulations with the full Euler system displayed the emergence of these features, whose existence does not seem to be immediately predictable from Euler dynamics alone. On the other hand, the
evolution of connection points, figure~\ref{eulercorn}, where the pycnoclines have discontinuous first derivatives show incipient shear instability which cannot be associated with hyperbolic-elliptic transition in the two-fluid model. 

Among the results above summarized, our work has identified possible scenarios in the evolution of initial data where internal and rigid confining boundary come into contact. In particular, filling a dry hole in a body of water may occur with a singularity of global kind, whereby an infinite derivative develops over a whole segment simultaneously, as opposed to the standard case of pointwise derivative divergence at first. Conversely, with a non-negligible upper fluid density, the same initial data develop a standard shock with an interface that closes the gap in a zipper-like fashion (though it remains to be seen whether a left over filament of upper fluid, while becoming increasingly thinner, nonetheless persists at all times, as possibly suggested by the numerical simulations). It is worth remarking, at this point, that so long as viscosity is neglected as in the parent Euler equations, the symmetry of the initial conditions we have considered can be viewed as enforcing the boundary condition of a {\it vertical} wall at $x=0$, since antisymmetry of the horizontal velocity
is equivalent to an impermeable boundary condition at the origin. Thus, we see that, according to the model, the impinging mass of water smashing into the wall produces a jet whose apex remains lower than the background state if the initial condition has a dry point at the wall, which is recovered only at large times after collision.

Our study opens future avenues of investigation which we are currently pursuing. In a separate paper, we will report on closed form expressions for the simple-wave wings, and also study how the evolution of the hyperbolic models can be continued after shocks develop.  Further, the effects we have studied can be expected to be altered by the orientation of the confining boundaries, for instance when the bottom is a sloping ``beach." The classical analysis in~\cite{Carrier} provides a hint of what could happen in this case, for a single-layer fluid, and would have to be extended to treat more general configurations and stratifications.  
These are challenging questions which we plan to address in future work. 

\section*{Acknowledgments}
We thank R.\ Colombo and M. Garavello for discussions and useful comments on the theory of quasilinear PDE's. 
R.C. and G.O. thank S.L. Gavrilyuk for bringing to their attention, after a first draft of this work had been submitted for publication, reference \cite{Ovs} during the Summer school
``Dispersive hydrodynamics and oceanography: from experiments to theory'' 27 August - 1 September 2017, Les Houches (France).
Support by NSF grants 
DMS-0908423, DMS-1009750, DMS-1517879, RTG DMS-0943851, CMG ARC-1025523, ONR grants N00014-18-1-2490, DURIP N00014-12-1-0749,  ERC grant H2020-MSCA-RISE-2017 PROJECT No.\ 778010 IPaDEGAN,  and the auspices of the GNFM Section of INdAM are all gratefully acknowledged.  
R.C. and M.P. thank  the {Dipartimento di Matematica e Applicazioni\/} 
of Universit\`a Milano-Bicocca  for its hospitality. 
In addition, the Istituto Nazionale di Alta Matematica (INdAM) and the program ``Singularities and Waves in Incompressible Fluids" in the Spring of 2017 at the Institute for Computational and Experimental Research in Mathematics (ICERM),  supported by the National Science Foundation under Grant DMS-1439786, are gratefully acknowledged
for hosting the visits of R.C. (INdAM, Summer 16) and R.C. \& C.T. (ICERM, Spring 17) while some of this work was carried~out. {\color{black} Last, but 
not least, we would like to thank the anonymous referees whose attentive reading of the manuscript greatly helped improve it.}

\appendix
\setcounter{equation}{0}
\renewcommand{\theequation}{A.\arabic{equation}}
\section*{Appendix A: The limiting case of air-water system}
\label{appendice:lim-Airy}
We have seen in section \ref{sezione:two-fluid} that the Hamiltonian density (\ref{hami-disp-adim})
\begin{equation}
H=\frac12
\left( \frac{\eta(1-\eta)\sigma^2}{1-r\eta} + \eta^2 \right)
\end{equation}
gives rise for $r=0$ to the 
Boussinesq two-layer model smoothly in $r$ (i.e., the Hamiltonian is continuous in $r$ at $r=0$).
The opposite limit $r\to 1$ (that is, the air-water system) is more subtle. 
In fact, as long as the interface 
does not get into contact with the top boundary $\eta=1$ of the channel, expression (\ref{hami-disp-adim}) coincides with the classical Airy 
Hamiltonian density
\begin{equation}
  H_{\mbox{\tiny Airy}}=\frac{\eta \sigma^2}{2} + \frac{\eta^2}{2}.
\end{equation}
However, the limit $r \to 1$ preserves the memory of the upper lid when the interface is close to the top boundary. Indeed, expanding (\ref{hami-disp-adim}) with respect to small $\rho_r=1-r$ yields 
\begin{equation}
   H=H_{\mbox{\tiny Airy}}-\frac{\eta^2 \sigma^2}{2(1-\eta)}\rho_r+ o(\rho_r).
\end{equation}
For every fixed value $\rho_r\ll 1$, if the interface gets sufficiently close to the upper lid, 
the correction term 
of the Hamiltonian density can no longer be assumed to be smaller than the leading order term. In other words, an asymptotic expansion in the small parameter $\rho_r=1-r$ fails in a neighborhood of~$\eta =1$.

\begin{figure}
\centering
\includegraphics[scale=.8]{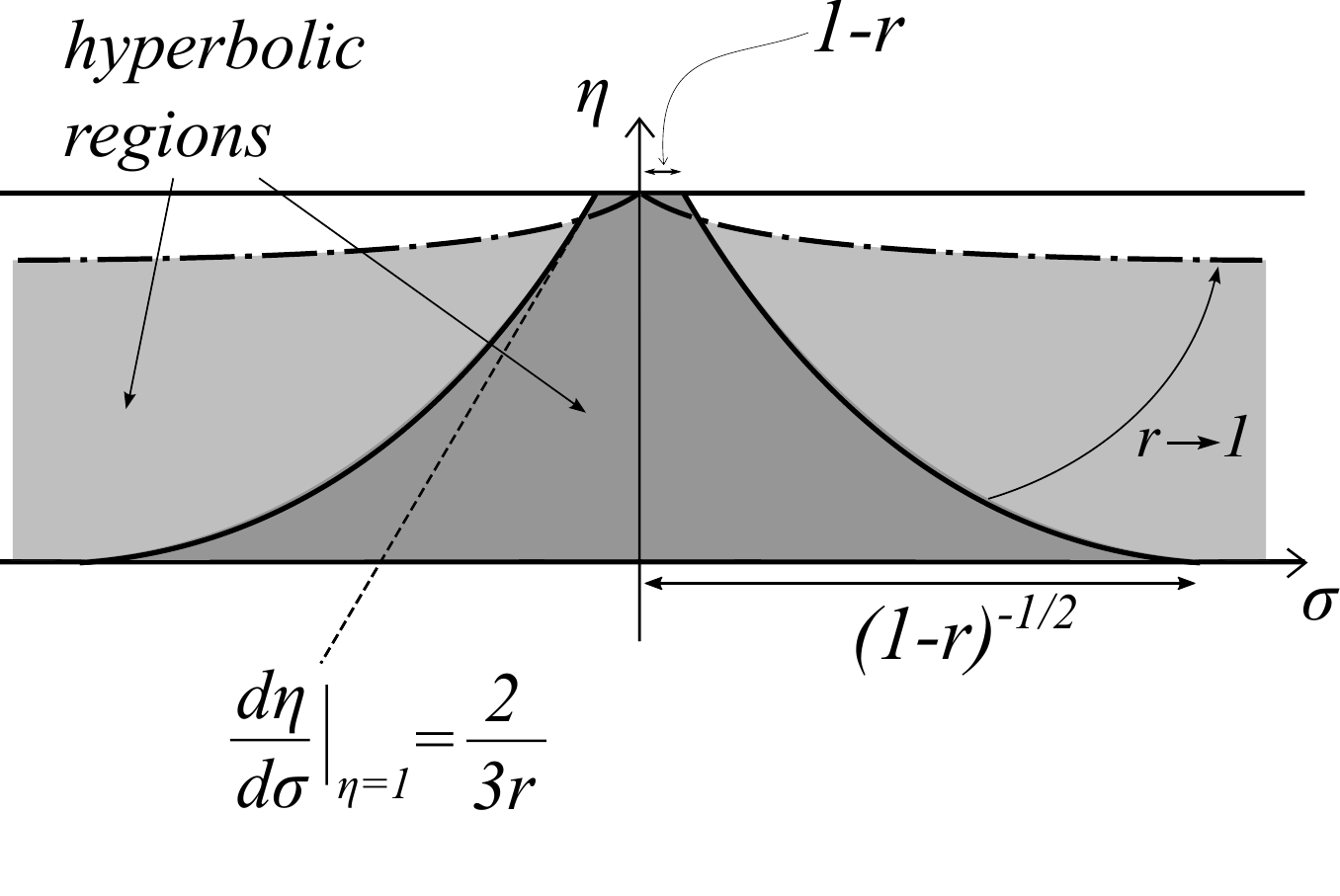}
\caption{Hyperbolic regions for the two-fluid model~(\ref{eq-disp}) as the density ratio $r$ varies, between the limiting cases of $r=0$ (Boussinesq) and $r=1$ (air-water). In this limit, for $ \sigma = o((1-r)^{-1/2})$ and $1-r/\sigma = o(1)$, the hyperbolic region 
tends to fill the whole strip $\eta\in[0,1]$ of hodograph plane $(\sigma,\eta)$, and the elliptic regions shrink to narrow slivers below $\eta=1$ for all finite values of $\sigma$.}
\label{aaa}
\end{figure}

Let us now consider the hyperbolicity region of the Hamiltonian system generated by~(\ref{hami-disp-adim}), i.e., the domain in the $(\sigma,\eta)$-plane where the characteristic velocities $\lambda_\pm$ 
are real and distinct (see figure~\ref{aaa}). It is easy to show that its boundary (the so-called transition line) is given by $H_{\sigma\sigma}=0$ 
and $H_{\eta\eta}=0$. The first equation gives the physical boundaries of the system: $\eta=0$ and $\eta=1$. The second equation is  
\begin{equation}
\frac{(1-r) \sigma^2}{(1-r\eta)^{3}}=  1\qquad\Rightarrow\qquad  \sigma=\pm\sqrt{\frac{(1-r\eta)^{3}}{1-r}}.
\end{equation}
It follows that 
\begin{equation}
\label{hyperbolic-corners}
\sigma(0)=\pm\frac{1}{\sqrt{1-r}},\qquad \sigma(1)=\pm (1-r),\qquad \frac{\d\eta}{\d\sigma}\Big|_{\eta=0}=\mp\frac{2}{3r}\sqrt{1-r},
\qquad \frac{\d\eta}{\d\sigma}\Big|_{\eta=1}=\mp\frac{2}{3r}.
\end{equation}
When $r\to 1$, the hyperbolic domain fills the whole configuration space $\mathbb{R}\times [0,1]$. However, the transition line $H_{\eta\eta}=0$ 
does not merge smoothly with the horizontal line $\eta=1$, since the last derivative in (\ref{hyperbolic-corners}) tends to $\mp{2}/{3}$. As a consequence, if the distance of 
some point of the interface is smaller than the infinitesimal $1-r$, the system 
will experience a hyperbolic-elliptic transition if the weighted vorticity $\sigma$ is larger than an infinitesimal of the same order. 

\section*{Appendix B: An example of simple-wave behavior for a smooth interface}
\label{appendice:smooth-behav-sw}
\setcounter{equation}{0}
\renewcommand{\theequation}{B.\arabic{equation}}
In this appendix we provide an explicit example of evolution from an analytic initial interface in the simple-wave setting for the Airy system. 

We consider the following family of interfaces at $t=0$,
\begin{equation}
 \eta_0 (x)= (\tanh x)^{2n} H(-x)  +K ,
\end{equation}
where $K$ is a nonnegative parameter, $n$ is a positive integer and $H(x)$ is the Heaviside function. 
For a simple wave, 
by choosing the minus sign in (\ref{simplewave}), the velocity field is 
\begin{equation}
 u_0 (x)= 2 \sqrt{(\tanh x)^{2n}H(-x)+ K} + c.  
\end{equation}
The Riemann constant value $c=-2\sqrt{1+K}$ is fixed by the asymptotic zero-value of the velocity at $x \to - \infty$. 
The simple wave equation is
\begin{equation}
 \eta_t 
 +\left(3 \sqrt{\eta} - 2\sqrt{1+K}\right) \eta_x=0,
\end{equation}
so that its hodograph solution is
\begin{equation}
 \label{hodo-nK}
\left( \tanh(x-(3\sqrt{\eta}-2\sqrt{1+K})t)\right)^{2n} H(-x+(3\sqrt{\eta}-2\sqrt{1+K})t ) +K=\eta\,  .
\end{equation}
In the region $x > (3 \sqrt{K} - 2\sqrt{1+K})t$ the solution remains constant, 
\begin{equation}
 \eta=K\, , \qquad u=2 \sqrt{K}-2\sqrt{1+K} \, ,
\end{equation}
while if $x < (3 \sqrt{K} - 2\sqrt{1+K})t$ the solution is implicitly given by 
\begin{equation}
   x=  (3 \sqrt{\eta} - 2\sqrt{1+K})t - \mathrm{arctanh}(\eta-K)^{\frac{1}{2n}}\, .  
\end{equation}
The velocity at the interface minimum is $u_0(0)=2 \sqrt{K}-2\sqrt{1+K}<0$ independently of $n$, 
therefore a shock will always form  there. The shock curve (\ref{catas-curve}) 
is given by
\begin{equation}
 \label{sln1simple}
 \left\{
 \begin{array}{l}
  x(\eta)=\displaystyle \frac{\left(3 \eta - 2 \sqrt{\eta } \sqrt{1+K}\right) (\eta -K)^{\frac{1}{2 n}-1}}{3 n \left(1-(\eta
   -K)^{\frac{1}{n}}\right)}-\mathrm{arctanh}(\eta -K)^{\frac{1}{2 n}} \\ 
   \noalign{\smallskip}
  t(\eta)= \displaystyle \frac{\sqrt{\eta } (\eta -K)^{\frac{1}{2 n}-1}}{3 n\left(1- (\eta -K)^{\frac{1}{n}}\right)}
 \end{array}
 \right. 
\end{equation}
As recalled in section~\ref{sezione:simpler-catastrophe} the shock value $\eta_c$ is 
a solution of ${\d t}/{\d \eta}=0$, so that 
\begin{equation}
 n (\eta_c +K) \left((\eta_c -K)^{\frac{1}{n}}-1\right)+\eta_c 
   \left((\eta_c -K)^{\frac{1}{n}}+1\right)=0.
\end{equation}
The solution of this equation obviously depends on the values of $K$ and $n$.  
For example, when  
$n=1$ 
we have
\begin{equation}
 \eta_c= \frac{1}{4} (K + \sqrt{9 K^2 + 8 K})
\end{equation}
and
\begin{equation}
 \begin{split}
& x_c= \frac{6 K+6 \sqrt{9 K^2+8 K} -8 \sqrt{K+1} \sqrt{K+\sqrt{9 K^2+8K}
   }}{3 \left(3 K+4-\sqrt{9 K^2+8K}\right) \sqrt{ \sqrt{9
   K^2+8K}-3 K}}-\mathrm{arctanh}\left(\frac{1}{2} \sqrt{ \sqrt{9 K^2+8K}-3 K}\right)\\
& t_c=\frac{4 \sqrt{K+\sqrt{9 K^2+8K} }}{3 \left(3 K+4-\sqrt{9 K^2+8K}\right) \sqrt{ \sqrt{9 K^2+8K}-3 K}} \, . 
\end{split}
   \end{equation}
We remark that  even if
\begin{equation}
\lim_{K \to \infty} (\eta_c-K) = \frac{1}{3},
\end{equation}
both $x_c$ and $t_c$ diverge when $K \to \infty$.

If $K=0$, the shock value is 
\begin{equation}
 \eta_c=\left(\frac{n-1}{n+1}\right)^n \, , 
\end{equation}
and therefore it is at the bottom of the channel only if $n=1$, see below.
The other critical values are
\begin{equation}
 \begin{split}
 & x_c=\frac1{6 n}(n-1)^{\frac{1-n}{2}} (n+1)^{\frac{1+n}{2}} \left(3 \left(\frac{n-1}{n+1}\right)^{\frac{n}2}-2\right)
 -\mathrm{arctanh}\left(\sqrt{\frac{n-1}{n+1}}\right) \\
 & t_c= \frac1{6 n}(n-1)^{\frac{1-n}{2}} (n+1)^{\frac{1+n}{2}}
 \end{split}
\end{equation}

Consider now the case when $K=0$ and $n=1$: 
the initial velocity 
\begin{equation}
u_0 (x)= -2H(-x)\tanh x  -2
\label{tanh_shape} 
\end{equation} is not differentiable at $x=0$, and the critical values of the 
field $\eta$ is $\eta_c=0$.
Hence the shock forms at the boundary of the domain, and one can check that $x_c=-2/3$ and $t_c=1/3$. 
We can conclude that for this example the smoothing does not introduce qualitatively new phenomena on the position of the shock formation with respect to the non-smooth cases considered in section~\ref{sezione:simpler-catastrophe}
for simple-waves solutions of the Airy equations.


\end{document}